\documentclass[english, 10pt]{book}
\usepackage{a4}
\usepackage{amsmath}
\usepackage{graphicx}
\usepackage{rotating}
\usepackage{epsfig}
\usepackage{xspace}
\begin{document}
\ifx\href\undefined\else\hypersetup{linktocpage=true}\fi
\newcommand{\lsim}   {\mathrel{\mathop{\kern 0pt \rlap
  {\raise.2ex\hbox{$<$}}}
  \lower.9ex\hbox{\kern-.190em $\sim$}}}
\newcommand{\gsim}   {\mathrel{\mathop{\kern 0pt \rlap
  {\raise.2ex\hbox{$>$}}}
  \lower.9ex\hbox{\kern-.190em $\sim$}}}
\def\be{\begin{equation}}
\def\ee{\end{equation}}
\def\ba{\begin{eqnarray}}
\def\ea{\end{eqnarray}}
\def\d{{\rm d}}
\def\numubar{\bar{\nu}_{\mu}}
\def\nutaubar{\bar{\nu}_{\tau}}
\def\nuxbar{\bar{\nu}_x}
\def\nue{\nu_e}
\def\numu{\nu_\mu}
\def\nutau{\nu_\tau}
\def\adm2{\Delta{{m}^2_{\text{atm}}}}
\def\ap{\approx}
\def\sdm2{\Delta{{m}^2_{\text{sol}}}}
\def\Dm2{\Delta{m}^2}
\def\s2t{\sin^2{2\theta}}
\def\eff{{\rm eff}}
\def\L{{\mathcal L}}
\def\Ue{|U_{e3}|}
\def\UeUe{|U_{e3}|^2}
\def\s2t13{\sin^2{(2\theta_{13}})}
\def\tsol{\theta_{\text{sol}}}
\def\tatm{\theta_{\text{atm}}}
\def\t13{\theta_{\text{13}}}
\newcommand{\ssqtt}{\ensuremath{\sin^2(2\theta_{13})}\xspace}
\newcommand{\chisq}{\ensuremath{\chi^2}\xspace}
\newcommand{\sabs}{\ensuremath{\sigma_{\text{abs}}}\xspace}
\newcommand{\srel}{\ensuremath{\sigma_{\text{rel}}}\xspace}
\newcommand{\sshp}{\ensuremath{\sigma_{\text{shp}}}\xspace}
\newcommand{\sscl}{\ensuremath{\sigma_{\text{scl}}}\xspace}
\newcommand{\sbtb}{\ensuremath{\sigma_{\text{b2b}}}\xspace}
\newcommand{\sbkg}{\ensuremath{\sigma_{\text{bkg}}}\xspace}
\newcommand{\sdmt}{\ensuremath{\sigma_{\Delta{m}^2}}\xspace}
\newcommand{\scfl}{\ensuremath{\sigma_{\text{cfl}}}\xspace}
\newcommand{\Enu}{\ensuremath{E_{\nu}}\xspace}
\newcommand{\dmgui}{\ensuremath{\Delta{m}^2}\xspace}
\newcommand{\nuebar}{\ensuremath{\overline{\nu}_\text{e}}\xspace}
\newcommand{\atom}[3]{\ensuremath{{\protect\vphantom{#1}}^{#2}_{#3}{\mbox{#1}}}}
\newcommand{\cf}{\emph{cf.}\xspace}
\newcommand{\ie}{\emph{i.e.}\xspace}
\newcommand{\dd}{\ensuremath{\text{d}}}
\long\def\symbolfootnote[#1]#2{\begingroup
\def\thefootnote{\fnsymbol{footnote}}\footnote[#1]{#2}\endgroup} 
\newcommand\fverb{\setbox\pippobox=\hbox\bgroup\verb}
\newcommand\fverbdo{\egroup\medskip\noindent
                        \fbox{\unhbox\pippobox}\ }
\newcommand\fverbit{\egroup\item[\fbox{\unhbox\pippobox}]}
\newbox\pippobox
\thispagestyle{empty}
\begin{titlepage}
\begin{center}
\vspace*{2.0cm}
\vskip 2cm
{\LARGE \bf Letter of Intent for Double-CHOOZ:\\
\vspace*{6mm}
a Search for the Mixing Angle~$\theta_{13}$} 
\end{center}
\vskip 1cm
\begin{figure}[h!]
\begin{center}
\includegraphics[width=\textwidth]{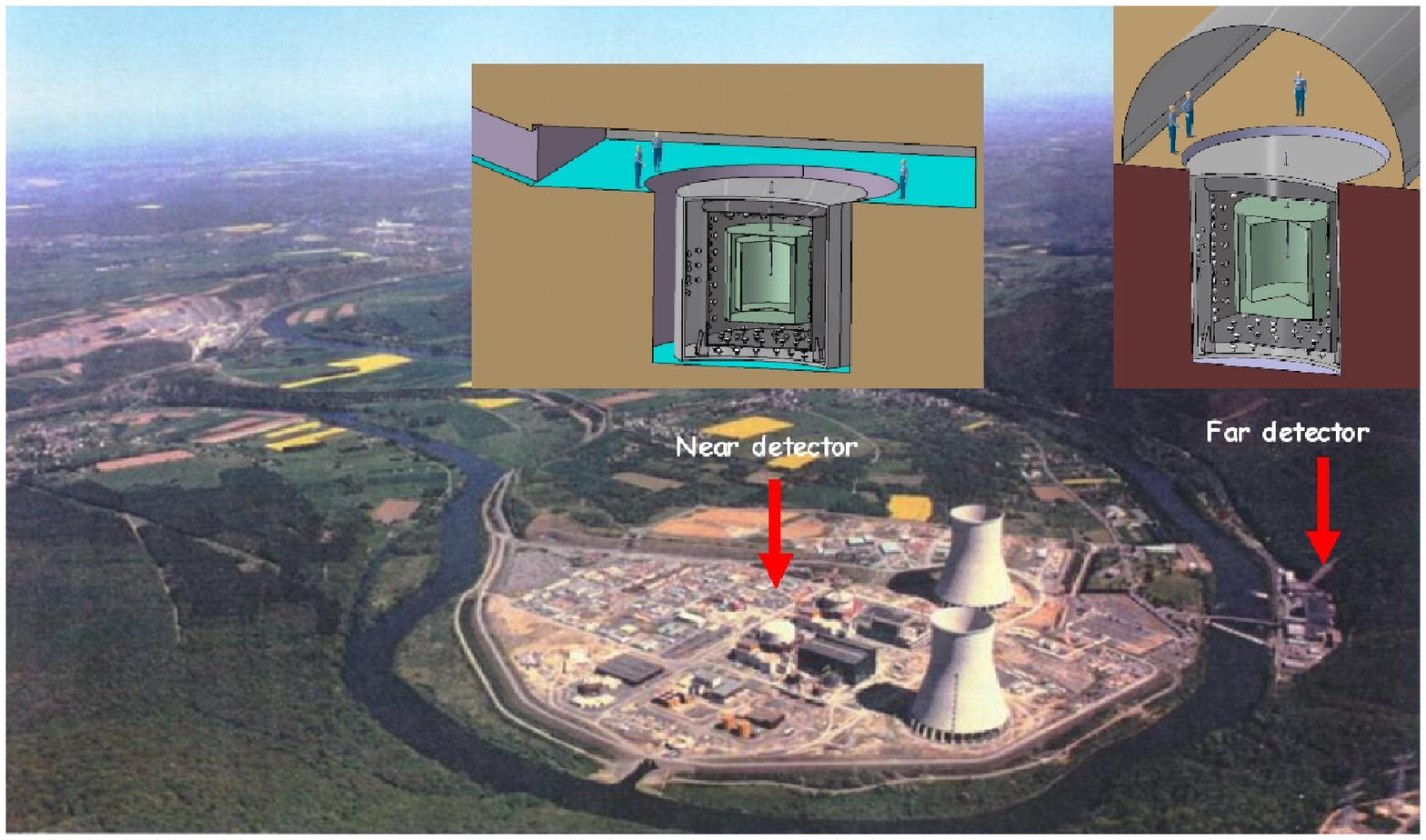}
\end{center}
\end{figure}
\begin{center}
APC,~Paris~-
~RAS,~Moscow~- 
~DAPNIA,~Saclay \\
~\mbox{EKU-T\"ubingen}~- 
~INFN,~Assergi~\&~Milano \\
~Insitute~Kurchatov,~Moscow- 
~MPIK,~Heidelberg \\
~Subatech, Nantes~-~\mbox{TUM,~M\"unchen} \\ 
~University~of~l'Aquila~-Universit\"at~Hamburg
\vskip 2cm
\vspace*{5mm}
{May 2004}\\
\end{center}
\end{titlepage}

\pagestyle{empty}
\cleardoublepage
\renewcommand{\thepage}{\arabic{page}}
\thispagestyle{empty}
\vspace*{5mm}
\noindent
{ \bf 
F.~Ardellier~$^3$,
I.~Barabanov~$^{7}$,
J.C.~Barri\`ere~$^3$,
M.~Bauer~$^{4}$,
L.~Bezrukov~$^{7}$,
Ch.~Buck~$^{8}$,
C.~Cattadori~$^{5,6}$,
B.~Courty~$^{1,9}$,
M.~Cribier~$^{1,3}$,
F.~Dalnoki-Veress~$^{8}$,
N.~Danilov~$^2$,
H.~de Kerret~$^{1,9}$,
A.~Di~Vacri~$^{5,13}$,
A.~Etenko~$^{10}$,
M.~Fallot~$^{11}$,
Ch.~Grieb~$^{12}$,
M.~Goeger~$^{12}$,
A.~Guertin~$^{11}$,
T.~Kirchner~$^{11}$,
Y.S.~Krylov~$^{2}$,
D.~Kryn~$^{1,9}$,
C.~Hagner~$^{14}$,
W.~Hampel~$^{8}$,
F.X.~Hartmann~$^{8}$,
P.~Huber~$^{12}$,
J.~Jochum~$^{4}$,
T.~Lachenmaier~$^{12}$,
Th.~Lasserre~$^{1,3,}$\symbolfootnote[2]{Corresponding author, thierry.lasserre@cea.fr},
Ch.~Lendvai~$^{12}$,
M.~Lindner~$^{12}$,
F.~Marie~$^{3}$,
J.~Martino~$^{11}$,
G.~Mention~$^{1,9}$,
A.~Milsztajn~$^{3}$,
J.P.~Meyer~$^{3}$,
D.~Motta~$^{8}$,
L.~Oberauer~$^{12}$,
M.~Obolensky~$^{1,9}$,
L.~Pandola~$^{5,13}$,
W.~Potzel~$^{12}$,
S.~Sch\"onert~$^{8}$,
U.~Schwan~$^{8}$,
T.~Schwetz~$^{12}$,
S.~Scholl~$^{4}$,
L.~Scola~$^{3}$,
M.~Skorokhvatov~$^{10}$,
S.~Sukhotin~$^{9,10}$,
A.~Letourneau~$^{3}$,
D.~Vignaud~$^{1,9}$,
F.~von~Feilitzsch~$^{12}$,
W.~Winter~$^{12}$,
E.~Yanovich$^{7}$}\\
\\

\noindent 
\vspace*{6mm}
{$^1$ \rm APC, 11 place Marcelin Berthelot, 75005 Paris, France} \\
\vspace{6mm}
{$^2$ \rm IPC of RAS, 31, Leninsky prospect, Moscow 117312, Russia}\\
\vspace{6mm}
{$^3$ \rm DAPNIA (SEDI, SIS, SPhN, SPP), CEA/Saclay, 91191 Gif-sur-Yvette, France}\\
\vspace{6mm}
{$^4$ \rm Eberhard Karls Universit\"at, Wilhelmstr. D-72074  T\"ubingen, Germany}\\
\vspace{6mm}
{$^5$ \rm INFN, LGNS, I-67010  Assergi (AQ), Italy}\\
\vspace{6mm}
{$^6$ \rm INFN Milano, Via Celoria 16, 20133 Milano, Italy}\\
\vspace{6mm}
{$^{7}$ \rm INR of RAS, 7a, 60th October Anniversary prospect, Moscow 117312, Russia}\\
\vspace{6mm}
{$^{8}$ \rm MPI f\"ur Kernphysik, Saupfercheckweg 1, D-69117 Heidelberg, Germany}\\
\vspace{6mm}
{$^{9}$ \rm PCC Coll\`ege de France, 11 place Marcelin Berthelot, 75005 Paris, France}\\
\vspace{6mm}
{$^{10}$ \rm RRC Kurchatov Institute, 123182 Moscow, Kurchatov sq. 1, Russia}\\
\vspace{6mm}
{$^{11}$ \rm Subatech (Ecole des Mines), 4, rue
  Alfred~Kastler, 44307 Nantes, France \\
\vspace{6mm}
{$^{12}$ \rm TU M\"unchen. James-Franck-Str., D-85748 Garching, Germany}\\
\vspace{6mm}
{$^{13}$ \rm University of L'Aquila, Via Vetoio 1, I-67010 Coppito,
  L'Aquila, Italy \\
\vspace{6mm}
{$^{14}$ \rm Universit\"at Hamburg, Luruper Chaussee 149, D-22761 Hamburg, Germany}\\
\cleardoublepage

\vspace*{\stretch{1}}
\begin{center}
{\bf Abstract}
\vspace*{5mm}
\end{center}
Tremendous progress has been achieved in neutrino oscillation physics
during the last few years. However, the smallness of
the $\t13$ neutrino mixing angle still remains enigmatic. The current best
constraint comes from the CHOOZ reactor neutrino experiment $\s2t13<0.2$
(at 90\%~C.L., for $\adm2=2.0 \, 10^{-3} \, \text{eV}^2$).  
We propose a new experiment on the same site, Double-CHOOZ, 
to explore the range of $\s2t13$ from 0.2 to 0.03, within three years of data taking.
The improvement of the CHOOZ result requires an increase in the statistics,
a reduction of the systematic error below one percent, and a
careful control of the cosmic ray induced background.
Therefore, Double-CHOOZ will use two identical detectors, one at $\sim$150~m
and another at 1.05~km distance from the nuclear cores.
The plan is to start data taking with two detectors
in 2008, and to reach a sensitivity for  $\s2t13$ of 0.05 in 2009,
and 0.03 in 2011.

\vspace*{\stretch{1}}

\cleardoublepage
\pagestyle{headings}
\tableofcontents
\cleardoublepage
\pagestyle{headings}
\chapter{Physics opportunity}
\label{sec:physicsopp}
Neutrinos play a crucial role in fundamental particle physics and
have a huge impact in astroparticle physics and cosmology. 
Before 2002, neutrino oscillation physics was still in a discovery
 phase, even though strong evidence for atmospheric 
\cite{imb, soudan, Fukuda:1998mi,Ambrosio:1998wu,Ronga:2001zw,SK_atm_nu2002} 
and solar neutrino oscillations have already been established since 1998. 
Thirty years after the discovery of the solar neutrino
anomaly~\cite{Cleveland:1998nv,Abdurashitov:2002xa,
Hampel:1998xg,Altmann:2000ft}, 
the combined SNO Super-Kamiokande discovery of the flavor conversion~
\cite{Fukuda:2002pe,Ahmad:2002jz} together with the
first reactor $\nuebar$ flux suppression observed by
KamLAND~\cite{Eguchi:2002dm}, is now moving  neutrino physics to a new
era of precision measurements.\\

In the Standard Model of electroweak interactions, neutrinos are massless 
particles, and there is no mixing between the leptons. There exists only a left-handed 
neutrino, and a right-handed antineutrino.
In the quark sector of the Model, the mixing between quark weak and mass eigenstates occurs 
among the three flavor families, and the amount of mixing is
determined by the CKM mixing matrix. In the lepton sector, the analogue of the CKM
matrix for quarks is just the identity matrix, and three conservation
laws have been empirically included, for the three lepton families. \\

The strong evidence for non-zero neutrino masses clearly indicates the existence of physics 
beyond the minimal Standard Model. The smallness of neutrino masses together
with the amounts of lepton flavor violation found in neutrino
oscillation experiments provide insights into possible modifications of the
current Standard Model of electroweak interactions, and open a new
window towards the Grand Unification energy scale~\cite{seesaw}. \\

In the current paradigm, the neutrino mass and weak eigenstates 
are related through the Pontecorvo-Maki-Nakagawa-Sakata (PMNS) neutrino mixing
matrix \cite{MNS, Pontecorvo}. 
A synthesis of atmospheric, solar, and reactor neutrino oscillation data requires 
the existence of (at least) three-neutrino mixing.
The PMNS mixing matrix can be parameterized by three mixing angles $\tsol$, $\tatm$, $\t13$, 
and one or three CP-violating phases, depending on the Dirac or Majorana nature of the massive 
neutrinos~\cite{PDG}. 
Although not favored by the current data, a scenario with more than three
neutrinos might be required to account for the LSND anomaly~\cite{LSND}. 
In this case, the mixing of the three active neutrinos with the additional
sterile neutrino(s) decouples from the oscillations described by the
PMNS matrix. The presently running MiniBoone experiment will settle
the controversy in the near future~\cite{Miniboone}.\\

A wide range of experiments using accelerator, atmospheric, reactor,
and solar neutrinos will be necessary to achieve a full
understanding of the neutrino mixing matrix. 

Solar neutrino experiments combined with KamLAND have measured the
so-called {\it solar parameters}\footnote{The intervals vary slightly
in the different analyzes, we give here the values quoted in~\cite{huberreactor2003}. 
Furthermore, we assume here the normal neutrino mass hierarchy case.}
$\sdm2  = \Dm2_{21} = 7_{-3}^{+2}\cdot 10^{-5} \,\text{eV}^2$ 
and $\sin^2 (2\tsol) = \sin^2 (2\theta_{12})  =  0.8_{-0.2}^{+0.2}$~
\cite{Maltoni2003, Cleveland:1998nv,Abdurashitov:2002xa,Hampel:1998xg,
  Altmann:2000ft,Fukuda:2002pe,Ahmad:2002jz, huberreactor2003}.
Future solar neutrino data as well as the forthcoming KamLAND results will 
undoubtedly improve the solar neutrino parameters determination. 
A new middle baseline (20-70 km) reactor neutrino
experiment could further improve $\sdm2$ or/and $\tsol$ if
necessary~\cite{PetcovHLMA,HLMA}. 

Atmospheric neutrino experiments such as Super-Kamiokande together with the 
K2K first long baseline accelerator neutrino experiment have measured
the so-called {\it atmospheric parameters}
$\adm2= | \Dm2_{32} | = 2_{-0.7}^{+1.0}\cdot 10^{-3} \,\text{eV}^2$ 
and $\sin^2 (2\tatm)=\sin^2 (2\theta_{32})  =  1.0_{-0.2}^{+0.0}$ 
~\cite{Fukuda:1998mi,Ambrosio:1998wu,Ronga:2001zw,SK_atm_nu2002}.
Experimental errors will slowly decrease with additional K2K and
Super-Kamiokande data, but a major improvement of the results is
expected from the currently starting MINOS long baseline neutrino 
experiment~\cite{Minos1, Minos2}. 
\begin{figure}
\begin{center}
\includegraphics[width=0.6\textwidth]{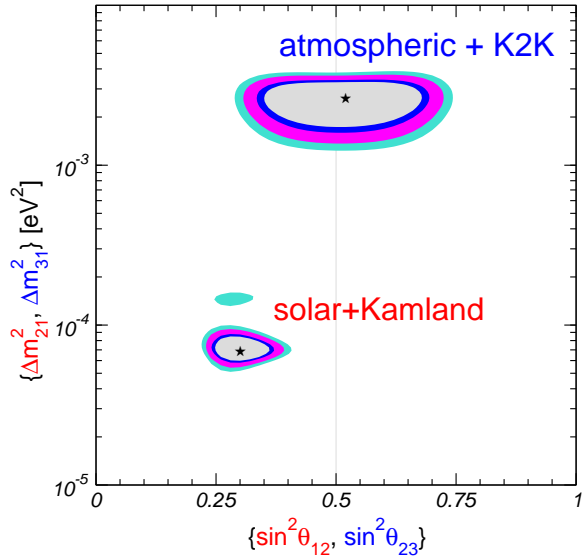}
\caption[Solar and atmospheric allowed regions from the global
  oscillation analysis]{Solar and atmospheric allowed regions from the
  global oscillation  data analysis at 90~\%, 95~\%, 99~\%, and
  3$\sigma$~C.L. for 2  degrees of freedom~\cite{Maltoni2003}.}
\label{fig:sol+atm}
\end{center}
\end{figure}
\\
The {\it third sector} of the neutrino oscillation matrix is driven
by the mixing angle $\t13$, currently best constrained by the CHOOZ
reactor neutrino experiment~\cite{chooz1,chooz2, chooz3, choozlast}. 
CHOOZ provides the upper bound
$\s2t13 < 0.20$ (90~\%~C.L.), assuming $\adm2=2.0 \ 10^{-3} \, \text{eV}^2$ 
(this upper limit is strongly correlated with the assumed value of
$\adm2$.) A weaker upper bound, $\s2t13 < 0.4$, 
has been obtained by the Palo-Verde experiment~\cite{PaloVerde}. \\

Concerning the determination of the PMNS mixing parameters, the
measurement of the angle $\t13$  is the next experimental step to accomplish. 
Knowing the value of $\t13$, or lowering the CHOOZ bound is
already fundamental, in itself, in order to better understand the
structure of the PMNS matrix. Both atmospheric and solar
mixing angles have been found to be maximal or large, thus the smallness of
$\t13$ remains a mystery. Moreover, any sub-leading
three-neutrino oscillation effects, such as the solar-atmospheric
driven oscillation interferences~\cite{PetcovHLMA,HLMA} or the CP-violation in
the lepton sector, could only be observable for non-vanishing $\t13$ values. \\

Which sensitivity is then relevant for the forthcoming projects dedicated to
$\t13$~? On the one hand, neutrino mass models predict $\s2t13$ values
ranging  from~0~to~0.18~\cite{Reactorwhitepaper}. 
Any neutrino experiment with a sensitivity of a few
percents, like Double-CHOOZ, has thus an important discovery potential.  
On the other hand, the neutrino mass models connect, in most cases, the
CP-$\delta$ phase to the leptogenesis mechanism~\cite{leptogenesis}. 
The search for CP violation effects in the lepton sector is
thus of great interest since the leptogenesis mechanism is one of
the best current explanation of the matter-antimatter baryon 
asymmetry observed in our Universe. 
The target sensitivity  to achieve is thus strongly driven by the
potential of future CP-$\delta$ neutrino appearance experiments. 
In the distant future, CP violation could be observed at neutrino factories 
if $\s2t13 > 0.001$. However, on a shorter time scale, a value of  $\s2t13$ of a few
percent might allow superbeam based experiments, possibly combined with a large
reactor neutrino detector, to probe part of the $\delta-\t13$
parameter space~\cite{huberreactor2003}. \\

Although they are not designed to measure $\t13$, a
marginal improvement of the CHOOZ constraint can be obtained with 
conventional neutrino beams. 
For instance, the MINOS experiment~\cite{Minos1} may achieve a sensitivity 
$\s2t13 < 0.1$, while the CNGS experiments, OPERA and
ICARUS~\cite{CNGS1, CNGS2, theta13globalana}, may improve the CHOOZ
bound down to $\s2t13 < 0.14$ and $\s2t13 < 0.09$, 
respectively,  if no excess of $\nue$ is observed after
five years of data taking\footnote{ICARUS and OPERA results could be combined, leading to
  a value very close to the ICARUS sensitivity (10~\% improvement).} 
($\adm2=2.0 \, 10^{-3} \, \text{eV}^2$, 90~\%~C.L.). 
The quoted values reduce to 0.05~\cite{Minos2}, 
0.08~\cite{nove03,  CNGS2},  0.04~\cite{Icarus}, respectively, by
neglecting matter effects, CP-$\delta$ phase (set to zero), and mass hierarchy
induced correlations and degeneracies~\cite{minakatareactor2002,
  huberreactor2003}. 
 
Concerning the future of neutrino physics, the next generation of accelerator
neutrino experiments coupled with powerful neutrino beams 
(the so-called Superbeam long baseline experiment) 
are  primarily dedicated to the determination of the PMNS mixing matrix
elements $\t13$ and CP-$\delta$, as well as the precise measurement of
the atmospheric mass splitting and mixing angle, and the
identification of the neutrino mass hierarchy (the sign of  $\Dm2_{32}$).
After five years of data taking, the T2K experiment aims to reach
the sensitivity $\s2t13 < 0.02$  (90~\%~C.L.)~\cite{jparc1, jparc2, huberreactor2003}; 
a similar sensitivity is foreseen by the NuMI Off-Axis
project\footnote{This value takes into account, in a very
  conservative manner, all correlations and degeneracies. At a fixed
  $\delta$ phase taken to be 0, the value quoted would be three times
lower.}~\cite{NuMi}. 

The observation of a $\nue$ excess in an almost pure $\numu$
neutrino beam at any accelerator experiment would be major evidence 
for a non-vanishing $\t13$.  
But unfortunately, in addition to the statistical and systematic 
uncertainties, correlations and degeneracies between 
$\t13$, $\tatm$, $\text{sgn}(\Dm2 _{31})$, and the CP-$\delta$
phase degrade the knowledge of $\t13$~\cite{minakatareactor2002, huberreactor2003}. 
Even though appearance experiments seem to be the easiest way to measure
very small mixing angles, as might be the case for $\t13$, it is of great interest
nevertheless to get additional information with another experimental method. \\

A reactor neutrino experiment, like Double-CHOOZ, is able to measure $\t13$
with an independent detection principle (inverse neutron beta decay), 
and thus different systematic uncertainties. 
Unlike appearance experiments, it does not suffer from parameter degeneracies
induced by the  CP-$\delta$ phase. In addition, thanks to the low
$\nuebar$ energy (a few MeV) as well as the very short baselines (a few
kilometers) the reactor measurement is not affected by matter effects.
As a consequence reactors provide a clean information on $\s2t13$. 
Double-CHOOZ  will use two identical detectors
at $\sim$150~m  and 1.05~km from the CHOOZ-B nuclear power plant cores. 
The near detector is used to monitor both the reactor $\nuebar$ flux and
energy spectrum, while the second detector is dedicated to the search
for a deviation from the expected $(1/\mathrm{distance})^2$ behavior, tagging an 
oscillation effect. For  $\adm2=2.0 \ 10^{-3} \, \text{eV}^2$ we expect a
sensitivity of $\s2t13 < 0.03$ (90~\%~C.L.) after three years of data taking. \\

In conclusion, due to the fundamental interest of $\t13$  as well
as the importance of its amplitude for the  
design of future neutrino experiments dedicated to CP-$\delta$, 
independent $\t13$-dedicated experiments are mandatory.  
To accomplish this goal, both reactor and accelerator programs should
provide the required independent and complementary results
~\cite{Reactorwhitepaper}.

\cleardoublepage
\cleardoublepage
\chapter{Searching for $\mathbf{\text{\bf sin}^2(2\theta _{13})}$ with reactors}
\label{sec:measurementreactor}
\section{Neutrino oscillations}
Neutrino flavor transitions have been observed in atmospheric, solar,
reactor and accelerator neutrino experiments.
To explain these transitions, extensions to the minimal Standard Model of
particle physics are required. The simplest and most widely accepted
extension is to allow neutrinos to have masses and mixing,
similar to the quark sector. The flavor transitions can then
be explained by neutrino oscillations.  
\subsection{Quark mixing}
The Wolfenstein parameterization of the CKM matrix~\cite{PDG} is
based on the very small mixing between the quarks. The mixing matrix
is almost the identity matrix with only small corrections for the off--diagonal elements. 
It uses the observed quark mixing angles 
hierarchy\footnote{$\theta_{12} \sim 0.1 > \theta_{23}  \sim 0.01 > \theta_{13} \sim 0.001$} 
to introduce an expansion parameter $\lambda$ describing the mixing
between $u$ and $s$ quarks. This leads to the parameterization \\
\begin{equation}  \begin{array}{c}\label{eq:wolf} 
V_{\rm CKM} \simeq 
\left( 
\begin{array}{ccc} 
1 - \frac{1}{2} \, \lambda^2 & \lambda & A \, \lambda^3 \, 
(\rho - i \eta) \\
-\lambda & 1 - \frac{1}{2} \, \lambda^2 & A \, \lambda^2 \\
A \, \lambda^3 \, (1 - \rho + i \eta) & -A \, \lambda^2 & 1 
\end{array}
\right) + {\cal{O}}(\lambda^4)~,
\end{array} 
\end{equation}
where $\lambda$ corresponds to the Cabibbo angle $\sin \theta_C \simeq
0.22$, and the other parameters are roughly 
$A \simeq 0.83$, $\rho \simeq 0.23$ and $\eta \simeq 0.36$~\cite{PDG}. 
The latter parameter describes  $CP$ violation in the quark sector; 
all such effects are proportional to~\cite{JCP}
\begin{equation}
J_{CP} \simeq -A^2 \, \lambda^6 \, \eta \sim -3 \cdot 10^{-5}~.
\end{equation}
Therefore, $CP$ violation in the quark sector is a small effect. 
\subsection{Neutrino mixing}
The neutrino oscillation data can  be described within a three
neutrino mixing scheme, in which the flavor states $\nu_\alpha$
($\alpha = e, \mu, \tau$) 
are related to the mass states $\nu_i$ ($i = 1,2,3$) through 
the PMNS (Pontecorvo-Maki-Nakagawa-Sakata) unitary lepton 
mixing matrix. \\
It can be parameterized as U$_{\text{PMNS}} = $ \\
\begin{eqnarray}
\left( \begin{array}{ccc}
  1       &    & \\
    & c_{23} &  s_{23} \\
  & - s_{23} & c_{23} 
\end{array} \right) \, 
%
\left( \begin{array}{ccc}
  c_{13} &  & s_{13} e^{-i\delta} \\
& 1 & \\
  - s_{13} e^{i\delta} & & c_{13} 
\end{array} \right) \, 
%
\left( \begin{array}{ccc}
  c_{12}       & s_{12}  & \\
- s_{12} & c_{12} & \\
& & 1 
\end{array} \right) \,
%
\left( \begin{array}{ccc}
  1       &    & \\
    & e^{i\alpha}&   \\
         &  & e^{i\beta}
\end{array} \right) \, 
& & \nonumber
\end{eqnarray}
\vspace{-0.35cm}
\begin{eqnarray}
& = & \\
& \hspace{-0.25cm}
\left( \begin{array}{ccc}
  c_{13} c_{12}       & c_{13} s_{12}  & s_{13} e^{-i\delta} \\
- c_{23} s_{12} - s_{13} s_{23} c_{12} e^{i\delta}
& c_{23} c_{12} - s_{13} s_{23} s_{12} e^{i\delta}
& c_{13} s_{23} \\
    s_{23} s_{12} - s_{13} c_{23} c_{12} e^{i\delta}
& - s_{23} c_{12} - s_{13} c_{23} s_{12} e^{i\delta}
& c_{13} c_{23} 
\end{array} \right) \, & \nonumber  
\left( \begin{array}{ccc}
  1       &    & \\
    & e^{i\alpha}&   \\
         &  & e^{i\beta}
\end{array} \right) \, 
\end{eqnarray}
where $c_{ij} = \cos\theta_{ij}$ and $s_{ij} = \sin\theta_{ij}$, 
$\delta$ is a Dirac $CP$ violating phase, $\alpha$ and $\beta$
are Majorana $CP$ violating phases, not considered in the following. 
Up to now, the angles $\theta_{12}$ and $\theta_{23}$ are probed via
the oscillations of solar/reactor and atmospheric neutrinos, while the angle 
$\theta_{13}$ is mainly constrained by the CHOOZ reactor experiment;
the Dirac phase $\delta$ has not been constrained yet.

The factorized form of this PMNS mixing matrix is often used to
identify the mixing angles reported by the 
experiments
\footnote{Thanks to the smallness of  $\frac{\sdm2}{\adm2}$ and $\sin^2 \theta_{\text{CHOOZ}}$.}
\begin{equation}
\theta_{23} \cong \tatm , \quad
\theta_{12} \cong \tsol ,
\quad {\rm and} \quad \theta_{13} \cong \theta_{\rm CHOOZ}.
\end{equation}

\noindent The relevant formula for the oscillation probabilities is 
\begin{equation} \label{eq:Pab}
P(\nu_\alpha \rightarrow \nu_\beta) 
= \delta_{\alpha \beta} - 2 \, {\rm Re} \, 
\sum_{j > i} U_{\alpha i} \, U_{\alpha j}^\ast  \, 
U_{\beta i}^\ast \, U_{\beta j} \, 
\left( 1 - \exp{\frac{i \Dm2_{ji} \, L}{2 \, E}} \right)~,
\end{equation}
where $\Dm2_{ji} = m_j^2 - m_i^2$.  \\

Since the identification of the MSW-LMA mechanism as the solution 
of the solar neutrino anomaly~\cite{Fukuda:2002pe,Ahmad:2002jz,Eguchi:2002dm}, 
we now know that the mass eigenstate with the larger electron neutrino component has the smaller
mass (state 1). Solar neutrino oscillations occur then mainly
together with the little heavier state 2: 
\begin{equation}
\Dm2_{21} =  m^2_2 - m^2_1 \equiv \sdm2 > 0.
\end{equation}
The large mass squared difference measured in the atmospheric sector
is therefore the splitting between the mass eigenstate 3 and the more 
closely spaced 1 or 2. In addition, the CHOOZ reactor neutrino
experiment shows that the mass eigenstate 3 has only a very
small  electron neutrino component. 
In this description, the sign of the splitting between
state 3 and states 1 and 2 is unknown; this leads to two
possibilities of mass ordering:
\begin{equation}
|\Dm2_{32}|=  |  m^2_3 - m^2_2 | \equiv \adm2 .
\end{equation}
Thus, one defines the normal hierarchy (NH) scenario
$m_3 > m_2 > m_1$, and the inverted hierarchy scenario (IH) 
$m_2 > m_1 > m_3$. 
The determination of the sign of $\Dm2 _{32}$ is one of the next
goals in neutrino oscillation physics.
\section[Measurement of $\text{sin}^2(2\theta _{13})$ with reactor $\nuebar$]{Measurement of $\mathbf{\text{\bf sin}^2(2\theta _{13})}$ with reactor $\nuebar$}
\subsection{Reactor $\nuebar$ flux}
The fissionable material in the CHOOZ pressurized water reactors (PWR)
mainly consists of $^{235}$U and $^{239}$Pu, which undergo thermal neutron
fission. The fresh fuel is enriched to about 3.5~\% in $^{235}$U. Fast
fission neutrons are moderated by light water pressurized to 150
bar. The dominant natural uranium isotope, $^{238}$U, is fissile only for
fast neutrons (threshold of 0.8 MeV) but it also generates fissile $^{239}$Pu
by thermal neutron capture, 
\begin{equation}
\text{n} + ^{238}\text{U} \rightarrow ^{239}\text{U} \rightarrow
^{239}\text{Np} \rightarrow ^{239}\text{Pu} \,\, \text{(T}_{1/2}= 24,100~\text{y)}.
\end{equation}
The  $^{241}$Pu isotope  is produced in a manner similar to $^{239}$Pu
\begin{equation}
\text{n} + ^{239}\text{Pu} \rightarrow ^{240}\text{Pu}~+~\text{n} \rightarrow ^{241}\text{Pu}
\,\,\text{(T}_{1/2} = 14.4~\text{y)}. 
\end{equation}
As the reactor operates, the concentration of
$^{235}$U decreases, while that of $^{239}$Pu and $^{241}$Pu
increases. After about one year, the  reactor  is stopped and one
third of the fuel elements are replaced. 
Typical numbers for an annual cycle are given in Table~\ref{tab:choozburnup}.
\begin{table}[h]
\begin{center}
\begin{tabular}{lrrr}
\hline
            &  \multicolumn{1}{c}{Mean energy per fission} & \multicolumn{2}{c}{Refueling cycle} \\
            & (MeV)                   & \multicolumn{1}{c}{beginning}  & \multicolumn{1}{c}{end} \\ 
\hline
$^{235}$U        & 201.7 $\pm$ 0.6  & 60.5~\% & 45.0~\% \\
$^{238}$U        & 205.0 $\pm$ 0.9  & 7.7~\%  & 8.3~\%\\
$^{239}$Pu       & 210.0 $\pm$ 0.9  & 27.2~\% & 38.8~\%\\
$^{241}$Pu       & 212.4 $\pm$ 1.0  & 4.6~\% & 7.9~\%\\
\hline
\end{tabular}
\caption[Typical fuel composition of a PWR reactor]
{Typical fuel composition for an annual cycle of a PWR power station,
  for the four main isotopes, normalized to 100~\%.
  There are also other isotopes, not included here, which contribute for
  a few percents. 
}
\label{tab:choozburnup}
\end{center}
\end{table}
Due to the threshold of the detection reaction at 1.8~MeV only the
most energetic antineutrinos
are detected; they correspond to the decay of fission products with the
highest Q-values and hence to the shortest lived. The detected
antineutrinos thus closely follow changes in power. In particular spent
fuel elements which are kept on site out of the core contain only long
lived emitters with a low Q-value; 
their contribution to the detected $\nuebar$ signal is negligible. 
Measurements of the neutrino rate per fission have been performed for
$^{235}$U, $^{239}$Pu and $^{241}$Pu by Borovoi et al.~\cite{Borovoi}
 and Schreckenbach et al.~\cite{Schreckenbach:1985ep,Hahn:1989zr}. 
The latter measurement includes the shape of the
energy spectrum, with a 2~\% bin-to-bin accuracy and an
overall normalization error of 2.8~\%. 
The measurement performed by~\cite{Schreckenbach:1985ep} 
can be compared with several computations and is found to be
in good agreement with that of~\cite{KM,vogel}. 
We will therefore use this computation for the $^{238}$U
neutrino rate, which has never been measured. The $^{238}$U contribution to
the total number of fissions is $\sim$10~\%, and is therefore not a
major source of error. The antineutrino spectra of the four dominant
fissioning isotopes are shown in Figure \ref{fig:nuspe}.
\begin{figure}[h!]
\begin{center}
\includegraphics[width=0.8\textwidth]{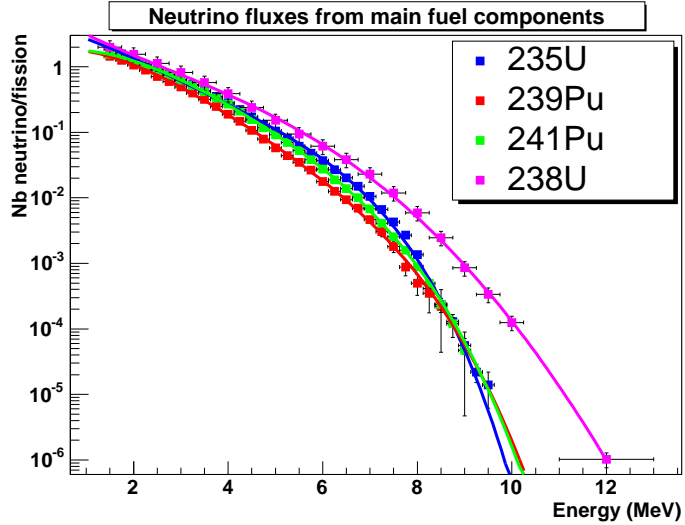}
\caption[$\nuebar$ spectra of the four dominant
fissioning isotopes]{$\nuebar$ spectra of the four dominant
isotopes with their experimental error bars ($^{238}$U
spectrum has not been measured but calculated). }
\label{fig:nuspe}
\end{center}
\end{figure}
During the cycle, the contributions of the different fissile isotopes to
energy production evolve.  For fresh fuel, $^{235}$U fissions
dominate, whereas  $^{238}$U fissions amount for a few times less. 
Quickly after the beginning of the cycle $^{239}$Pu gives an important
 contribution (see Figure \ref{fig:burnup}). 
\begin{figure}
\begin{center}
\includegraphics[width=0.8\textwidth]{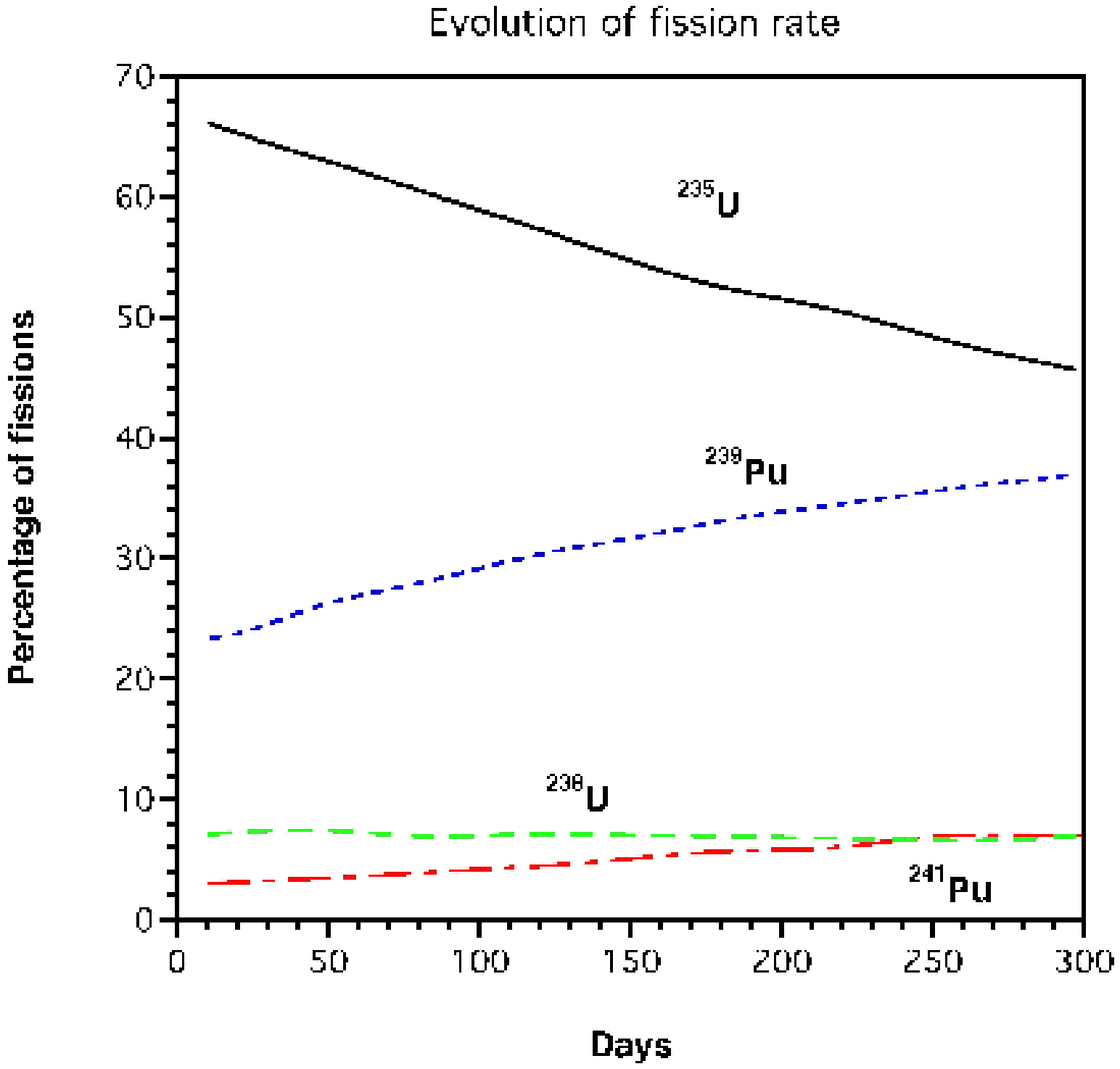}
\caption[Percentage of fissions of the main fissile elements during a
  fuel cycle]{Percentage of fissions of the four dominant fissile 
 isotopes during 300 days of a typical fuel cycle.}
\label{fig:burnup}
\end{center}
\end{figure}
%
%
%
\subsection{$\nuebar$ detection principle}
Reactor antineutrinos are detected through their
interaction by inverse neutron decay (threshold of 1.806~MeV) 
\begin{equation}
\bar\nu_e + p \rightarrow e^+ + n~.
\end{equation}
The cross section for inverse $\beta$-decay has approximately the form
\begin{equation}
\sigma(E_{e^+})\simeq\frac{2\pi^2\hbar^3}{m_e^5 f \tau_n}p_{e^+}E_{e^+}\,,
\label{eq:ibdecaycrosssection}
\end{equation}
where $p_{\text{e}^+}$ and $E_{\text{e}^+}$ are the momentum and the energy of the
positron\footnote{$E_{\text{e}^+}$ is the sum of the rest mass and kinetic energy of the positron.}, 
$\tau_\text{n}$ is the lifetime of a free neutron and $f$ is the free
neutron decay phase space factor.
As an approximation, we use an averaged
fuel composition typical during a reactor cycle corresponding to  
$^{235}$U (55.6~\%), $^{239}$Pu (32.6~\%), $^{238}$U  (7.1~\%) and 
$^{241}$Pu (4.7~\%). The mean energy release per fission  W is  
then 203.87~MeV and the energy weighted cross section amounts to
\begin{equation}
<\sigma>_{\rm fission} = 5.825 \cdot 10^{-43}~\text{cm}^2 \, \, {\rm per} \,
\, {\rm fission}~. 
\end{equation}
The reactor power $P_\text{th}$ is related to the number of fissions per
second $N_\text{f}$ by 
\begin{equation}
N_f = 6.241 \cdot 10^{18} {\rm sec}^{-1} \cdot (P_{th}[{\rm MW}]) / (W[{\rm MeV}])~. 
\end{equation}
The event rate at a distance $L$ from the source, assuming
no oscillations, is thus 
\begin{equation}
R_\text{L}~=~N_\text{f}~\cdot~<\sigma>_{\rm fission} \cdot \, n_\text{p} \cdot~1/(4\pi L^2)~,
 \end{equation}
where $n_p$ is the number of protons in the target. 
For the purpose of simple scaling, a reactor with a power of
1~GW$_{th}$ induces a rate of $\sim$450 events per year in a detector
 containing $10^{29}$~protons, at a distance of 1~km. \\

Experimentally one takes advantage of the coincidence signal of the
prompt positron followed in space and time by the delayed neutron
capture. This very clear signature allows to strongly reject the accidental
backgrounds. The energy of the incident antineutrino is then related to the
energy of the positron by the relation
\begin{equation}
\label{eq:energy}
E_{\nuebar}=E_{\text{e}^+} +(m_\text{n}-m_\text{p})+O(E_{\nuebar}/m_\text{n})~.
\end{equation}
Experimentally, the visible energy seen in the detector is given by  
$E_\text{vis}=E_{\text{e}^+} + 511~\text{keV}$, where the
additional $511\,\mathrm{keV}$ come from the annihilation of the
positron with an electron when it stops in the matter.  
\subsection{$\nuebar$ oscillations}
Reactor neutrino experiments measure the survival probability 
$P_{\nuebar \rightarrow \nuebar}$ of the electron antineutrinos 
emitted from the nuclear power plant\footnote{The low neutrino energy 
(a few MeV) does not allow any appearance measurement.}.
This survival probability does not depend 
on the $\delta$-CP phase. Furthermore, because of the low energy as well
as the short baseline considered, matter effects are negligible~\cite{minakatareactor2002}. 
Assuming a ``normal'' mass hierarchy scenario, $m_{1} < m_{2} <
m_{3}$, the $\nuebar$ survival probability 
can be written~\cite{CHOOZU13,PetcovNHIH}
\begin{eqnarray}
P_{\nuebar\to\nuebar} & = &
1-2\sin^2\theta_{13}\cos^2\theta_{13}\sin^2\left(\frac{\Delta{m}^2_{31}L}{4E}\right)
\\ \nonumber
& - & 
\frac{1}{2}\cos^4\theta_{13}\sin^2(2\theta_{12})\sin^2\left(\frac{\Delta{m}^2_{21}L}{4E}\right)\nonumber\\ 
& + & 
2\sin^2\theta_{13}\cos^2\theta_{13}\sin^2\theta_{12}\left(\cos\left(\frac{\Delta{m}^2_{31}L}{2E}-\frac{\Delta{m}^2_{21}L}{2E}\right)-\cos\left(\frac{\Delta{m}^2_{31}L}{2E}\right)\right) \nonumber
\end{eqnarray}
\label{3nuSP}
%
%
The first two terms in Eq.~\ref{3nuSP} contain respectively the
atmospheric driven ($\Dm2_{31} = \adm2$) and solar driven 
($\Dm2_{21} = \sdm2$, $\theta_{12} \sim \tsol$)
contributions, while the third term, absent from {\it any } 
two-neutrino mixing model, is an interference between solar and
atmospheric driven oscillations whose amplitude is a function of
$\t13$.
%
Thus, up to second order in  $\sin 2\theta_{13}$ and $\alpha =
\frac{\sdm2}{\adm2}$ the survival probability can be
expressed as
\be
P_{\nuebar \rightarrow \nuebar}  \simeq 1 - \sin^2 2 \theta_{13} \, \sin^2 (\Dm2 _{31} L/4E) +  
\alpha^2 \, (\Dm2 _{31} L/4E)^2 \, \cos^4 \theta_{13} \, \sin^2 2\theta_{12}~, 
\label{equ:reactorsurvivalprob}
\ee
where the third term on the right side can safely
be neglected given the current range (90~\% error intervals) 
of mixing parameters found in neutrino oscillation 
experiments\footnote{Two different best fit values for the atmospheric mass splitting have been released by the
Super-Kamiokande collaboration, based on two different analyzes of the
same data.}~\cite{SK_atm_nu2002, SkAtmLoverE2004}:
\begin{eqnarray}
( \adm2 )_\text{SK-I}    & = & 2.0_{-0.7}^{+1}\cdot
  10^{-3}~\mathrm{eV}^2 \, \nonumber \\ 
( \sin^2 2\theta_{23})_\text{SK-I} & = & 1_{-0.1}^{+0}\, \nonumber \\
( \adm2 )_\text{SK-L/E} & = &  
  2.4_{-0.5}^{+0.6}\cdot10^{-3}~\mathrm{eV}^2 \, \nonumber \\ 
( \sin^2 2\theta_{23})_\text{SK-L/E} & = & 1_{-0.1}^{+0}\, \nonumber \\
 \sdm2 & = & 7.0_{-3}^{+2}\cdot10^{-5}~\mathrm{eV}^2 \, \nonumber \\
\sin^2 (2\theta_{12}) & = & 0.8_{-0.2}^{+0.2}~.  \nonumber 
\label{equ:oscilfitparam}
\end{eqnarray}

Reactor experiments thus provide a clean measurement of the
mixing angle $\t13$, free from any contamination coming from matter
effects and other parameter correlations or degeneracies~\cite{minakatareactor2002, huberreactor2003}. 
Therefore they are exclusively dominated by statistical and systematic errors.
\section{Complementarity with Superbeam experiments} 
A very detailed comparison of reactor antineutrino
experiments with superbeams is described in~\cite{minakatareactor2002,
  huberreactor2003}.
Forthcoming accelerator neutrino experiments, or superbeams, will
 search for a $\nue$  appearance signal. 
The appearance probability $P_{\numu \rightarrow \nue}$  with terms up
to second order, {\it e.g.}, proportional to $\sin^2 2\theta_{13}$, $\sin
2\theta_{13} \cdot \alpha$, and $\alpha^2$, can be written as:
\begin{eqnarray}
P_{\numu \rightarrow \nue} & \simeq & \sin^2 2\theta_{13} \, \sin^2 \theta_{23}
\sin^2 {(\Dm2 _{31} L/4E)} \nonumber \\
&\mp&  \alpha\; \sin 2\theta_{13} \, \sin \delta  \,  \sin2\theta_{12} \sin 2\theta_{23} \nonumber  \\
& & (\Dm2 _{31} L/4E) \sin^2{(\Dm2 _{31} L/4E)} \nonumber \\
&-&  \alpha\; \sin 2\theta_{13}  \, \cos\delta \, \sin 2\theta_{12} \sin 2\theta_{23} \nonumber  \\
& &  (\Dm2 _{31} L/4E) \cos {(\Dm2 _{31}L/4E)} \sin {(\Dm2 _{31} L/4E)} \nonumber  \\
&+& \alpha^2 \, \cos^2 \theta_{23} \sin^2 2\theta_{12} (\Dm2 _{31}L/4E)^2,
\label{equ:propappearance}
\end{eqnarray}
where the sign of the second term refers to neutrinos (minus) or
antineutrinos (plus). 
From Equation \ref{equ:propappearance} one sees that superbeams
 suffer from parameter correlations and degeneracies coming from the 
different combinations of parameters. Many of the degeneracy problems 
originate in the summation of the four terms in
 Equation \ref{equ:propappearance}, since changes in one parameter
 value often can be  compensated by adjusting another one in a
 different term. This leads to the
$(\delta, \theta_{13})$~\cite{Burguet-Castell:2001ez},
$\mathrm{sgn}(\Dm2 _{31})$~\cite{Minakata:2001qm}, and
$(\theta_{23},\pi/2-\theta_{23})$~\cite{Fogli:1996pv} degeneracies,
{\it e.g.} an overall ``eight-fold'' degeneracy~\cite{Barger:2001yr,
  Minakata:2001qm}.  Table~\ref{tab:summary} summarizes the
sensitivity of accelerator and Double-CHOOZ experiments. 
\begin{table}[t!]
\begin{center}
{\small
\begin{tabular}{lcccc}
\hline
& Chooz & Beams & Double-CHOOZ & T2K  \\  
\hline
\multicolumn{5}{l}{$\sin^2(2\theta _{13})$ sensitivity limit (90~\%
  CL)}\\
\hline
\\[-0.2cm]
$ \sin^2(2\theta _{13})$ &  0.2 & 0.061 & 0.032 &  0.023 \\[0.1cm]
$\sin^2(2\theta _{13})_{\mathrm{eff}}$ & 0.2 & 0.026 & 0.032 & 0.006  \\[0.05cm]
\hline
\multicolumn{5}{l}{Measurements for large $\sin^2(2\theta _{13})=0.1$ (90~\% CL)} \\
\hline
\\[-0.2cm]
$\sin^2(2\theta _{13}) $ & $-$ & 0.1$^{+0.104}_{-0.052}$ & 0.1$^{+0.034}_{-0.033}$ & 0.1$^{+0.067}_{-0.034}$ \\[0.1cm] 
\hline
\end{tabular}
} 
\end{center}
\caption[Comparison of the sensitivity of
  reactor and accelerator based future neutrino experiments sensitive
  to $\t13$]{\label{tab:summary} Comparison of the sensitivity of
  reactor and accelerator based future neutrino experiments. 
  The results of the table have been extracted
  from~\cite{theta13globalana}. ''Beams'' is the combination of the
  forthcoming MINOS, ICARUS, and OPERA experiments. Results for
  accelerator experiments are given for five years
  of data taking. Results for Double-CHOOZ are given for three years
  of operation. The line starting by ``$\sin^2(2\theta _{13})$''
  provides the results of the computation taking into account all
  correlation and degeneracy effects, while the line starting by 
$\sin^2(2\theta _{13})_{\mathrm{eff}}$  give the results of a similar
  computation performed after ``switching off'' those effects.}
\end{table}
\cleardoublepage
\cleardoublepage
\chapter{Overview of the Double-CHOOZ experiment}
\label{sec:overview}
This section is an overview of the Double-CHOOZ experiment.
The Double-CHOOZ technology of reactor neutrino detection is based 
on experience obtained in numerous experiments:
Goesgen~\cite{GOE86}, Bugey~\cite{Bugey} (at short distances), 
CHOOZ \cite{chooz1, chooz2, chooz3, choozlast}, 
Palo Verde~\cite{PaloVerde} (at km scale distance) 
CTF~\cite{CTF}, Borexino~\cite{BorexReactor}, 
Kamland~\cite{Eguchi:2002dm} (distances of a few hundred km).
Therefore, no long term R\&D program has to be conducted prior to designing and
building the new detector. Nevertheless, in order to be a precision
experiment, the Double-CHOOZ design has to be improved with respect to
CHOOZ. The liquid scintillator is described in
Chapter~\ref{sec:scintillator}, 
the calibration in Chapter~\ref{sec:calibration}, and the backgrounds 
in Chapter~\ref{sec:backgrounds}. The systematic error handling is presented 
in Chapter~\ref{sec:systematics}. To conclude, the sensitivity of
Double-CHOOZ, taking into account the overall set of systematic
errors, is presented in Chapter~\ref{sec:sensitivity}.
\section{The $\nuebar$ source}
To fulfill the aim of the Double-CHOOZ experiment, precise
knowledge of the antineutrino emission in the nuclear core is not
crucial thanks to the choice of comparing two similar detectors at
different distances, where the near detector measures the 
flux without $\nuebar$ losses consequent to oscillations. Nevertheless this
information is available and will be used for cross checks and
other studies of interest (see Appendix~\ref{sec:IAEA}). More details
on the $\nuebar$ source will be necessary to do those specific studies 
with the near detector, such as the $\nuebar$ spectrum and flux expected
for a given fuel composition and burn up.
\subsection{The CHOOZ nuclear reactors}
The antineutrinos used in the experiment are those produced by the pair
of reactors located at the CHOOZ-B nuclear power station operated by
the French  company Electricit\'e de France (EDF) in partnership
with the Belgian utilities Electrabel S.A./N.V. and Soci\'et\'e Publique
d'Electricit\'e.  They are located in the Ardennes region, northeast of
France, very close to the Belgian border, in a loop of the Meuse
river (see Figures \ref{fig:chooznearfarsites} and \ref{fig:chooznearfarsitemap}).
At the CHOOZ site, there are two nuclear reactors, both are of
the  most recent  N4 type (4 steam generators) with a thermal power of
4.27~GW$_{\text{th}}$, and recently upgraded from 1.45~GW$_{\text{e}}$ to 1.5~GW$_{\text{e}}$.
These reactors are of the Pressurized Water Reactor type (PWR) and are fed with
UOx type fuel. They are the most powerful reactor type in operation in
the world. One unusual characteristic of the N4 reactors is their
ability to vary their output from 30~\% to 95~\% of full power in
 less than 30~minutes, using the so-called gray control
 rods  in the reactor core. These rods are referred to as gray
 because  they absorb fewer free neutrons than conventional black
 rods. One advantage is a bigger thermal homogeneity.
205 fuel assemblies are contained within each reactor core.
 The entire reactor vessel is a cylinder of 13.65~meters high and 
4.65~meters in diameter.
The first reactor started operating at full power in May 1997,
and the second one in September of the same year.
\section{Detector site}
The Double-CHOOZ experiment will run two almost identical detectors of medium size,
containing 12.7 cubic meters of liquid scintillator target doped with
0.1~\% of Gadolinium (see Chapter~\ref{sec:scintillator}).  
The neutrino laboratory of the first CHOOZ experiment, located 1.05~km
from the two cores of the CHOOZ nuclear plant will be used again
(see Figure \ref{fig:choozfarfoto}). This is the main advantage of
this site compared  with other French locations. 
\begin{figure}[h]
\begin{center}
\includegraphics[width=\textwidth]{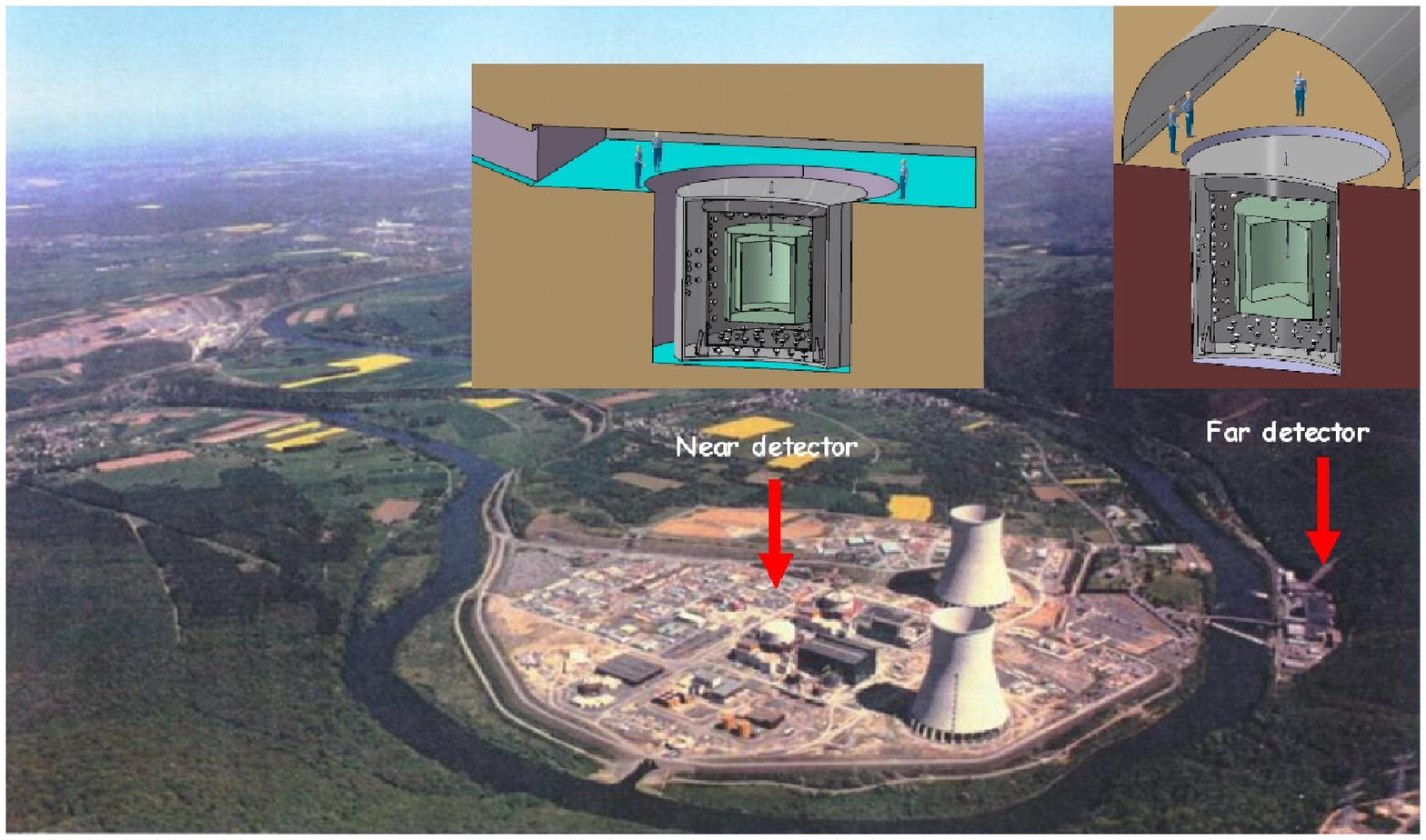}
\caption[Overview of the experiment site]{Overview of the experiment site.}
\label{fig:chooznearfarsites}
\end{center}
\end{figure}
\begin{figure}[h]
\begin{center}
\includegraphics[width=\textwidth]{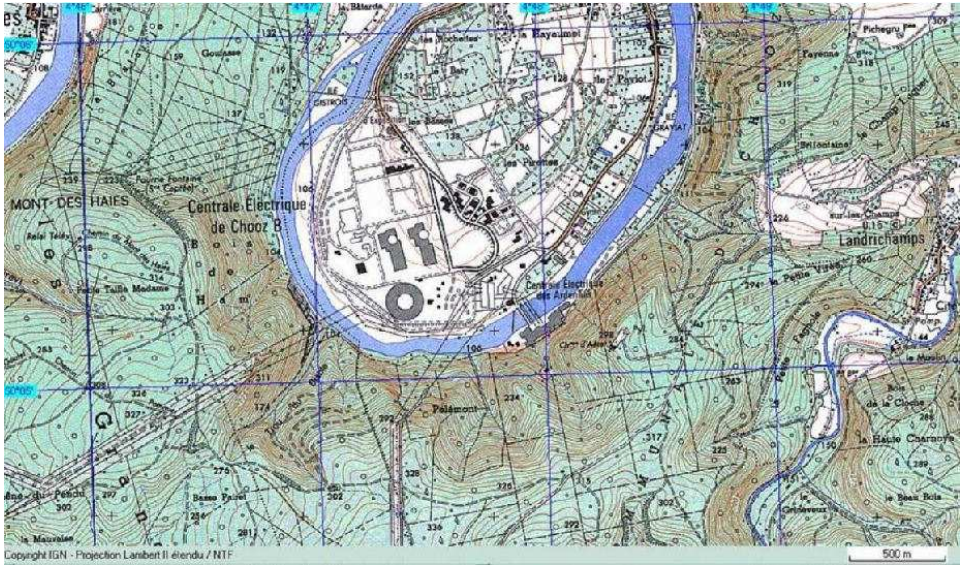}
\caption[Map of the experiment site]{Map of the experiment site. The
  two cores are separated by a distance of 100~meters. The far
  detector site is located at 1.0 and 1.1~km from the two cores.}
\label{fig:chooznearfarsitemap}
\end{center}
\end{figure}
\begin{figure}[h]
\begin{center}
\includegraphics[width=\textwidth]{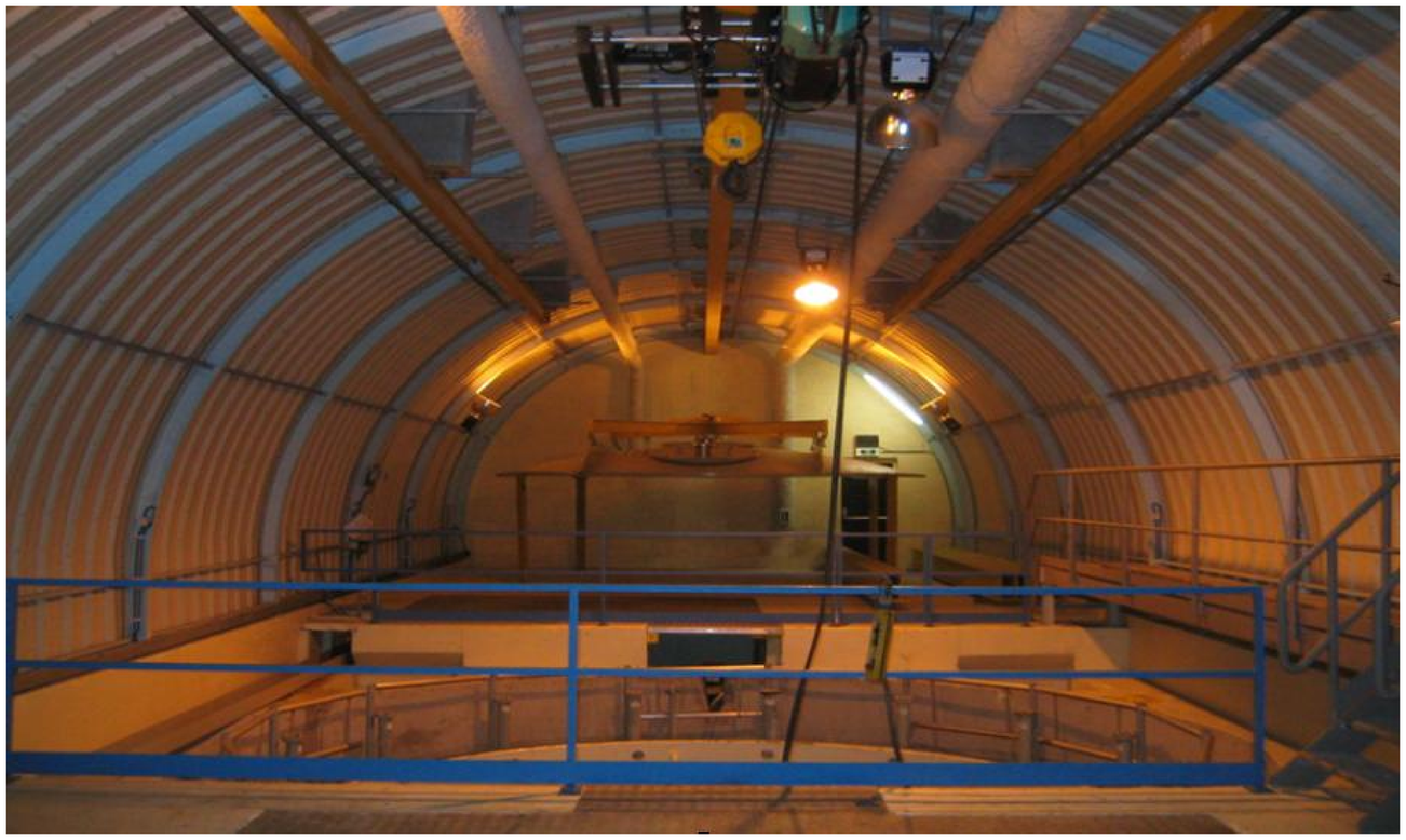}
\caption[Picture of the CHOOZ-far detector site]
{Picture of the CHOOZ-far detector site taken in September
 2003. The original CHOOZ laboratory hall constructed by EDF, 
 located close the the old CHOOZ-A underground power plant, is still in
 perfect condition and could be re-used without additional civil engineering 
 construction.}
\label{fig:choozfarfoto}
\end{center}
\end{figure}
We label this site the far detector site or {\bf CHOOZ-far}. 
A sketch of  the CHOOZ-far detector is shown in Figure~\ref{fig:choozfar}. 
The CHOOZ-far site is shielded by about 300~m.w.e. of 2.8~$\text{g/cm}^3$ rocks. 
Cosmic ray measurements were made with Resistive
Plate Chambers and compared with the expected angular
distributions. A geological study revealed the existence of several very
high density rock layers (3.1~$\text{g/cm}^3$ whose
positions and orientations were in agreement with 
the cosmic ray measurements~\cite{choozlast}).
It is intended to start taking data at CHOOZ-far at the beginning of the year 2007. \\

In order to cancel the systematic errors originating from the nuclear
reactors (lack of knowledge of the $\nuebar$ flux and spectrum), as well as to
reduce the set of systematic errors related to the detector and to the
event selection procedure, a second detector will be installed  close
to the nuclear cores. We label this detector site the near site or
{\bf CHOOZ-near}.
Since no natural hills or underground cavity already exists
at this location, an artificial overburden of a few tens of meters
height has to be built. The required overburden ranges from 53 to
80 m.w.e. depending on the near detector location, between 100 and 200
meters away from the cores (see Table~\ref{tab:edf}). 
A sketch of this detector is shown in Figure~\ref{fig:chooznear}. 
\begin{table}[h]
\begin{center}
\begin{tabular}{rrr}
\hline
\multicolumn{1}{c}{Distance} & \multicolumn{1}{c}{Minimal overburden} & \multicolumn{1}{c}{Required overburden} \\
          & \multicolumn{1}{c}{(m.w.e.)}           & \multicolumn{1}{c}{(m.w.e.)} \\
\hline
 100  & 45    & 53 \\
 150  & 55    & 65 \\
 200  & 67.5  & 80 \\
\hline
\end{tabular}
\caption[Overburden required for the near detector]
{Overburden required for the near detector. The second column is the
estimation of the minimal overburden required for the
experiment. The third column is minimal overburden added to a 
safety margin.}
\label{tab:edf}
\end{center}
\end{table}
After first discussions, this construction has been considered as
technically possible by the power plant company authorities. 
An initial study has been commissioned by the French electricity
power company EDF to determine the best combination of
location-overburden and to optimize the cost of the project.
Plan is to start to take data at CHOOZ-near at the beginning of the year 2008. \\
\section{Detector design}
The detector design foreseen is an evolution of the detector of the
first experiment~\cite{choozlast}. 
To improve the sensitivity of Double-CHOOZ with respect to CHOOZ 
it is planned to increase statistics and to reduce the systematic
errors and backgrounds.\\
%
\begin{figure}[h]
\begin{center}
\includegraphics[width=0.6\textwidth]{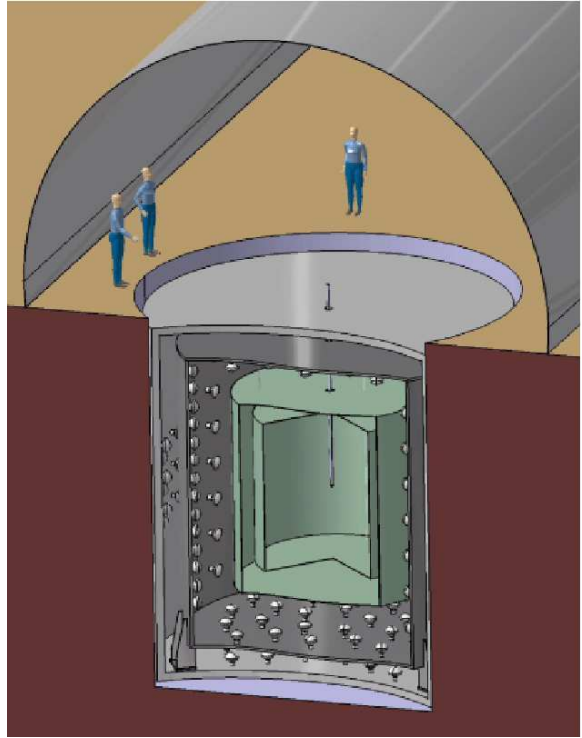}
\caption[CHOOZ-far detector]
{The new CHOOZ-far detector, at the former CHOOZ underground site. 
The detector is located in the tank used for the CHOOZ experiment (7 meters high and 7
meters in diameter) that is still available. About 12.7~$\text{m}^3$
of a dodecane+PXE based
liquid scintillator doped with gadolinium is contained in a 
transparent acrylic cylinder surrounded by the $\gamma$-catcher 
region and the buffer. The design goal is to achieve a light yield of 
about 200 pe/MeV (see Chapter~\ref{sec:fulldetsimul}) which requires an optical coverage of about 15~\%, 
provided by the surrounding PMTs.
The PMTs are mounted on a cylindrical structure which separates
optically the outer part of the detector, which is used as a muon veto.}
\label{fig:choozfar}
\end{center}
\end{figure}

In order to increase the exposure to 60,000~$\nuebar$  events at
CHOOZ-far  (statistical error of 0.4~\%) it is planned to use a target cylinder of 120~cm radius 
and 280~cm height, providing a volume of 12.7$~\text{m}^3$, $\sim$2.5 
larger than in CHOOZ. In addition, the
data taking period will be extended to at least three years, and the
overall data taking efficiency will be improved. The global load
factor of the reactor, i.e. the average reactor efficiency, is
about 80~\%, whereas it was significantly lower for the CHOOZ
experiment performed during the power plant commissioning. In
addition, the detector efficiency will be slightly improved.  
The background level at CHOOZ-far will be decreased to have a
signal to noise ratio over 100 (about 25 in CHOOZ). \\

The near and far detectors will be identical inside the PMTs supporting
structure. 
This will allow a relative normalization systematic error
 of  $\sim$0.6~\% (see Chapter~\ref{sec:systematics}). 
However, due to the different overburdens (60-80 to 300 m.w.e.), 
the outer shielding will not be identical since the
cosmic ray background varies between CHOOZ-near and CHOOZ-far. 
The overburden of the near detector has been chosen in order to keep
the signal to background ratio above 100.
Under this condition, even a knowledge of the backgrounds within a
factor two keeps the associated systematic error well below the percent
(assuming that its energy distribution is known). \\

The detector design has been intensively studied and tested with
Monte-Carlo simulations, using two different softwares derived from   
the simulation of the CHOOZ and Borexino experiments
(see Chapter~\ref{sec:fulldetsimul}). 
In order to increase the width of the liquid buffers
protecting the $\nuebar$ target, the 1~meter low radioactivity sand
shielding of CHOOZ will be replaced by a 15 cm metal shielding, steel or iron
(this is used to reduce the external gamma rays coming from the rock.) 
This will increase the size of the liquid active buffer and will thus
improve the rejection of muon induced backgrounds (see Chapter~\ref{sec:backgrounds}). 
Starting from the center of the target the detector elements are as
follows (see Figures~\ref{fig:choozfar} and~\ref{fig:chooznear})
\begin{itemize}
\item{\bf $\nuebar$ target}\\
A 120 cm radius, 280 cm height, 6-10~mm width acrylic cylinder,
filled with 0.1~\% Gd loaded liquid scintillator target
(see Chapter~\ref{sec:scintillator}).
\item{\bf $\gamma$-catcher} \\
A 60 cm buffer of non-loaded liquid scintillator with the same
optical properties as the $\nuebar$ target 
(light yield, attenuation length), in order to get the full
positron energy as well as most of the neutron energy released after
neutron capture.
\item{\bf Buffer}\\
A 95 cm buffer of non scintillating liquid, to decrease the
level of accidental background (mainly the contribution from
 photomultiplier tubes radioactivity).
\item{\bf PMT supporting structure}
\item{\bf Veto system}\\
A 60 cm veto region filled with liquid scintillator for the far
detector, and a slightly larger one (about 100~cm) for the near detector.
\end{itemize}
\begin{figure}[h]
\begin{center}
\includegraphics[width=0.9\textwidth]{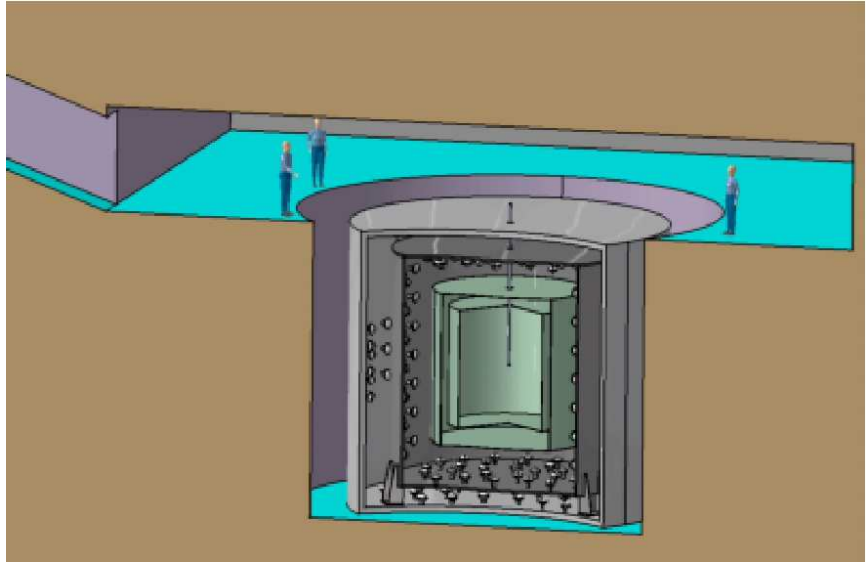}
\caption[CHOOZ-near detector]
{The CHOOZ-near detector at the new underground
site, close to the reactor cores. 
This detector is identical to the CHOOZ-far detector up to and
including  the PMT  surface. 
The veto region will be enlarged to better reject the cosmic 
muon induced backgrounds (see Chapter~\ref{sec:backgrounds}).}
\label{fig:chooznear}
\end{center}
\end{figure}
Compared to previous scintillator $\nuebar$ detectors, the
Double-CHOOZ experiment will use cylindrical targets; Monte-Carlo 
simulation shows that the spatial reconstruction in a cylinder is
suitable for the experiment. A spherical configuration gives
slightly better results, however. 
Each parameter of the detector is being studied by Monte-Carlo
simulation  in order to define the tolerance on the differences
between  the two detectors (see Chapter~\ref{sec:fulldetsimul}). 
The inner volume dimensions as  well as the shape of the target
vessels are still preliminary, within a few tens of percents, and
could change prior to the publication of the proposal. 
\subsection{Experimental errors and backgrounds}
In the first CHOOZ experiment, the total systematic error amounted to
2.7~\%~\cite{choozlast}.
Table~\ref{choozto2chooz_bis} summarizes the control of the
systematic uncertainties that had been achieved in the first CHOOZ experiment
as well as the goal of Double-CHOOZ. The main
uncertainties at CHOOZ came from the 2~\% only knowledge of the
antineutrino flux coming from the reactor. 
This systematic error vanishes by adding a near detector 
to monitor the power plant antineutrino flux and energy spectrum. 
A complete description of the systematic uncertainties is given 
in Chapter~\ref{sec:systematics}. 
The main challenge of the Double-CHOOZ experiment is  to decrease
the overall systematic error from 2.7~\% to 0.6~\%.
The strategy is to improve the detector design, to rely on the
comparison of the two detectors, and to reduce the number of analysis cuts.
The non-scintillating buffer will reduce the singles rates in each
detector by two orders of magnitude with respect to CHOOZ. This
allow to lower the positron threshold down to $\sim$500~keV, well below the
1.022~MeV physical threshold of the inverse beta decay reaction.
A very low threshold has three advantages:
\begin{itemize}
\item{The systematic error due to this threshold is suppressed.
It was one of the largest source of systematic error,  0.8~\% in CHOOZ~\cite{choozlast}.}
\item{The background below the physical 1~MeV threshold can be measured.}
\item{The onset of the positron spectrum provides an additional
  calibration point between the near and far detectors.}
\end{itemize}
This reduction of the singles events relaxes or even suppresses
the localization cuts, such as the distance of an event to the PMT
surface and the distance between the positron and the neutron.  
These cuts, used in CHOOZ~\cite{choozlast}, are difficult to calibrate
 and have to be avoided or relaxed in Double-CHOOZ.
The remaining event selection cuts will have to be calibrated between the two 
detectors  with a very high precision.  
Most important will be the calibration of the energy selection of the delayed neutron
after its capture on a Gd nucleus (with a mean energy release of 8 MeV
gammas). The requirement is $\sim$100~keV on the precision of this cut
between both detectors, which is feasible with standard techniques using
radioactive sources (energy calibration) and lasers (optical
calibrations) at different positions throughout the detector active
volume (see Chapter~\ref{sec:calibration}). The sensitivity of a reactor
experiment of Double-CHOOZ scale ($\sim$300~GW$_{\text{th}}$.ton.year) is mostly
given by the total number of events detected in the far detector. The requirement on the
positron energy scale is then less stringent since the weight of the
spectrum distortion is low in the analysis. (This is being studied by
simulation.)
\begin{table}[h]
\begin{center}
\begin{tabular}{lrr}
\hline
 & \multicolumn{1}{c}{CHOOZ} & \multicolumn{1}{c}{Double-CHOOZ}  \\
\hline
 Reactor cross section & 1.9~\% & ---   \\
 Number of protons     & 0.8~\% & 0.2~\%  \\
 Detector efficiency   & 1.5~\% & 0.5~\%  \\
 Reactor power         & 0.7~\% & ---    \\
 Energy per fission    & 0.6~\% & ---    \\
\hline
\end{tabular}
\caption[Systematic errors in CHOOZ 
and Double-CHOOZ goals]{Summary of the systematic errors
  in CHOOZ and Double-CHOOZ (goal). 
 The first line, ``Reactor cross section'', accounts for the
 uncertainties on the neutrino flux as well as the
 inverse neutron beta decay cross section.
 A two $\nuebar$ detectors concept makes the experiment largely
 insensitive to the ``Reactor cross section'' and the reactor power
 uncertainties.
 The number of protons in the first acrylic vessel targets as well as
  the detection efficiencies  have then to be calibrated  between the 
two detectors, but only in a relative sense.}
\label{choozto2chooz_bis}
\end{center}
\end{table}
\begin{figure}[h]
\begin{center}
\includegraphics[angle=-90 , width=\textwidth]{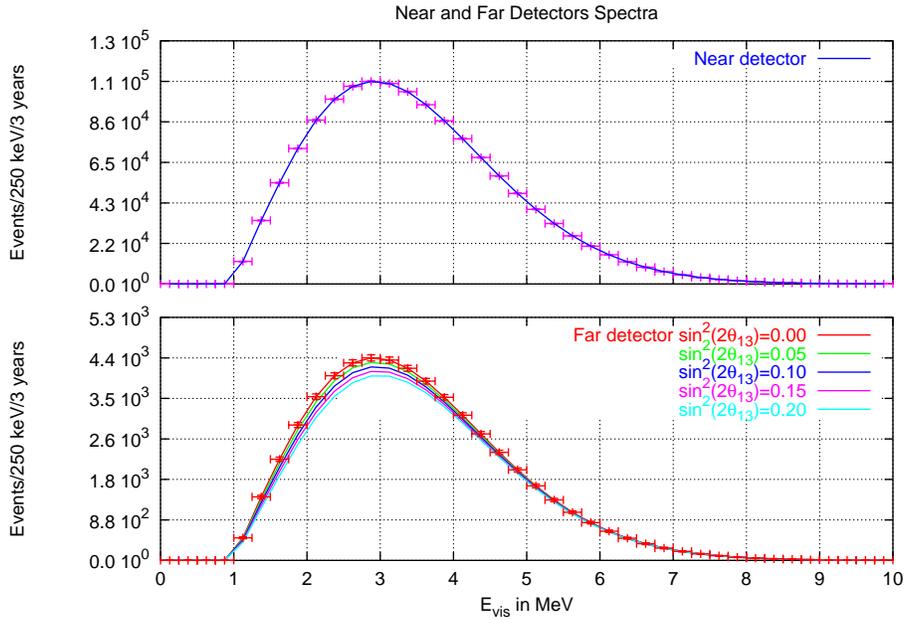}
\caption[Positron spectrum expected in both near and far detectors]
{Positron spectrum (visible energy, MeV) simulated for the CHOOZ-near and far detectors}
\label{fig:spectrumfar}
\end{center}
\end{figure}
A detailed background study is presented in Chapter~\ref{sec:backgrounds}.
In CHOOZ the dominant correlated proton recoil background was measured 
to be about one event per day~\cite{choozlast}. 
At CHOOZ-far the active buffer will be increased, with a solid 
angle for the background being almost unchanged. This together with
a signal increased by about a factor of 3 will fulfill the 
requirement of a signal to noise ratio greater than 100. 
At CHOOZ-near, due to the shallow depth between 60 and 80 m.w.e, the 
cosmic ray background will be more important. If, for instance, the CHOOZ-near
detector is located 150 meters from the nuclear cores, the signal will be
a few thousand events per day, while the muon rate is expected to be a
factor of ten less. A dead time of about 500~$\mu$sec will be applied to
each muon\footnote{This is a conservative number that could be
  reduced.}, leading to a global dead time of about 30~\%. A few tens
of recoil proton events per day, mimicking the $\nuebar$ signal, are
expected while the estimate of the muon induced cosmogenic events 
($^{9}$Li and $^{8}$He) is less than twenty per day with a large uncertainty (this
last point is being carefully studied). 
This fulfills the requirement of a signal to noise ratio greater than 
100 at CHOOZ-near.
\subsection{Sensitivity}
A detailed study of the Double-CHOOZ sensitivity is presented in 
Chapter~\ref{sec:sensitivity}.
From the simulations, we expect a sensitivity of 
$\s2t13 < 0.03$ 
at 90~\%~C.L. for $\adm2 =  2.0 \, 10^{-3}~\text{eV}^{2}$
(best fit value of Super-Kamiokande~\cite{SK_atm_nu2002}), 
after three years of operation. According to the
latest  Super-Kamiokande L/E analysis the best mass splitting is 
$\adm2=  2.4 \, 10^{-3}~\text{eV}^{2}$~\cite{SkAtmLoverE2004}. 
The Double-CHOOZ sensitivity would then be 
$\s2t13 < 0.025$
\footnote{The sensitivity of the other experiments is as well
  better for higher $\adm2$.}.
A study of the evolution of the sensitivity with respect to the
  luminosity is presented in Figure~\ref{fig:lscaling}~\cite{theta13globalana}.
\begin{figure}[h]
\begin{center}
\includegraphics[angle=-90 , width=\textwidth]{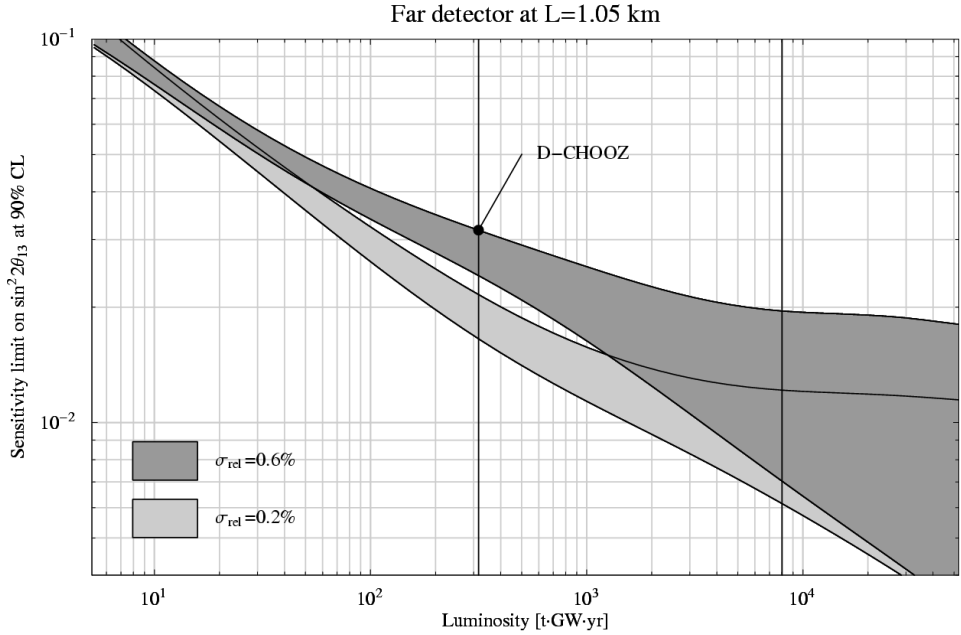}
\caption[Luminosity scaling of the Double-CHOOZ $\s2t13$ sensitivity at the
  90~\%~C.L..]
{Luminosity scaling of the Double-CHOOZ $\s2t13$ sensitivity at the
  90~\%~C.L.. Here, $\adm2 =  2.0 \, 10^{-3}~\text{eV}^{2}$ is assumed
to be known within 5~\%. The relative normalisation error between the
two detectors is taken to be (0.2~\%)0.6~\% for the light (dark)
shaded regions. Correlated backgrounds with known shapes account for
1~\% and are supposed to be known within~50~\%. A 0.5~\%
``Flat''bin-to-bin uncorrelated background component as been accounted
as well (known within 50~\%). A luminosity of 300~GW$_{\text{th}} \cdot
  \text{ton} \cdot \text{year}$ (left vertical line) correspond approximately to the setup of
  the Double-CHOOZ experiment as described in this Letter of
  Intent ($\s2t13 < 0.03$ within 3 years of data taking). 
However, a luminosity of  8000~GW$_{\text{th}} \cdot
  \text{ton} \cdot \text{year}$ (right vertical line) would describe
  a~$\sim$300~tons next detector generation at Chooz~\cite{theta13globalana}} 
\label{fig:lscaling}
\end{center}
\end{figure}
A sensitivity of $\sim$0.05 is reachable within the first
year of operation with 2 detectors.
These estimates are based on the assumptions that the
relative normalization error between the near and far detectors could
be kept at 0.6~\%, and that the backgrounds at both sites
amount to  1~\% of the $\nuebar$ signal (we assume
those backgrounds to be known within a factor of two).\\

The effect of $\nuebar$ oscillations on the positron spectrum
 is displayed in Figure~\ref{fig:ratiospectra1},
for different values of $\adm2$ and  $\s2t13$. 
For  $\adm2 \gsim 2.0 \, 10^{-3}~\text{eV}^{2}$
a significant shape distortion is expected at the onset of the energy
spectrum. Assuming $\sin ^2 (2 \theta_{13})= 0.15$, the ratio of the
 near and far detector spectrum is presented in
 Figure~\ref{fig:ratiospectra2}, with the expected statistical error
 bars (1~$\sigma$) after three years of data taking. 
\begin{figure}[h]
\begin{center}
\includegraphics[angle=-90, width=\textwidth]{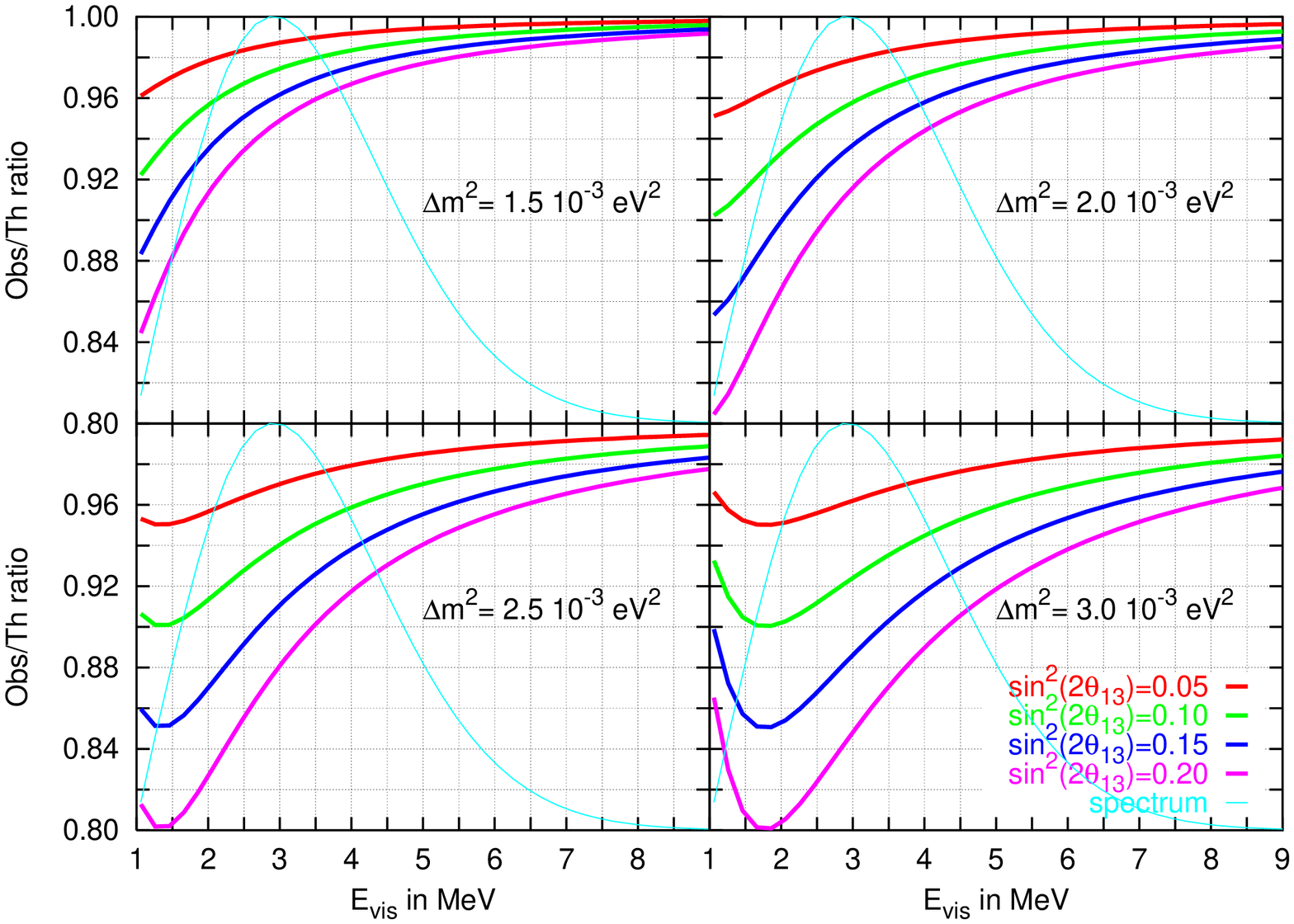}
\caption[Ratio of the expected number of $\nuebar$ events in the far and near detector]
{Ratio of expected number of $\nuebar$ events in the far detector with respect
  to the no oscillations scenario, after 3 years data taking,  
for different values of $\adm2$ \ and $\ssqtt$.
\label{fig:ratiospectra1}}
\includegraphics[angle=-90, width=\textwidth]{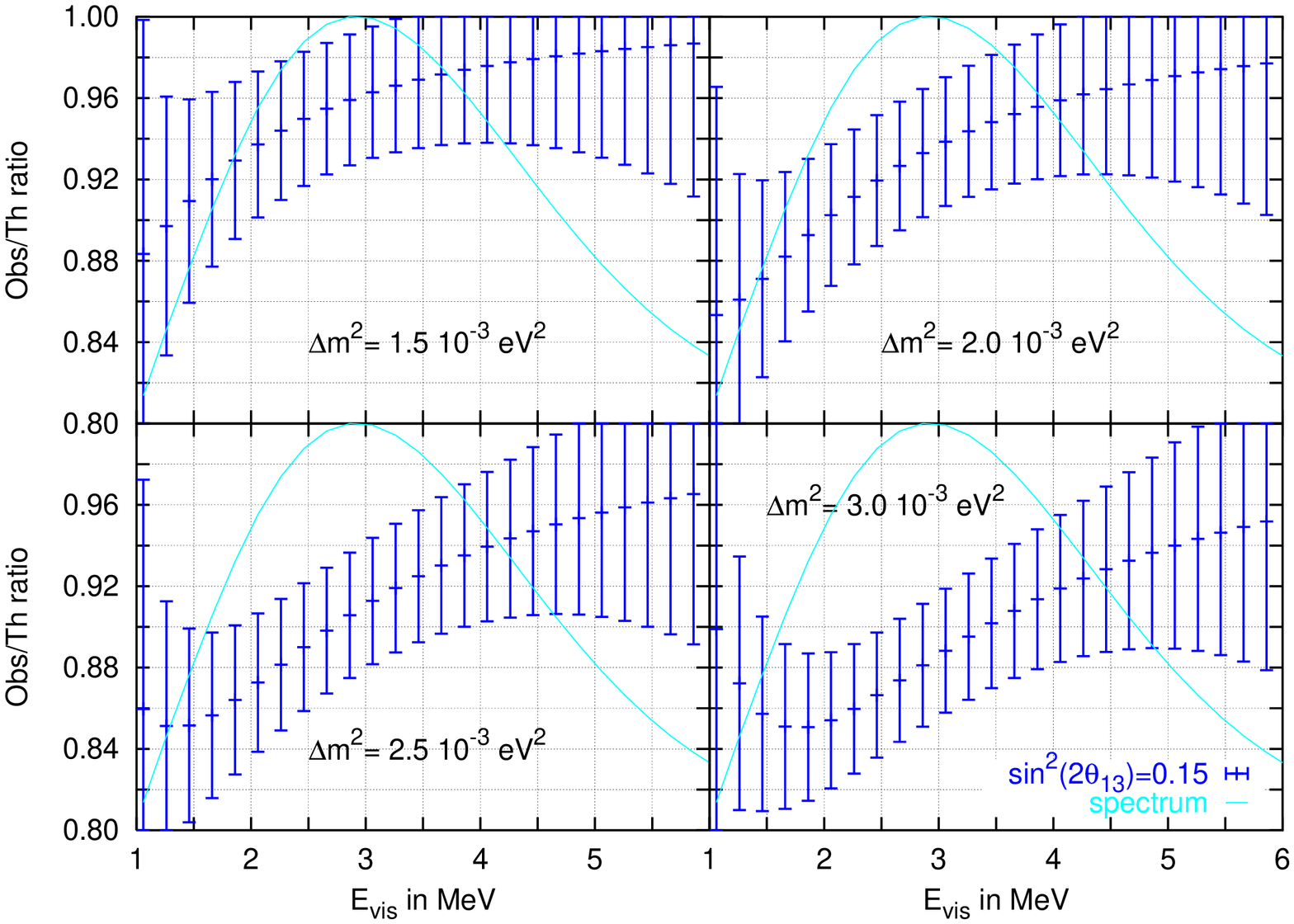}
\caption[Ratio of the expected number of $\nuebar$ events in the far and near detector: shape information]
{Ratio of observed number of events in the far detector  with respect
  to the no oscillations scenario, after 3 years data taking,
for $\ssqtt=0.15$ and different values of $\adm2$. The error bars
plotted here are only statistical ($1\sigma$). The positron spectrum shape is
also displayed in the background. The potential of the experiment to exclude $\ssqtt=0$ may be
seen as a deviation from unity in the ratio.
Note that in some cases spectral information may be important. 
The largest spectrum deviation effect is located
at the onset of the spectrum, below 4~MeV. \label{fig:ratiospectra2}}
\end{center}
\end{figure}
It is worth mentioning that the 1.05 km average baseline at CHOOZ is
not optimal (the optimal distance should be roughly 1.5 km) 
compared to the first oscillation maximum if the
atmospheric mass splitting is  $\adm2 = 2.0 \, 10^{-3}~\text{eV}^2$. 
Nevertheless, a new Super-Kamiokande analysis of
the data indicates  $1.9 \, 10^{-3}~\text{eV}^{2} <
\adm2 < 3.0 \, 10^{-3}~\text{eV}^{2}$ (90~\%~C.L.), 
with a best fit at $\adm2 = 2.4 \, 10^{-3}~\text{eV}^{2}$~\cite{SkAtmLoverE2004}. 
A shorter baseline is compensated by higher statistics for a 
fixed size detector. 
However, a value of $\adm2 < 1.5 \, 10^{-3}~\text{eV}^{2}$
would restrict the absolute potential of the Double-CHOOZ experiment
(see Chapter~\ref{sec:sensitivity}).
\cleardoublepage
\cleardoublepage
\chapter{Detector design and simulation}
\label{sec:fulldetsimul}
In this section we describe the detector design envisaged for the
Double-CHOOZ experiment. Although the generic design is almost
complete, some specific technical solutions are still preliminary and could
evolve prior to publication of the proposal.
\section{Detector design}
Detector dimensions are shown in Figure~\ref{fig:detectorsize}.
\begin{figure}[h]
\begin{center}
\includegraphics[width=0.8\textwidth]{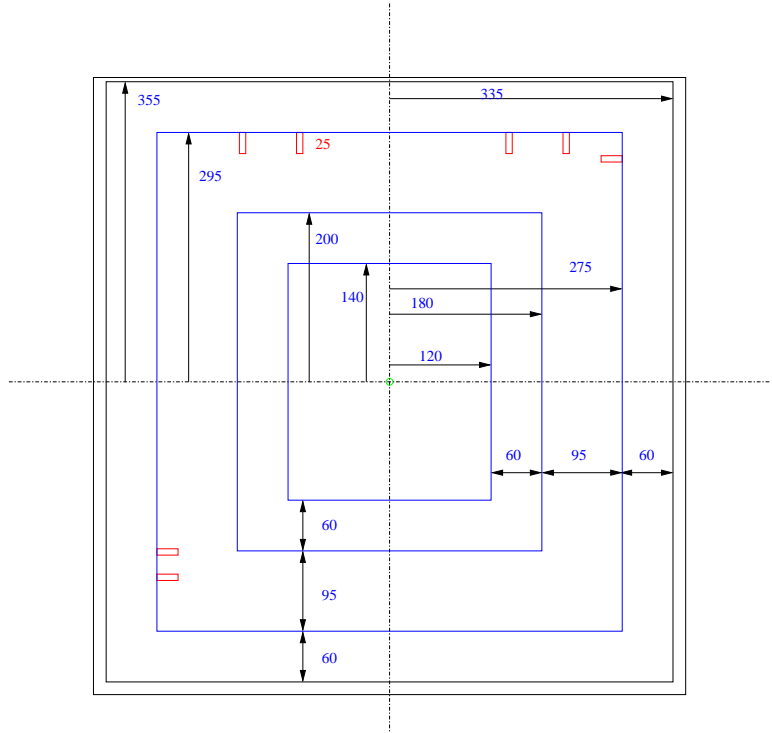}
\caption[Dimensions of the CHOOZ-far detector]
{Dimensions of the CHOOZ-far detector (in cm). Starting from the
  center we have: the neutrino target region composed of Gd doped liquid
  scintillator (12.7~m$^3$), the $\gamma$-catcher region composed of
  unloaded liquid scintillator (28.1~m$^3$), 
 the non scintillating buffer region (100.0~m$^3$),
  and the veto (110.0~m$^3$). The CHOOZ-near detector is
  identical up to and including the PMT support structure; however, 
  its external muon veto is slightly larger to better reject the
  cosmic muon induced backgrounds. The exact PMT positioning has not
  been chosen yet.}
\label{fig:detectorsize}
\end{center}
\end{figure}
\subsection{The $\mathbf{\nuebar}$ target acrylic vessel (12.67~$\mathbf{\text{m}^3}$)}
The neutrino target is a 120~cm radius 280~cm height transparent acrylic  cylinder,
filled with 0.1~\% Gd loaded liquid scintillator (see Chapter~\ref{sec:scintillator}).
Wall thicknesses (under study) range from 6 to 10~mm. The inner acrylic vessel
is depicted in Figure~\ref{fig:acrylics} (left). Since the relative
volume between the two inner acrylic vessels has to be controlled at a
very accurate level (0.2~\%), we plan to build both acrylic vessels at
the manufacturer site and to move them as single units
into the detector sites. This is possible for the far site, thanks to
the size of the underground tunnel. The near detector site will be
designed in order to allow this operation. With this strategy, no
acrylic welding or gluing has to be done on site, thus reducing the
uncontrolled differences between the two envelopes.
Furthermore, a very precise calibration of both inner vessels is  foreseen
at the manufacturer (filling test).
Current R\&D focuses on the chemical compatibility between acrylic and
liquid scintillator (see Chapter~\ref{sec:scintillator}). 
Preliminary stress calculations have been done
for this purpose (see Figure~\ref{fig:acrylicstress}). \\
\begin{figure}
\begin{center}
\includegraphics[width=0.6\textwidth]{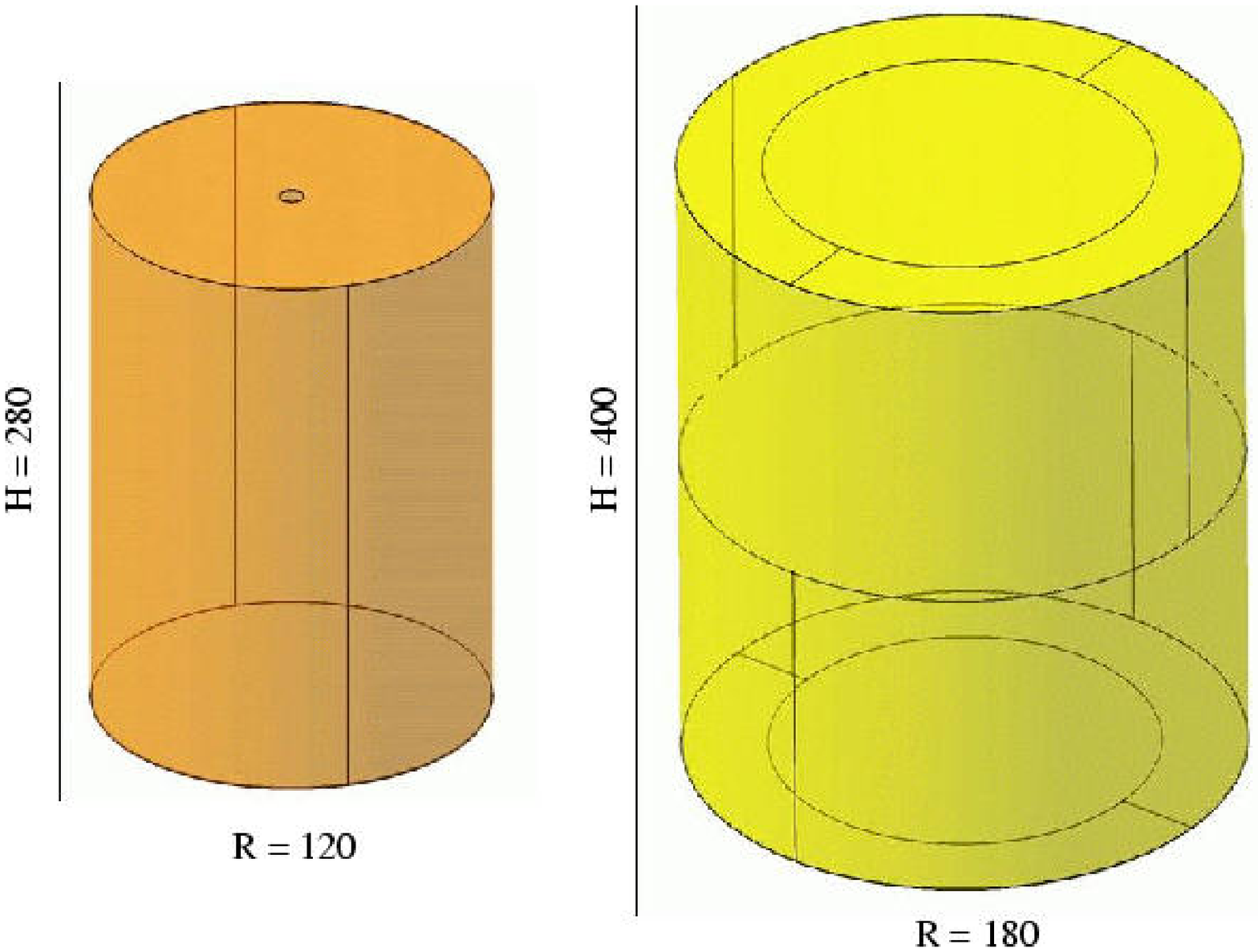}
\caption[Sketch of the two acrylic vessels containing the Gd doped and
  undoped scintillators]{The two acrylic vessels containing the Gd doped and
  unloaded scintillators. The lines drawn on the cylinders show the
  preliminary positioning of the welding joint between the acrylic pieces.
The inner envelopes will be constructed at the manufacturer and
  transported as single units to the detector sites while the outer
  envelopes will have to be assembled on sites.}
\label{fig:acrylics}
\end{center}
\end{figure}
\begin{figure}
\begin{center}
\includegraphics[width=0.5\textwidth]{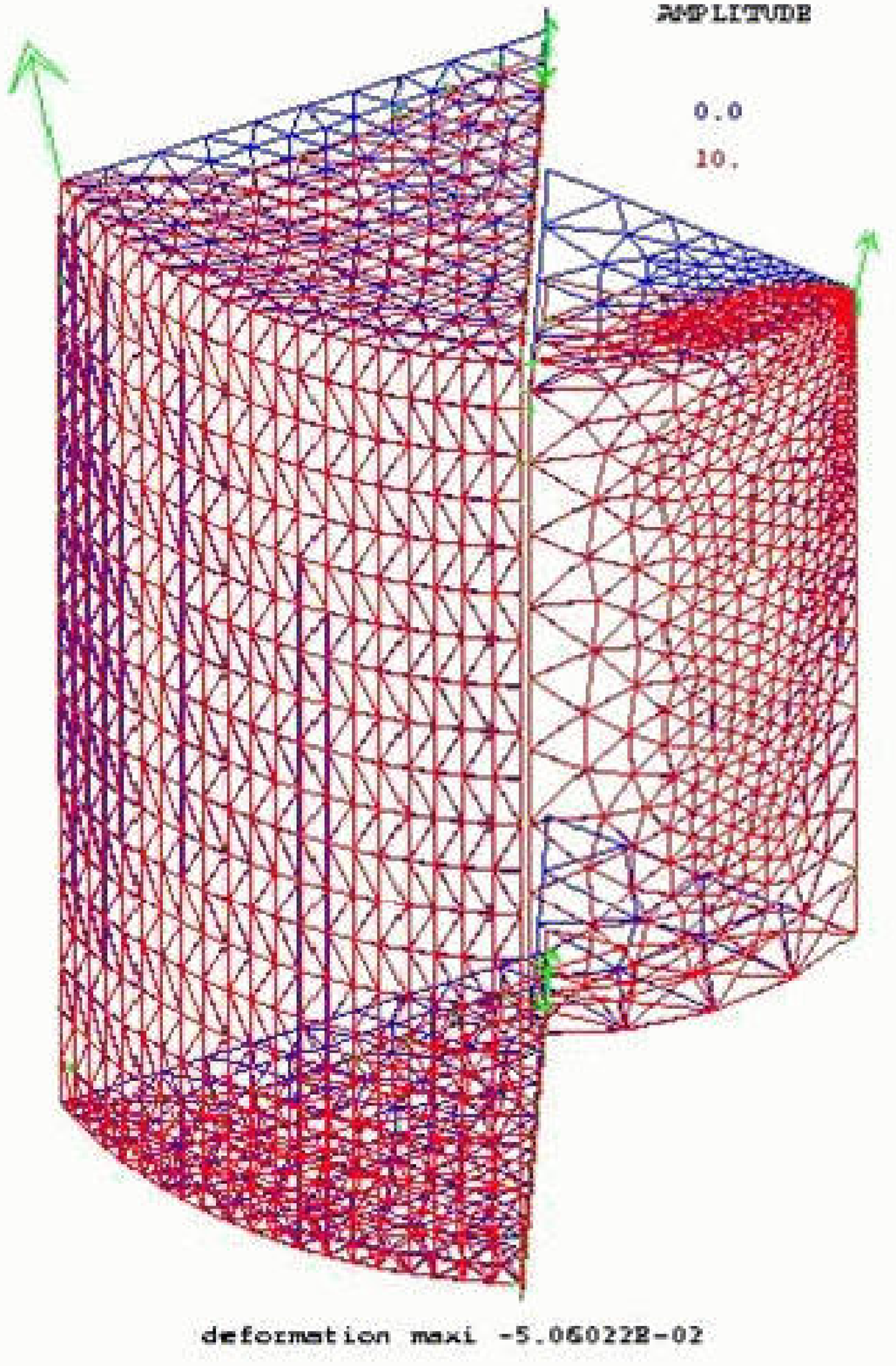}
\caption[Preliminary evaluation of the stress applied on an empty acrylic
  cylinder suspended with three kevlar ropes]{Preliminary evaluation
  of the stress applied on an empty acrylic cylinder suspended with
  three kevlar ropes (set at 120 degrees from each other). 
 The  maximum stress has been estimated at 12~MPa,
 while acrylic supports a maximum of $\sim$24~MPa in the elastic regime.}
\label{fig:acrylicstress}
\end{center}
\end{figure}
\subsection{$\gamma$-catcher  acrylic vessel (28.1~$\mathbf{\text{m}^3}$)}
The $\gamma$-catcher is a 180~cm radius and 400~cm
height acrylic cylinder filled with non-loaded
liquid scintillator, which has the same optical properties 
as the $\nuebar$ target (light yield, attenuation length). 
Unlike the inner envelope, this second acrylic vessel will have to be
partially assembled on site. Nevertheless, the shape and dimensions
between far and near $\gamma$-catcher are less critical than for the
inner vessels. Therefore, small differences between
the near and far $\gamma$-catcher  acrylic vessels could be tolerated
(a Monte-Carlo study is being done to provide the construction 
tolerance). \\

\noindent
This scintillating buffer around the target is necessary to:\\

\noindent
   {\bf Measure the gammas from the neutron capture on Gd. \\}
   The total released energy is 8~MeV, with a mean gamma multiplicity
   of 3 to 4. But there are also some 8~MeV single gammas. The buffer must be
   thick enough to reduce the gamma escape out of the sensitive
   volume, i.e. the target and the $\gamma$-catcher. 
   This escape creates a tail below the 8~MeV peak. Since we must
   apply an energy cut to define the neutron capture on Gd, 
   the tail of the energy spectrum has to be small enough
   to keep the systematic error negligible (if there is an energy scale
   mismatch between both detectors). Monte-Carlo simulations with a
   60~cm buffer and a 100(150)~keV energy
   error gives 0.2(0.3)~\% difference in the neutron counting, which is tolerable.\\

\noindent
   {\bf Measure the positron annihilation. \\ }
   To have a clean
   threshold at 500~keV, it is mandatory to have very few events with an
    energy below 1~MeV. From the simulation, a thickness of 35~cm is adequate.\\

\noindent
   {\bf Reject the background. \\}
   This is the most demanding constraint.
   One of the most severe background in Double-CHOOZ comes from very fast neutrons, 
   created by muons crossing rocks near the detector (see Chapter
   \ref{sec:backgrounds}). To be able to reach
   the target traveling through the 2~m buffers, these neutrons must
   have an energy greater than 20~MeV.
   So when arriving at the scintillating buffer, they often deposit more
   than 8~MeV in the sensitive volume. This provides a useful 
   rejection, by a factor of $\sim$2. In the simulation, this rejection was
   seen to be stable for large buffer thickness, and to decrease when
   this thickness is reduced below 60~cm. Another advantage of this thickness
   is to allow to scale the result of the first experiment, since the
   sensitive volume around the target will be the same in both experiments
   (the veto volume was not sensitive to low energy events in the first
   experiment).
\subsection{Non scintillating buffer (100~$\mathbf{\text{m}^3}$)}
  The non-scintillating region aims to decrease the level of the
  accidental backgrounds, mainly due to the contribution from
  the photomultiplier tubes (see Chapter~\ref{sec:backgrounds}).
  To define the size of this region, we have to consider the
  following constraints:
\begin{enumerate}
  \item{The fast neutron background implies to keep the distance from 
  the rock to the neutrino target at least as it was for the CHOOZ
  experiment case\footnote{The target vessel is seen from outside of the
  detector under a similar solid angle in both experiments.}.
   Scaling from the CHOOZ experiment, we thus need at least 215~cm of liquid
  from the rock to the target.}
  \item{The size of the target has been chosen to be 120~cm to
  decrease the statistical error down to 0.4~\%, after three years of operation.}
\end{enumerate}
  From those constraints, the total thickness of the veto and the
  non-scintillating buffer has to be smaller than 155~cm. 
  Accounting for the size of the laboratory (mechanical constraints)
  and  the requirement   to have an efficient veto, we choose the    
  thickness of the veto to be around 60~cm. 
  From those considerations, the non-scintillating buffer region
  reduces to 95~cm. 
  The simulation shows that this configuration fulfills our
  requirement on the accidental background level tolerated
 (which is mainly driven by PMT radioactivity). \\    
\subsection{PMTs and PMT support structure}
The PMT support structure is a 275~cm radius and 590~cm
height cylinder (material under study) filled with the same liquid as
the $\gamma$-catcher, mixed with a quencher (DMP for instance). \\

\begin{figure}
\begin{center}
\includegraphics[width=0.65\textwidth]{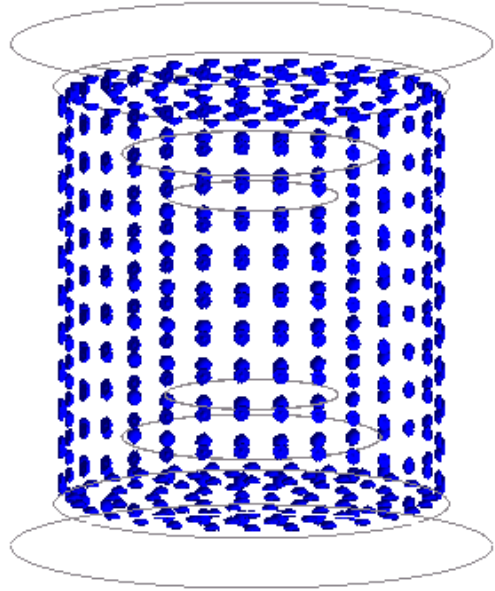}
\caption[Surface of PMTs mounted on the support structure]
{Surface of PMTs mounted on the support structure of the    
detector as  described in the GEANT4 simulation. About 500~PMTs are
displayed.}
\label{fig:pmtsurface}
\end{center}
\end{figure}
From the simulation, 500~PMTs of 8" are necessary to achieve a
photoelectron yield of $\sim$200~photoelectrons per MeV. 
Another possibility would be to  use a smaller number of larger 
PMTs, 10", 12" or 13" for instance.
The reference PMT is the photomultiplier 9354KB of ETL
\cite{ETL}. The glass used has a very low activity 
(60~ppm in K, 30~ppb in Th and U), and the quantum efficiency peaks 
at about 28~\% at 430~nm. 
For those PMTs, the peak-to-valley ratio of the single photoelectron 
signal is typically~2 (1.5~guaranteed by the manufacturer). 
Since we expect 600~photoelectrons for a medium energy signal of 3~MeV 
(visible energy), there will always be an important fraction of the 
PMTs working in the single photoelectron regime. 
The electronics gain is in the $10^{6}-10^{-7}$~range, hence some additional 
amplification is required in the front-end electronics system to
obtain a good single photoelectron peak definition (additional dynodes could
also be a way to increase the gain).
Photonis as well as Hamamatsu PMTs are under study. The final
photomultiplier choice will be made in 2004, during the design phase.
\subsection{Veto (110~$\mathbf{\text{m}^3}$)}
The external veto is contained in a steel cylinder of 350 cm radius and 710 cm
height. The veto thickness is 60~cm for the far detector. It can
be enlarged for the near detector, to better reject the cosmic muon induced
backgrounds, since the laboratory has to be
build. This tank is shielded by 15~cm of steel in order to reduce 
the external backgrounds.
\section{Fiducial volume}
\subsection{Definition of the fiducial volume}
A neutrino interaction in this detector will be tagged
by the  neutron capture on gadolinium (as was the case 
in the first CHOOZ experiment \cite{choozlast}). 
This is the main advantage of using a gadolinium loaded scintillator. However,
there is an additional effect to consider, the {\it spill in/out}, 
that leads to a compensation between two kinds of $\nuebar$
interactions:
\begin{itemize}
\item{The $\nuebar$ interacts in the inner acrylic target, near the vessel, but the
  neutron escapes the target, and is captured on hydrogen in the
  $\gamma$-catcher. In that case, there is no Gd capture to
  characterize the neutrino interaction, and this is thus 
  not selected as a neutrino event.}
\item{The $\nuebar$ interacts in the $\gamma$-catcher, not too far
  from the target, but the neutron enters the target and is captured on Gd.
  The neutrino interaction vertex is not in the target, but there is a well
  measured positron event followed by a Gd capture signal.
  This interaction is thus selected as a neutrino event.}
\end{itemize}
These two kinds of events do not compensate
exactly. However, the simulation shows that the difference is of the
order of $\sim$1~\% of the total neutrino interaction rate (the 
software used for this simulation is a low energy neutron Monte-Carlo 
that was extensively used and checked for the Bugey experiments~\cite{Bugey}). 
This imperfect compensation is due to the presence of
gadolinium in the target only. But, since this corresponding 
cross section is high only at epithermal neutron energies, the
neutrons slow down identically in both media. The difference of
behavior happens only in the last few centimeters of the neutron
path, before its capture. 
This spill in/out effect would lead to an irreducible $\sim$1~\% 
systematic uncertainty in a new single detector experiment. However,
it will cancel in the Double-CHOOZ  oscillation analysis
since two identical detectors will be used. 
Nevertheless, a second order spill in/out difference will remain
in Double-CHOOZ since the neutrino direction with respect to the
neutrino target boundary changes slightly between the two detectors. 
Indeed, this small effect comes from the correlation of the 
$\nuebar$ and the neutron directions \cite{choozlast}.\\

In conclusion, the method used to identify a neutrino interaction 
 allows a very good definition of the number of target atoms. The
 major concern is the design, the construction and the monitoring 
of the inner acrylic cylinders.
\subsection{Measurement of the fiducial volume} 
\label{subsec:volumemes}
We have to measure the volume of the inner acrylic vessels with an
uncertainty below 0.2~\%. 
With standard commercial materials such as dosing pumps, it is hard to
have an absolute volume determination better than 0.5~\%. We thus
suggest to use a combination of direct and indirect measurements 
to obtain the required precision.\\

A possible solution is to use weight measurements. For this, an
intermediate vessel close to the acrylic target is necessary
(in the experimental hall). We plan to measure first the weight of the empty
intermediate vessel, then fill the target vessel and re-measure its
weight. 
The difference of the two measurements indicates very accurately the
mass of liquid used to fill the target. 
Associated with a density measurement, this could provide  the volume
measurement with uncertainty below 0.1~\% 
(below 10 kg on the mass determination, and around $10^{-4}$ on
the density measurement).\\

A second method under study consists to use pH measurement. This
measurement has to be done with an acid/water mixture. 
It seems that this method can reach an 0.2~\% accuracy.\\

Independently of the volume of liquid used to fill the vessels, both
detector have to be kept at the same temperature. We
will thus have to monitor and control it. 
A simple regulation loop in the external veto is foreseen. 
\section{Light collection}
We consider in the following a concentric cylindrical model of the
 Double-CHOOZ detectors consisting of the target, the $\gamma$-catcher
 and the outer buffer. 
The target volume of the detector is filled with organic  liquid
 scintillator (LS)  loaded with Gadolinium (Gd) consisting
  of a mixture of 
\begin{itemize}
\item{PXE as solvent with small amount of Gd (1~g/l),}
\item{PPO (2,5-diphenyloxazole) as first fluor with a concentration of
  6.0~g/l, } 
\item{bis-MSB with concentration  of 0.02~g/l as second fluor or 
  wavelength shifter.}
\end{itemize}
The volume of the $\gamma$-catcher enclosing the target is filled with the
 same LS but without admixture of the Gd salt.   
The tank containing the non scintillating buffer is covered by a
 reflecting material, and about 500 PMTs are installed on this surface
(later called the PMT surface). Figure~\ref{fig:pmtsurface} 
displays the PMTs mounted on the support structure. \\

We consider the reflection coefficient (k) as a free parameter; it can be
changed within the interval from 0 (absolutely black surface) up 
to 0.98 (mirror reflection by VM2000 film \cite{lensdario, 3M}).
Charged particles deposit energy in the LS medium, mostly due to their
interaction with the solvent molecules. PXE excited molecules
transfer their energy to the PPO molecules via non-radiative
processes. Then, an energy transfer occurs between the PPO and the
bis-MSB, mainly by radiative transitions (100~\%~probability).
Therefore the primary (observed) fluorescence of LS is connected with
the radiative decay of the bis-MSB excited molecules. The energy
spectrum of the photons emitted by the shifter is shown in Figure~\ref{fig:bismsbemission}. 
\begin{figure}[h]
\begin{center}
\includegraphics[width=0.8\textwidth]{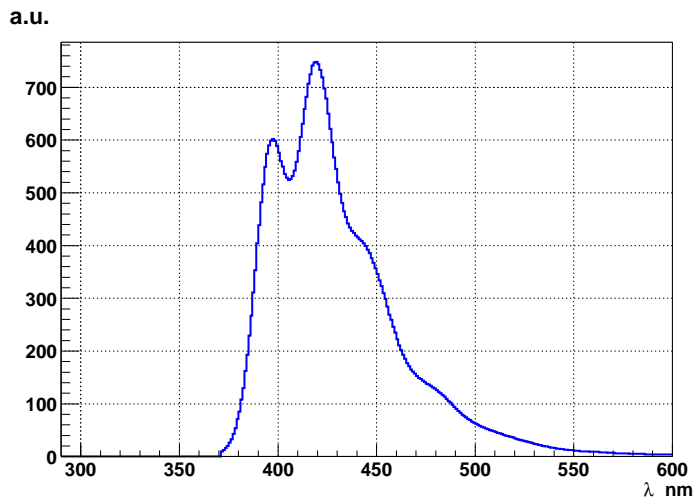}
\caption[Emission spectrum of the bis-MSB wavelength shifter]
{Emission spectrum of the bis-MSB wavelength shifter.}
\label{fig:bismsbemission}
\end{center}
\end{figure}
The radiative transport from the light emission vertex to the PMTs is
described  by a GEANT4 Monte-Carlo simulation. Borexino-like PMTs
cover between 12.5~\% to 17.5~\%  of the surface of the supporting cylinder. 
The quantum efficiency of the photocathode is shown on
Figure~\ref{fig:pmtefficiency}.
The Monte-Carlo simulation developed for this work is based on the light
propagation model described in \cite{nima440} and \cite{jbbirks}, and
has been used for the Borexino experiment. 
The time decay of the emitted bis-MSB photons is described
 phenomenologically by the sum of few exponentials having time
constants $\sim$5~ns.  
The light yield of the LS is taken to be 8,000~photons per~MeV
\footnote{This is 2/3 of the standard unloaded pure PC
  scintillator.}, both for the target and for the $\gamma$-catcher
of the detector.
The photons emitted by the bis-MSB  propagate through the target volume,
and interact with PXE, PPO, bis-MSB and Gd salt molecules.
Two physical processes have been taken into account: 
\begin{itemize}
\item{(Rayleigh) elastic scattering,}
\item{absorption.}
\end{itemize}
\begin{figure}[h]
\begin{center}
\includegraphics[width=0.8\textwidth]{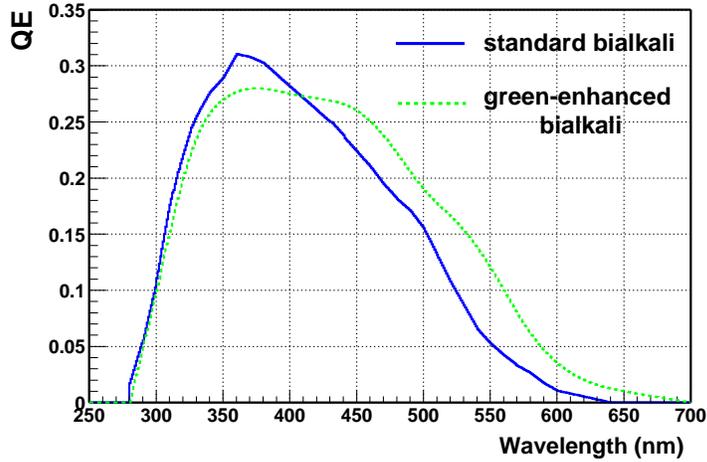}
\caption[Quantum efficiency of the PMT photocathode]
{Quantum efficiency of the PMT photocathode.}
\label{fig:pmtefficiency}
\end{center}
\end{figure}
The light attenuation was described by an exponential function with
the extinction coefficient $\mu (\lambda) = \mu _\text{a} (\lambda) + \mu _\text{s} (\lambda)$, 
where $\mu _\text{a}(\lambda)$ is the absorption coefficient, $\mu _\text{s}(\lambda)$ 
the scattering coefficient and $\lambda$ the wavelength of the light. 
The mean free path of the photon is equal to 
$\Lambda(\lambda) = 1 / (\log{( m \times \mu (\lambda))})$, where m 
is the molar concentration of the relevant scintillation component.
The cross sections for these interactions have been extracted from the
experimental data, obtained by usual spectroscopy methods.
\begin{figure}[h]
\begin{center}
\includegraphics[width=0.8\textwidth]{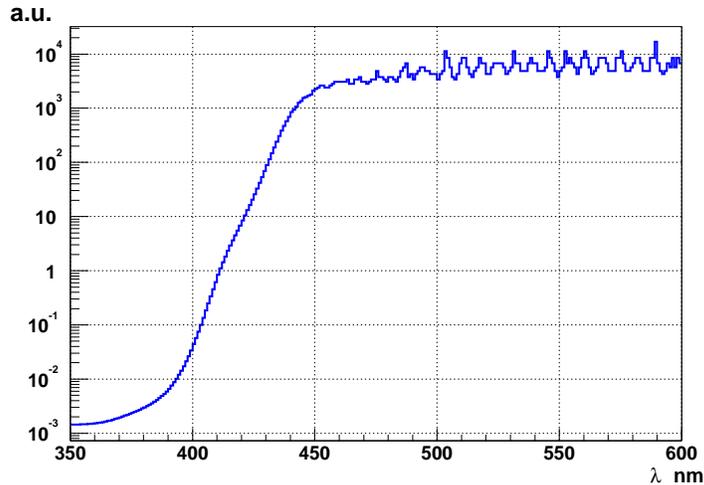}
\caption[Absorption spectrum of the bis-MSB wavelength shifter]
{Absorption spectrum of the bis-MSB wavelength shifter.}
\label{fig:bismsbabsorption}
\end{center}
\end{figure}
An example of $\Lambda(\lambda)$ variation for bis-MSB is presented in 
Figure~\ref{fig:bismsbabsorption}. Two different behaviors can be seen.   
At wavelengths longer than 450~nm the absorbance drops rapidly and the
measured extinction coefficients are practically equal to the coefficient for
Rayleigh scattering, while at wavelengths shorter than 450~nm
photons absorption is the main interaction process. 
Elastically scattered photons have an angular distribution described as
 $1 + \cos ^2{\theta}$, independent of the wavelength. 
The process of light absorption can be accompanied by an isotropic 
re-emission of the photons. 
The spectrum of re-emitted photons and time of the
re-emission process were taken equal to the fluor primary spectrum and
light time decay (1.3~nsec for bis-MSB). The re-emission probability was 
assumed to be equal to the absorbing molecule quantum efficiency taken 
around 0.36 for PXE, 0.8 for PPO and 0.96 for bis-MSB. 
This absorption/re-emission process can occur several times until
either the photon is  absorbed in the scintillator volume (its
energy disappears due to the non-radiative processes) or its wavelength
 falls in a region where the absorption probability is negligible. 
Photon reflection (or absorption)  near the wall of  the buffer is
described by the reflection coefficient. 
As a result of the transport process, a part of the photons reaches
photocathode surface of the PMTs. The spectrum of these photons 
is shown in Figure~\ref{fig:wavelengthatpmt}. 
It can be seen that the left part of
 the spectrum decreases more rapidly with respect to the emitted
 spectrum of bis-MSB; this is connected with the self-absorption of 
bis-MSB molecules.  
\begin{figure}[h]
\begin{center}
\includegraphics[width=0.8\textwidth]{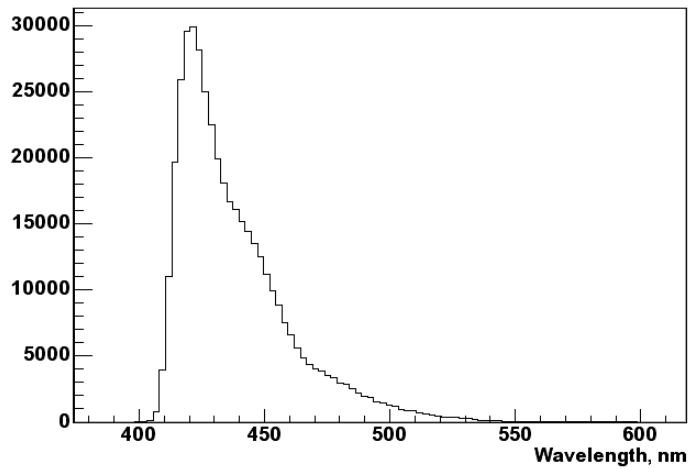}
\caption[Spectrum of the photons as they arrive at the PMT surface]
{Spectrum of the photons as they arrive at the PMT surface.}
\label{fig:wavelengthatpmt}
\end{center}
\end{figure}
The results of the simulation are presented as a number of
photoelectrons per MeV of energy deposit, from point like events
generated inside the target and the $\gamma$-catcher to PMT-visible 
photons that propagate to the photocathode of PMTs.
If the buffer wall is black (reflection coefficient k=0) the
number of photoelectrons was found to be around 300 for events in the
target center (for a 17.5~\% coverage). This number increases up
to 40~\% if the buffer wall is reflective. The light collection time distribution
 is shown in Figure~\ref{fig:lightcollection}. Obviously, the reflected light
increases the tail of the time distribution. During the first 30~ns, 
all photoelectrons arise from the photons that directly reach the
PMT surface.
The simulation shows a very good light collection homogeneity. 
The dependence of light collection from the event position inside 
the target was found to be within 5~\% and increased up to 10~\% at the
 position near the walls of the $\gamma$-catcher. The collection of the
 reflected light improves the homogeneity and for a reflection
 coefficient of 0.8, the light collection is very homogeneous (+2-3~\% at the target 
border, +5~\% at the scintillating buffer border).\\
\begin{figure}[h]
\begin{center}
\includegraphics[width=0.8\textwidth]{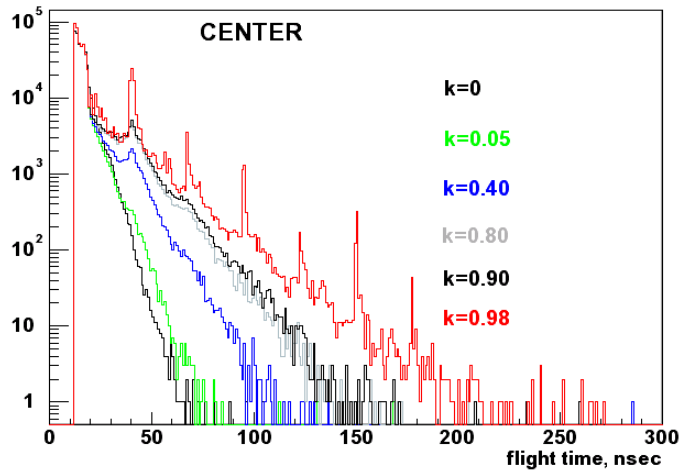}
\caption[Light collection for different reflectivity coefficients of the PMT
support structure]{Light collection for different reflectivity
coefficients  of the PMT support structure ranging from k=0
(black paint) to k=0.98 (VM2000 foil \cite{lensdario, 3M}).}
\label{fig:lightcollection}
\end{center}
\end{figure}
Detector design, the number of PMTs and their positioning is now
being optimized based on the Monte-Carlo simulation presented here. 
\section{Electronics}
\subsection{Data recording}
The following data have to be recorded:
\begin{itemize}
\item{Charge and time for each PMT.}
\item{Pulse shape for PMT clusters, to identify recoil protons due to fast neutrons. }
\end{itemize}
Neutrino events are made of two light pulses, separated by a delay
of a few~$\mu$s to 200~$\mu$s. The single trigger rate, although lower
than $10$~Hz can still be reduced using the delayed coincidence, as
explained below. This imply to store the data of the two pulses
before the trigger decision. Furthermore, for calibration
with Cf sources, several neutrons are detected after the fission signal,
with an average multiplicity of 4, extending beyond 8. The
dead time of the system must be kept low, stable and simple to control,
since it contributes to systematic error. \\

The front-end electronics will have to:
\begin{itemize}
\item{Separate the signal from the high voltage, if positive.}
\item{Amplify the signal by a factor $\sim$50 to use the single
photoelectron range (PMTs stability monitoring).}
\item{Add the analog signals (the total sum over the detector PMTs will be 
used in the trigger).}
\item{Include a discriminator per PMT (to monitor the trigger stability).}
\end{itemize}
The minimal solution for digitization is to use multihit charge ADCs, shapers
and multihit TDCs for all channels, completed by Flash-ADCs for a few tens
of PMT groups. The alternative would be to use Flash-ADCs for all
channels, build PMT clusters and emulate ADCs and TDCs. A new a model
of Waveform recorder with a smart memory management is being developed
for Double-CHOOZ. It will provide a multihit capability and
digitization with zero dead time. This is an upgrade
of an existing model used for Borexino (the prototype will be ready in 2004).
\subsection{Trigger logic}
The plan is to keep the trigger logic as simple as possible. It will
be based on a rough energy measurement made by the analog sum of all
PMT signals. A first level (single pulse) trigger will feature two channels:
\begin{enumerate}
\item The ``particle'' channel: a pulse of 0.5-50~MeV, which
will cause the recording of all channels. 
\item The ``muon'' channel: a pulse above 50~MeV {\it or} a signal
in the veto which will cause the recording of time and energy
information in a digital LIFO.
\end{enumerate}
The data are read out for all first level triggers and a second level
trigger (final) is made online with the coincidence of two
``particle'' triggers,  within 200~$\mu$s. A final event is composed of two singles,
including information about the last muons.
The data for each ``particle'' will be composed of
\begin{itemize}
\item{the charge and time for all PMTs,}
\item{the pulse shapes for $\sim$16 PMT groups (multiplicity tunable by software).} 
\end{itemize}
In addition, during data taking, some artificial light pulse patterns
will be generated inside the target, using laser or LEDs. 
These artificial events will mimic the physical $\nuebar$, in order to 
monitor the trigger system efficiency. Of course, each event triggered will
carry a specific tag and serial number, for its identification in
the offline data analysis. Table~\ref{tabelectronics} summarizes the
expected rates for neutrino like triggers.
\begin{table}[h]
\begin{center}
\begin{tabular}{lr}
\hline 
Neutrinos&
0.04~Hz\\
Artificial&
0.05~Hz\\
Multineutron after a muon&
0.3~Hz\\
Cosmogenic&
0.001~Hz\\
Fast neutrons&
0.001~Hz\\
Accidental coincidence&
0.001~Hz\\
\hline
\end{tabular}
\caption[Summary the expected trigger rates for neutrino like events
  at CHOOZ-near]{\label{tabelectronics}Summary the expected trigger rates for neutrino like events
  at CHOOZ-near. Trigger rates at CHOOZ-far will be smaller.}
\end{center}
\end{table}
The resulting data flow at CHOOZ-near will be around 20~kB/event,
dominated by pulse shape data. With a trigger rate lower than 1~Hz,
the amount of data remains below 2~GB/day.

\cleardoublepage
\cleardoublepage
\chapter{Liquid scintillators and buffer liquids}
\label{sec:scintillator}
\section{Liquid inventory}
The Double-CHOOZ detector design requires 
different liquids in the separate detector volumes as shown in 
Figures~\ref{fig:choozfar} and \ref{fig:chooznear}.
The inner most volume of 12.7~m$^3$, the $\nuebar$-target, 
contains a proton rich liquid 
scintillator mixture loaded with gadolinium (Gd-LS) at a concentration 
of approximately 1~g/liter. The adjacent volume, the $\gamma$-catcher,
has a volume of 28~m$^3$ and is filled with an unloaded liquid 
scintillator. The photomultipliers are immersed in a non-scintillating 
buffer in order to shield the active volume from the gamma rays
emitted by them. The volume of the buffer liquid is  
approximately 100~m$^3$. Last, an instrumented volume of approximately
110~m$^3$ encloses the whole setup serving as a shield against
external radiation and as a muon veto system. Table~\ref{tab:liquids}
summarizes the liquid inventory of a single detector system.  
\begin{table}[h]
\begin{center}
\begin{tabular}{lrc}
\noalign{\bigskip}
\hline
 \multicolumn{1}{c}{Labeling}        & \multicolumn{1}{c}{Volume [m$^3$]} & \multicolumn{1}{c}{Type} \\
\hline
$\nuebar$-target & 12.7           & Gd loaded LS (0.1~\%) \\
$\gamma$-catcher    & 28.1           & unloaded  LS \\
Buffer           & 100            & non-scintillating organic liquid\\
Veto             & 110            & low-scintillating organic liquid\\
\hline
\end{tabular}
\caption[Overview of liquid inventory for a single detector]{Overview
  of liquid inventory for a single detector. Alternatively we consider
  as well the use of a water Cherenkov detector for the veto.}
\label{tab:liquids}
\end{center}
\end{table}
The selection of the organic liquids are guided by physical and 
technical requirements, as well as by safety considerations. 
In particular, the solvent mixtures or their 
components have a high flash point 
(e.g. phenyl-xylylethane (PXE): flash point (fp) 145~$^o$C, 
dodecane: fp 74~$^o$C, mineral oil: fp 110~$^o$C).  
The $\nuebar$-target and $\gamma$-catcher have as solvent a  
mixture of 80~\% dodecane and 20~\% PXE, or alternatively
trimethyl-benzene (PC). Mineral oil is under study 
as an alternative to dodecane. 
A similar solvent mixture matching the density of the 
$\gamma$-catcher and $\nuebar$-target, will be used as the buffer 
liquid, however with the addition of a scintillation light quencher 
(e.g. DMP). Alternatively, pure water is under investigation 
provided the buoyancy forces can be contained, or a 
density matched water-alcohol mixture. 
The veto volume contains low-scintillating organic liquid and will be 
equipped with PMTs. Alternatively, we also consider to fill the veto
with water and to operate it as a water Cherenkov detector.    
\section{Status of available scintillators}
Metal loading of liquid scintillators have been comprehensively 
studied in the framework of the LENS (Low Energy Solar Neutrino
Spectroscopy) R\&D phase \cite{lens}.
The key groups involved in this research, MPIK and LNGS/INR, are
contributing their expertise to the Double-CHOOZ project. 
The challenge of the  LENS project was to produce stable 
liquid scintillators loaded with 
ytterbium as well as indium  at 5-10~\% in weight while 
simultaneously achieving attenuation lengths of several meters 
and high light yields. Novel scintillator formulations
\cite{In-acac1,In-acac2,In-carb1, In-carb2}
have been developed successfully. 
The scintillators have surpassed longterm tests on the 
scale of up to several years. Several prototype detectors filled with 
different scintillator samples are continuously
measured in the LENS low-background facility at Gran Sasso
since October 2003 to study the stability of the scintillator 
as well as backgrounds. No change in light yield nor in attenuation
length has been observed and backgrounds are extremely low.      

Research with gadolinium loaded scintillator at MPIK and LNGS/INR 
indicates that suitable gadolinium loaded scintillators can be 
produced using the chemistry of beta-diketone complexes as well
as using a single carboxylic acid stabilized by careful control 
of pH. Furthermore, research is being carried out to achieve stability 
with respect to interaction with detector container materials, through
the adjustment of inert solvent components of the 
scintillator while simultaneously retaining high scintillation yields.\\

\noindent
{\bf Beta-diketonate (BDK) Gd-LS:}\\
The studies of the synthesis and properties of beta-diketonates of rare 
earths and their relevant chemistry, especially stability 
at high temperatures, is illustrated
in \cite{Ho1,Ho2}. First
results of Gd-betadiketonate loaded liquid scintillators have been 
reported in \cite{Gd@TUM}. Figure~\ref{fig:pxe-dodecane} displays
the scintillation yield of the unloaded PXE \cite{pxe} based
scintillator as a function 
of dodecane concentration. A scintillation yield
of 78~\% with respect to pure PXE is observed 
at a volume fraction of 80~\% dodecane.   
Figure~\ref{fig:gdacac-ppo} shows the light
yield of a scintillator system with a solvent mixture
of  80~\% dodecane and 20~\% PXE  with varying PPO fluor
concentration. The observed light yield corresponds to 80~\% of the 
unloaded scintillator mixture, or to 60~\% of a pure PC based scintillator. 
Attenuation  length of the Gd-betadiketonate
is being studied and values greater than 10~m at 430 nm have been 
observed after optimizing the synthesis steps. Figure~\ref{fig:gdacac-att}
compares the spectral attenuation length of commercial 0.1~\% Gd-acetylacetone 
(Gd-acac) with that synthesized by us. A secondary fluor(bis-MSB, emission 
spectrum peaked between 420 to 450 nm) at 20-50~mg/l is
used to match the emission to the absorption spectrum (wavelength shifter)\\
\begin{figure}[h]
\begin{center}
\includegraphics[width=0.8\textwidth]{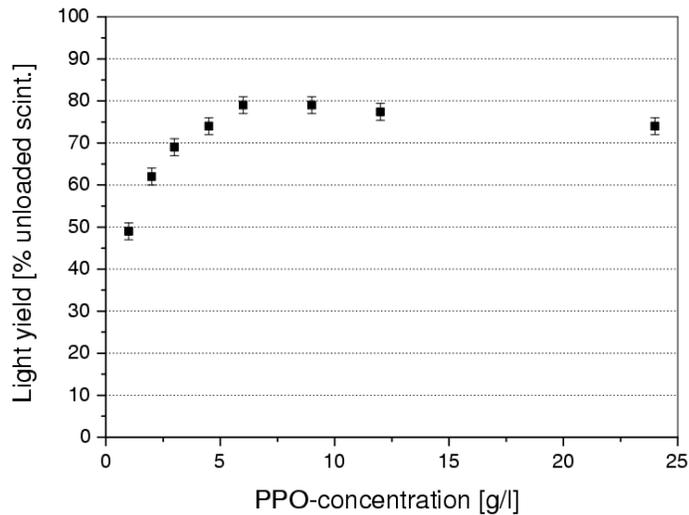}
\caption[Scintillation light yield of 80~\% dodecane 20~\% PXE 0.1~\% Gd
  beta-diketonate  LS with varying PPO concentration]{Scintillation
  light yield of 80~\%  dodecane 20~\% PXE 0.1~\% Gd~beta-diketonate LS
  with varying PPO concentration relative to the unloaded 80~\%
  dodecane  20~\% PXE mixtures with PPO at 6~g/l.}
\label{fig:gdacac-ppo}
\end{center}
\end{figure}
\begin{figure}[h]
\begin{center}
\includegraphics[width=0.8\textwidth]{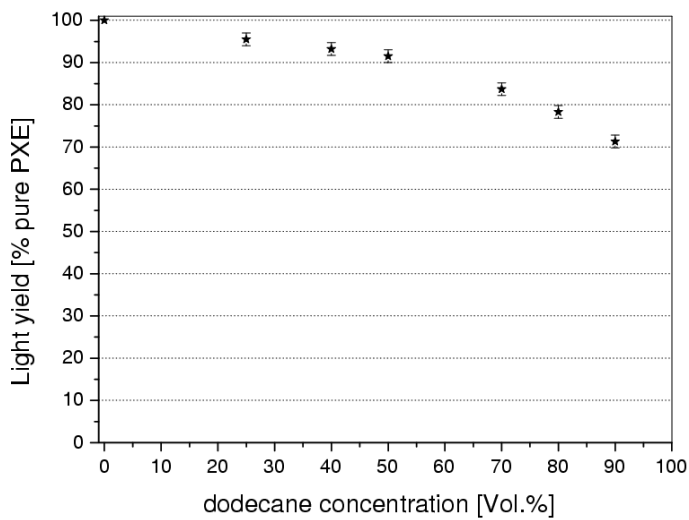}
\caption[Scintillation light yield of PXE/dodecane mixture with varying 
dodecane concentration]{Scintillation light yield of PXE/dodecane
  mixture with varying dodecane concentration. The PPO concentration
  is kept constant at 6~g/l.}
\label{fig:pxe-dodecane}
\end{center}
\end{figure}
\begin{figure}[h]
\begin{center}
\includegraphics[width=0.8\textwidth]{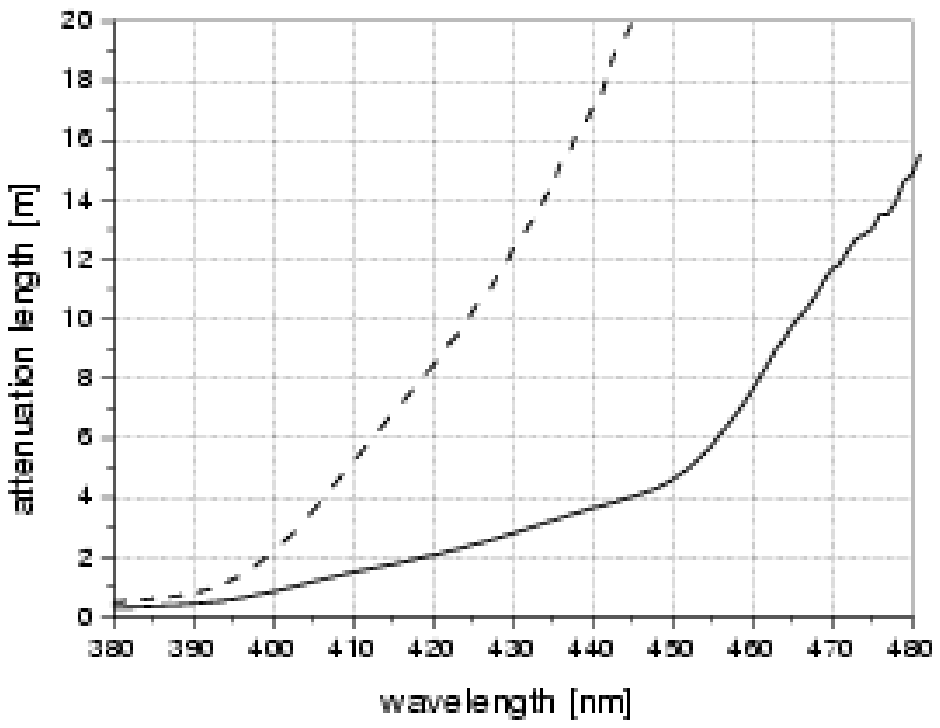}
\caption[Spectral attenuation length of Gd-acac (1~g/l) from an optimized 
synthesis compared with a commercial purchased product]{Spectral
  attenuation length of Gd-acac from an optimized synthesis compared
  with a commercial purchased product. Attenuation length of
  approximately  12~m is achieved at wavelength of 430~nm,
  corresponding  to the emission peak of the secondary shifter.}
\label{fig:gdacac-att}
\end{center}
\end{figure}

\noindent
{\bf Carboxylate (CBX) Gd-LS:}\\
The chemical preparation of Gd~loaded carboxylic acid based
scintillators (single 
acid, pH controlled) has been established and demonstrated to be sound
in our laboratories. These results have been submitted for 
publication, are in preparation
for submission and are presented in publications 
\cite{In-carb1, In-carb2,Yb-carb1,FXH-carb,Gd-carb-FXH}. 
Progress has been swift
towards the definition of scintillator specifics and quantitative 
performance. The main aspects are summarized below. 

A variety of Gd~carboxylate scintillators have been produced,
 using methyl-valeric (C$_6$), ethyl-hexanoic (C$_8$) 
as well as trimethyl-hexanoic (C$_9$) acids.
The possible solvents are trimethyl-benzene (PC) or PXE, 
mixed with either dodecane or mineral oil. 
The Gd~scintillator can be synthesized by adding a 
crystalline material or by direct extraction into the liquid.
Proper control of pH during the synthesis is important.

The solubility of two candidate Gd-carboxylate compounds namely 
Gd-2MeVA and Gd-EtHex, have been measured in a 65~\% PC and 35~\% 
Dodecane solvent mixture and found to be respectively 16.0 and 3.2~g/l. 
Light yields of 60~\% with respect to pure PC and attenuation length of 
15 m have been achieved with Gd~concentrations of 4~g/l and BPO 
(the primary fluor) concentration of 4~g/l in the same solvent mixture.
A C$_9$ CBX version in 50~\% PC and 50~\% dodecane and 1~g/l Gd
gave 87~\% of light with respect to the unloaded mixture.
Good optical properties have been achieved.

The first stability tests at elevated temperatures have been carried out
successfully with the carboxylate systems. Sample mixtures
of PC, mineral oil and Gd~salt were heated to 40~$^o$C during 18 days and
mixtures with dodecane instead of mineral oil to 50~$^o$C during 7 days.
Figure~\ref{fig:temp-test} shows the absorption spectra of the PC/dodecane
based Gd-carboxylate LS before and after the temperature test. Both
the light yield and the attenuation length are stable under the test 
conditions.
\begin{figure}[h]
\begin{center}
\includegraphics[width=\textwidth]{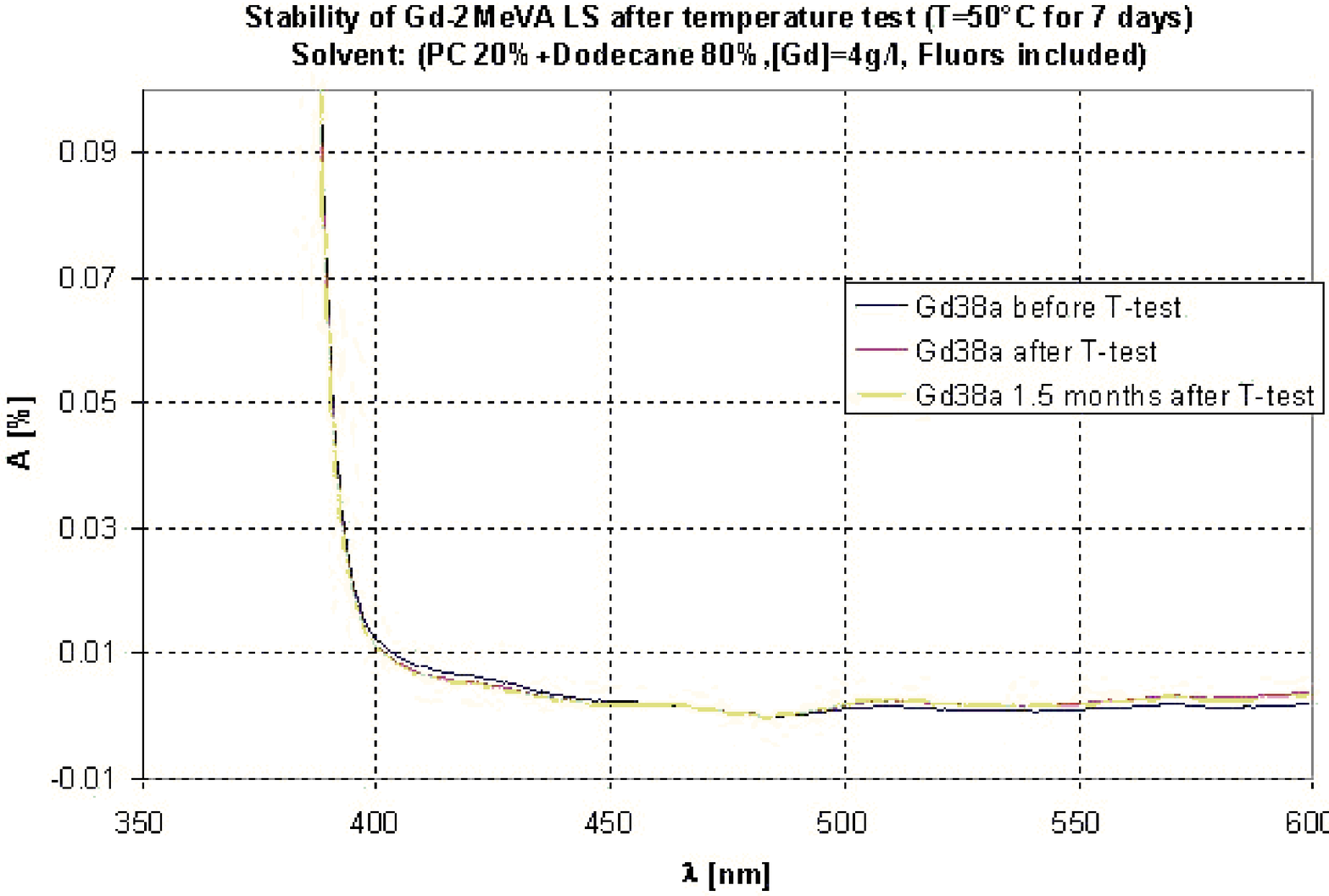}
\caption[Absorption spectra of carboxylate Gd~LS prior and after temperature
test]{Absorption spectra of carboxylate Gd~LS prior and after temperature
test. The sample was kept at 50~$^oC$ for 7 days. The scintillator
composition consists of PC (20~\%), dodecane (80~\%), [Gd]=4~g/l and fluors.}
\label{fig:temp-test}
\end{center}
\end{figure}
\section{Scintillator definition phase}
Both the beta-diketonate and the carboxylate based Gd-LS show 
excellent performances and are viable candidate liquid scintillators for 
the $\nuebar$-target. The research on these LS shifts now from
the R\&D phase to the definition phase and to qualification 
test of their use in Double-CHOOZ. Both  Gd-LS types
have to undergo long term tests to verify no changes 
in the optical performance in contact with detector 
materials. Backgrounds from radioactive trace contaminations
will be studied in the Lens Low Background Facility (LLBF) 
at Gran Sasso \cite{lensdario}. Work specific to the different 
scintillator formulation are listed below.   \\

\noindent
{\bf BDK Gd-LS:} \\
The nominal BDK GD-LS candidate is based on 
a mixture of PXE (20~\%),  dodecane (80~\%), PPO (6~g/l) and bis-MSB
(50~mg/l) with a Gd~loading of 0.1~\% by weight. Future 
laboratory work will concentrate on 
further optimization of the chemical synthesis with special focus on 
questions related to the solubility and purity of Gd-acac.
The solubility has an impact on the engineering of the Gd-LS
production scheme. Moreover, the optimization of energy transfer
properties will be studied. A further increase in light yield by 
fluor optimization appears possible. Mineral oil (MO) will
be studied in more detail as an alternative to dodecane 
since the density range of MO provides the possibility to 
adjust buoyancy forces applied to the scintillator containment 
vessel. A PXE (20~\%) / MO (80~\%) based scintillator can be designed
matching a density in the range from 0.8 to 0.9~g/l
compared to 0.80~g/l for the PXE (20~\%) / dodecane (80~\%) mixture.\\

\noindent
{\bf CBX Gd-LS:} \\
Work on the CBX Gd-LS formulation will concentrate on 
the selection of the carboxylic acid to use 
in the synthesis and on determining the chemical parameters
relevant for the chemical stability of the solution.
Possible surface induced chemical reactions will be investigated.
Optimization of light yield and attenuation length are 
being further pursued by optimizing the synthesis as well as 
the solvent and fluor composition. The same delineations
concerning solvents and densities described previously 
also apply here.\\

From the results of the laboratory research, we now have two 
working Gd-LS formulations and we expect that
both the BDK and CBX systems will comply with the design goals 
of Double-CHOOZ. The designation of the default and backup LS formulation 
will be one of the milestones during the definition phase.
A further outcome of this phase is the detail engineering
of the Gd-LS production scheme. This will be a critical input
for the finalization of the scintillator fluid systems discussed in 
the next section. The final selection of the buffer and veto liquids will 
be done contingent upon the mechanical design of the containment 
vessels and the definition of the Gd-LS formulation.

\section{Scintillator fluid systems}
The scintillator fluid systems (SFSs) include the {\bf off-site SFS} for 
production, purification and storage of the Gd-LS, as well as the 
$\gamma$-catcher LS. A possible location for the off-site SFS is 
MPIK. The {\bf on-site SFS} will be on the reactor area,  
close to the experimental location.

The SFSs scheme envisions the production and storage of the 
complete Gd-LS for both the near and 
far detector, in order to assure identical {\it proton
per volume} concentrations.
The off-site SFS will include ISO-containers for storage and subsequent
transport to the experimental site. Moreover, it will include
a purification column, a nitrogen purging unit, a mixing chamber,
nitrogen blankets 
and auxiliary systems. A similar system, known as Module-0 \cite{mod0}, 
has been constructed by groups in this LOI associated with
Borexino. Since the specifications for Module-0 are more demanding
than required for Double-CHOOZ, no problems are anticipated.

The on-site SFSs will consist of an 
area above ground close to the detector sites for the transport tanks 
which will be connected to the detector by a tubing system.
The purpose of the on-site SFS is to transfer the different liquids
from their transport container into the detector volumes in a safe and 
clean way. The different detector volumes will be filled 
simultaneously and kept at equal hydrostatic
pressures to guarantee the integrity of the detector vessels; 
this will require several parallel lines.
Details of the SFSs will be worked out during the definition phase.
\cleardoublepage
\cleardoublepage
\chapter{Calibration}
\label{sec:calibration}
The main goal of the calibration effort is to reach maximum
sensitivity to neutrino oscillations by comparing the positron
energy spectra measured by the CHOOZ-far and CHOOZ-near detectors.
This is necessary for reaching the desired sensitivity to neutrino
oscillations in Double-CHOOZ. 
Calculations show that a relative difference both in geometry
(construction) and in response of detectors slightly distorts the
ratio of the spectra in both detectors. Therefore, appropriate
corrections and errors obtained on the basis of absolute and
relative calibration measurements should be administered to the
data. This should be the result of detailed Monte-Carlo
simulations (see Chapter~\ref{sec:fulldetsimul})
backed up by an extensive program of source calibrations. 
The calibration sources (See Table~\ref{t:techniques})
must be deployed regularly throughout
the detector active volume to simulate and monitor the detector
response to positrons, neutron captures, gammas and
the backgrounds in the Double-CHOOZ experiment. This requires
a dedicated mechanical system in order to introduce calibration
sources into the different regions of the detector.
\begin{table}[h]
\begin{center}
 \begin{tabular}{ll}
 \hline
 Technique & Calibrations \\
 \hline 
 Optical Fibers, Diffusive Laser ball & Timing and Charge Slopes
 and Pedestals, \\
                   & attenuation length of detector components \\
 \hline
 Neutron Sources: Am-Be, $^{252}$Cf & Neutron response, relative
 and \\
                                    & absolute efficiency, capture time \\
 \hline
 Positron Sources: $^{22}$Na, $^{68}$Ge & $e^{+}$ response,
 energy scale, trigger thresh. \\
 \hline
 Gamma Sources: & Energy linearity, stability, resolution, \\
                & spatial and temporal variations. \\
 $^{137}$Cs & $\beta^{-}$, 0.662 MeV \\
 $^{22}$Na & $\beta^{+}$, 1.275 MeV + annih \\
 $^{54}$Mn & EC, 0.835 MeV \\
 $^{65}$Zn & 1.35 MeV \\
 $^{60}$Co & EC, 1.173, 1.33 MeV \\
 $^{68}$Ge & EC, $\beta^{+}$ 1.899 MeV + annih \\
 $^{88}$Y  & EC, 0.898, 1.836 MeV \\
 H neutron capture & 2.223 MeV \\
 $^{241}$Am-$^{9}$Be & ($\alpha$,n) 4.44 MeV ($^{12}$C)\\
 Gd neutron capture & Spectrum in 8 MeV window \\
\hline
 $^{228}$Th & 2.615 MeV \\
 $^{40}$K & EC, $\beta^{+},\beta^{-}$, 11~\% 1.46 MeV \\
\hline
 \end{tabular}
 \caption[Techniques available to calibrate the Double-CHOOZ experiment]{Table showing the different techniques that are available to calibrate the Double-CHOOZ experiment.}
 \label{t:techniques}
\end{center}
\end{table}
There are a number of specific tasks for a successful calibration of
the detectors. These include optical calibrations (single
photoelectron (PE) response, multiple PE response, detector
component optical constants), electronic calibrations (trigger
threshold, timing and charge slopes and pedestals, dead time), 
energy (energy scale and resolution), and neutron and positron detection
efficiency and response. In addition, detector calibrations must
test the Monte-Carlo and analysis code to verify the accuracy of
the simulations, throughout the detector (spatially), and during
the lifetime of the experiment.
\section{Optical and electronic calibrations}
The optical calibrations are based on the experience with CHOOZ
 and the CTF-Borexino experiments. In CTF-Borexino the
optical calibration consists of a UV pulsed-laser (jitter less than 1~ns)
coupled to an optical fiber illuminating separately each PMT. This
allows the single PE response to be measured since the amplitude
of the pulse is tuned to approximately a single PE. This technique
allows the gain, timing slope, charge slope and pedestals to be
determined relative to individual PMTs and to the triggers. In
addition to the optical fiber calibration, the light attenuation
in the liquid scintillator is monitored using a diffusive laser
ball source, as has been successfully used by the SNO experiment
\cite{Ahmad:2002jz}.
This source illuminates all the PMTs isotropically and allows the
attenuation length of the detector components and the PMT angular
response to be measured as a function of photon wavelength.
Finally, to ensure that we are able to veto muons with high
efficiency, we must also calibrate the PMTs mounted on the
stainless steel tank. This is done by also connecting optical
fibers to these PMTs. The attenuation length of the water (or~oil)
shielding is measured by deploying the laser ball in this region.
\section{Energy calibration}
The specific signature for the detection of an electron
antineutrino through inverse beta decay is the detection of prompt
gammas from the annihilation of the positron and the delayed
capture of a neutron several tens of $\mu$s later. While direct
calibration with an antineutrino source is impossible, it is
possible to simulate each of the components of the antineutrino
signal, such as the prompt positron and delayed neutron by
deploying positron, neutron, and gamma sources. \\

The standard calibration system will include a permanent vertical
tube, entering the detector until the center of the inner acrylic
target. This open tube will allow frequent and safe calibration with
radioactive sources.
\subsection{Gamma ray sources}
Positron annihilates at rest and
produces 2 back-to-back gammas. Thus, for a high detection
efficiency we must be able to calibrate the detector energy
response to gammas from 1~MeV to $\sim$10~MeV corresponding to the
endpoint of the fission product beta decays. In addition, a
neutron is detected by its capture on the Gd additive to the
liquid scintillator and produces a gamma cascade of approximately
8~MeV. For this reason, it is necessary to also know the energy
scale in the high energy window of 6-10~MeV to be able to identify
the delayed second trigger as a neutron. Specifically, it will be
necessary to know the gamma energy corresponding to the neutron
detection threshold for both the near and far detector with a 100~keV 
accuracy. This is accomplished by deploying various higher
energy gamma calibration sources (see Table~\ref{t:techniques})
and by detailed Monte-Carlo simulations in the energy region where
there are no calibration sources.

The overall energy scale can be determined from the position of
the 0.662~MeV peak of the $^{137}$Cs source, and then verified by
calibration with several gamma sources (see
Table~\ref{t:techniques}) in different energy ranges: $^{54}$Mn
(0.835~MeV), $^{22}$Na (1.275~MeV), $^{65}$Zn (1.351~MeV),
$^{60}$Co, and $^{228}$Th (2.614~MeV). These gammas allow the
energy response to the positron annihilation photons to be
determined for different positron energies. The capture of
neutrons from an Am-Be source scintillator (to be discussed later)
can also be used as a high energy gamma source as it produces
prompt 4.4~MeV gammas. We will also use the natural sources from
radioactive impurities of the detector materials ($^{40}$K,
$^{208}$Tl …) and neutrons produced by cosmic muons for energy
calibration. Since these sources are present permanently, they are
useful for monitoring the stability of the energy response. Thus,
the primary purpose of the gamma sources are to determine and
monitor the energy scale for both the far and near detectors as a
function of position and time during which the experiment is
conducted.
\subsection{Positron response}
Positron detection can be simulated experimentally by means of the
$^{22}$Na source. A $^{22}$Na source emits a 1.275~MeV primary
gamma accompanied by a low energy positron which annihilates
inside the source container. The primary and annihilation gammas
from the source mimic the positron annihilation resulting from an
antineutrino event inside the detector. An alternative positron
source is a $^{68}$Ge source which produces positrons with higher
energies, and therefore calibrates higher energy positrons.
$^{68}$Ge decays by EC to $^{68}$Ga and $\beta^{+}$-decays to
stable $^{68}$Zn with an endpoint of 1.9~MeV. This isotope also
has the advantage that it produces only low energy gammas in
coincidence with the nuclear decay, and the $\beta^{+}$ has an
endpoint of 1.889~MeV 89~\% of the time. A second purpose of this
source (if a source is constructed so that the beta is absorbed by
the shielding surrounding the source) is to tune the trigger
threshold to be sensitive to annihilation gammas and to monitor
its stability. A $^{68}$Ge source has been successfully used in
the Palo Verde reactor neutrino experiment \cite{paloverdenim1}.
\subsection{Neutron response}
Coincident with the production of a positron in inverse beta
decay, a neutron is produced. The neutron then quickly thermalizes
and is captured on the Gd ($^{155}$Gd or $^{157}$Gd, with cross
sections of 60,900 and  254,000 barns, respectively) loaded in the
central target. The neutron capture is accompanied by the emission
of a cascade of gamma-rays with the summed energies of 8.536 and
7.937~MeV, respectively. Thus, neutrons are selected by cutting on
gammas with energies exceeding 6~MeV. However, a fraction of the
gamma-rays can escape detection, especially events that occur near
the boundary of the fiducial volume. Therefore, it is expected
that the neutron detection efficiency decreases for events near
the borders of the acrylic vessel that contains the Gd loaded
liquid scintillator. Calibration of this effect must be quantified
by deployment of neutron calibration sources throughout the
detector and comparing the detector response to Monte-Carlo. In
addition to measuring the neutron response, neutron calibration is
also a very sensitive method for determining in-situ various
liquid scintillator properties, such as the hydrogen and
gadolinium concentration in the liquid scintillator.

There are two suitable and accessible neutron sources for neutron
calibration: the Am-Be source and $^{252}$Cf spontaneous fission
source. These sources emit neutrons with different energy spectra
from what is expected from inverse beta decay, and thus the
importance of these differences needs to be quantified. To
decrease the background during neutron source deployment, neutrons
from Am-Be should be tagged by the 4.4~MeV gamma emitted in
coincidence with the neutron. This allows the neutron capture
detection efficiency to be determined independent of knowing the
precise rate of the neutron source, because every time a 4.44~MeV
gamma is detected a neutron is released \cite{SCroft}. 
The absolute neutron detection efficiency can also be determined 
with a $^{252}$Cf source by
using the known neutron multiplicity (known to 0.3~\%). For the
source placed into the center, the size of the Gd region is larger
than the neutron capture mean free path, so that the neutron
capture is studied independent of the presence of the acrylic
vessel. In order to tag the neutron events, a small fission
chamber is used to detect the fission products.
Therefore, neutron source calibrations provide us with the
relevant data to calibrate the detector response to neutrons. In
particular, neutron sources allow us to measure the absolute
neutron efficiency, to determine and monitor the appropriate
thresholds of neutron detection, and to measure the neutron
capture time for both the far and near detectors.
\subsection{The Calibration source deployment system}
A mechanical system must be developed to introduce calibration
sources throughout the detector active volume. The system must be
easy to set up so that calibration can be done frequently without
loss of neutrino live time. The suggestion is to use a system of
ropes and pulleys similar to the SNO experiment (see
Figure~\ref{f:calibsetup} for a conceptual design). However,
unlike the SNO experiment we must be able to deploy sources
throughout the active volume, rather than in a plane as is done in
the SNO experiment. The reason for this is that because during the
lifetime of the detector, PMT mortality might result in an
anisotropy in the detector response. Moreover, the effect will
manifest as a anisotropy relative to Chooz-Near and Chooz-Far
which will impact on the energy resolution and scale of the two
detectors. The system of ropes and pulleys must be designed so
that the calibration sources sample a large fraction of the active
volume and can calibrate this effect.

\subsection{Map of the Gd-LS target}
The starting point of the design is to introduce sources through a
glove box situated at the center and top of the cylinder housing
the Gd-LS target.
\begin{figure}[ts!]
\centering
\includegraphics[angle=90,scale=0.7]{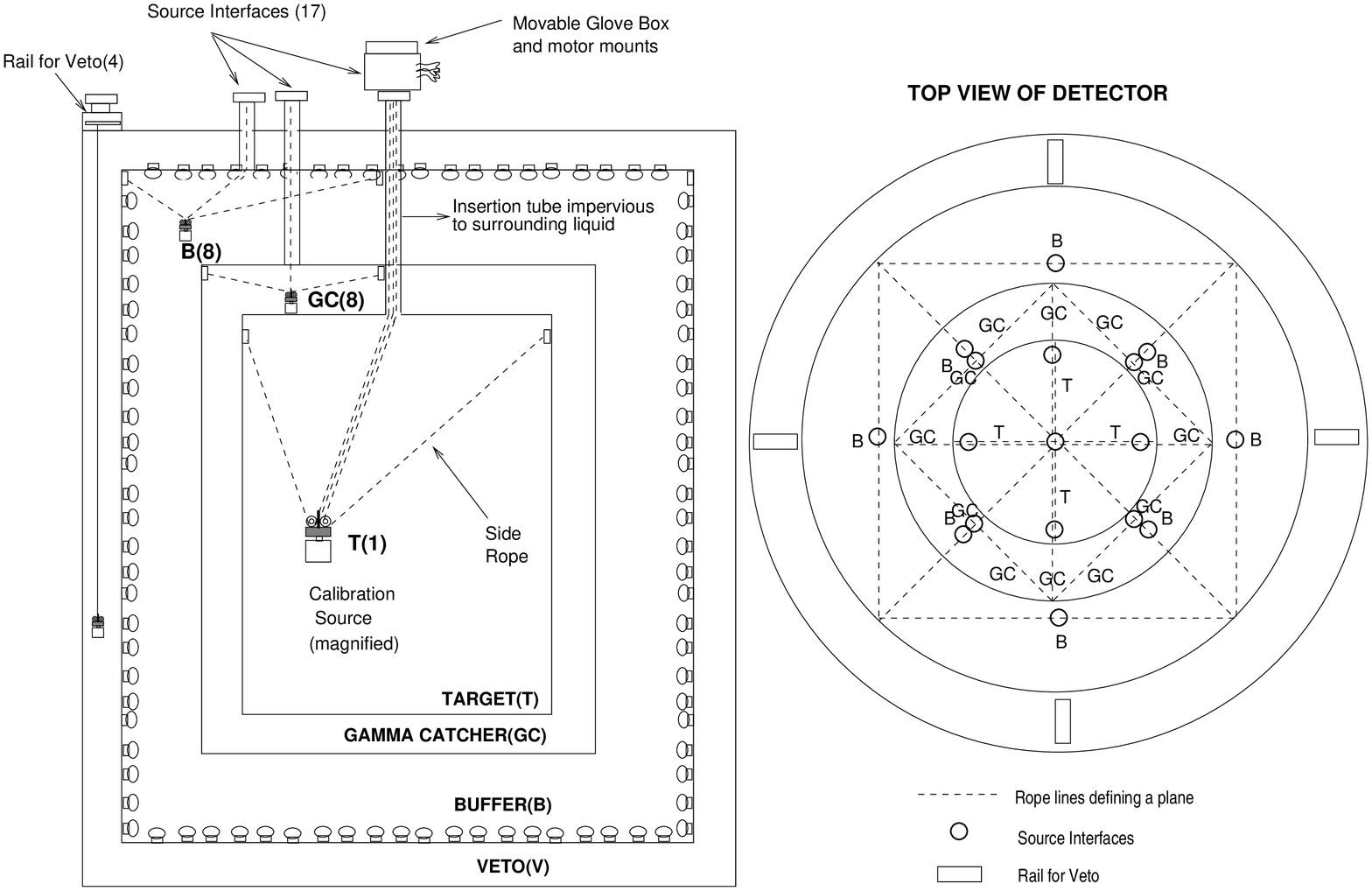}
\caption{A possible scenario for a calibration source deployment
system adapted from the SNO experiment calibration system.}
\label{f:calibsetup}
\end{figure}
In the glove box sources can be prepared for deployment without
introducing contaminants into the active volume. The glove box can
also be evacuated and flushed with LN$_{2}$ before deploying
sources to prevent Radon from entering the active volume. The
sources are then suspended from a rope and lowered straight down
from the glove box to the bottom of the cylinder using a stepper
motor. In this way we can calibrate the variation of the detector
response along the axis of symmetry of the cylinder (z-axis). To
calibrate the detector response off of the z-axis the
calibration sources must be physically moved away from the
z-axis such that the detector response as a function of radius
can be determined. To achieve this, the idea is to fasten two
ropes on opposite sides of the cylinder. Then, to feed the rope
through two pulleys (one for each side rope) attached to top of
the calibration sources, then to a stepper motor located near the
glove box. The tension on either of the ropes can then be
independently adjusted by carefully controlling stepper motors.
When the tension is changed of one of the side ropes the source
will move away from the z-axis. This will allow the sources to
be deployed throughout most of the area of a plane defined by the
central axis of symmetry (z-axis) and the line connecting the
places that the ropes are attached. Including as second set of
ropes perpendicular to the first will allow the source to be moved
not only away from the z-axis but throughout most of the active
volume of the target. The SNO experiment has been able to attain a
deployment accuracy of 5 cm using this method when the source is
moved in a plane~\cite{nimcalib}.
\subsection{Calibrating the gamma-catcher, buffer, and veto}
A deployment mechanism must also be devised to deploy sources
outside of the main central target. Sources must be deployed in
the gamma catcher region as well as the buffer and the veto. A
further requirement of the components inside the active volume is
that the system must not block the scintillation light, nor change
the detector response, and they must be impervious to liquids. The
suggestion is again to use a system of pulleys and side ropes to
cover most of the volume. A possible scenario for such a system is
shown in Figure~\ref{f:calibsetup}. The system would sample
calibration source positions in a plane from the z-axis outward
through the cylinders to the veto cylinder. The sources will be
accessed through a glove box which will be movable so that it can
be mounted on top of all the source interfaces. The calibration of
the veto can be done with a rail deployment system (only 1 central
rope but movable along the radius of the cylinder), since here the
mechanism can be constructed without blocking the scintillation
light. The right panel of Figure~\ref{f:calibsetup} shows a
possible configuration of the ropes, specifically the figure shows
how the top (ropes outward from the center) and sides (ropes in
shape of squares) of the detector will be calibrated. However,
calibration of the bottom of both detectors is more difficult
since access to this region is limited. Calibration of the bottom
portion of the detectors still needs to be investigated.

\cleardoublepage
\cleardoublepage
\chapter{Backgrounds}
\label{sec:backgrounds}
The signature for a neutrino event is a prompt signal with a minimal energy of about 
1~MeV and a delayed 8~MeV signal after neutron capture in gadolinium. 
This may be mimicked by background events which can be divided into two classes: 
accidental and correlated events. The former are realized when a neutron like event 
by chance falls into the time window (typically few 100~$\mu$s)  after an event in the 
scintillator with and energy of more than one MeV. 
The latter is formed by neutrons which slow down by scattering 
in the scintillator, deposit  $> 1$~MeV visible energy  and are captured in 
the Gd~region. 
In this chapter we first discuss possible sources and fluxes for background events and later
estimate their rates. 
With these numbers we find  criteria for the necessary overburden of the near 
detector and we will extract purity limits for detector components.\\
%
%
\section{Beta and gamma background}

\subsection{Intrinsic beta and gamma background}

In this section the intrinsic background due to beta and gamma events 
above $\sim$1~MeV is discussed. 
It can be produced in the scintillator or in the acrylic vessels which
contain the liquid. 
The contribution from the Uranium and Thorium chains is reduced to a few elements, 
as all alpha events show quenching with visible energies well below 1~MeV. 
Furthermore the short delayed Bi-Po coincidences in both chains can be
detected event by event, 
and hence rejected. In the end, only the decays of $^{234}$Pa (beta decay, $Q = 2.2$~MeV), 
$^{228}$Ac (beta decay, $Q = 2.13$~MeV) and $^{208}$Tl (beta decay, $Q = 4.99$~MeV) 
have to be considered. 
Assuming radioactive equilibrium the beta/gamma background rate due to both chains can be 
estimated by 
$b_1 \simeq M_\text{U} \cdot 6\cdot 10^3~{\mathrm s^{-1}} +M_\text{Th} \cdot 4\cdot 10^3~{\mathrm s^{-1}}$,
where the total mass of U and Th is given in gram. Taking into account the total 
  scintillator mass of the neutrino target plus the $\gamma$-catcher, this rate can be expressed by  
 $b_1\simeq 3~{\mathrm s^{-1}} (c_\text{U,Th}/10^{-11})$ , where $c_\text{U,Th}$ is the 
  mass concentration of Uranium and Thorium in the liquid. 
The contribution from $^{40}$K can be expressed by
$b_2\simeq 1~{\mathrm s^{-1}} (c_\text{K}/10^{-9})$, where $c_\text{K}$ is the mass concentration of natural 
   K in the liquid. 

The background contribution due to U, Th and K in the acrylic vessels can be written as 
   $b_3 \simeq 2~{\rm s^{-1}} (a_\text{K}/10^{-7}) + 5~{\rm s^{-1}} (a_\text{U,Th}/10^{-9})$,
   where $a_\text{K}$ and $a_\text{U,Th}$ describe the mass concentrations of K, U and Th in the acrylic. 
In total, the intrinsic beta/gamma rate is the sum $b=b_1+b_2+b_3$.
In the CTF of the Borexino experiment at Gran Sasso, concentration values of  $c_\text{U,Th}<10^{-15}$ and
$c_\text{K}<10^{-12}$    
have been measured for two liquid scintillators (PC and PXE) with volumes of about 4~m$^3$. 
 Upper limits on radioactive trace elements in acrylic have been reported to 
 be $a_\text{U,Th}<3 \cdot 10^{-12}$  by the SNO collaboration \cite{Ahmad:2002jz}. 
 Gamma spectroscopy measurements show upper limits of $a_\text{K}< 1 \cdot  10^{-9}$. 
 This shows that in principle the beta/gamma rate in the detector due to intrinsic 
 radioactive elements can be kept at levels well below 1~s$^{-1}$. 
 The aimed concentration values for this goal are given in
 Table~\ref{t1}. \\
\begin{table}[h]
\begin{center}
\begin{tabular}{lc}
\hline
\multicolumn{1}{c}{Element} &       allowed concentration (g/g) \\
        &       for $b < 1 $\,s$^{-1}$       \\
\hline
Uranium, Thorium in scintillator        &$\sim 10^{-12}$ \\
Potassium in scintillator       &$\sim 10^{-10}$ \\
Uranium, Thorium in acrylic vessels     &$\sim 10^{-10}$ \\
Potassium in acrylic vessels    &$\sim 10^{-8}$ \\
\hline
\end{tabular}
\caption[Upper limits on U, Th and K concentrations in the liquid
  scintillator and acrylic vessel]{\label{t1} Upper limits on U, Th
  and K concentrations in the liquid scintillator and acrylic vessels
  to achieve a beta/gamma rate below  1\,s$^{-1}$}
\end{center}
 \end{table}

\subsection{External gamma background}

According to the experience gained in the CTF of Borexino the dominant contribution to 
the external gamma background is expected to come from the photomultipliers (PMTs) and 
structure material. 
Again contributions from U, Th and K have to be considered. 
However, because of the shielding of the buffer region only the 2.6~MeV gamma 
emission from $^{208}$Tl has to be taken into account. 
The activity of one PMT in the CTF (structure material included) 
is known to be $\sim 0.4~\mathrm s^{-1}$. 
The shielding factor $S$ due to the buffer liquid can be calculated to be $S \sim 10^{-2}$. 
Hence, the resulting gamma background in the neutrino target plus the $\gamma$-catcher 
can be written as $b_{ext} \simeq 2~{\rm s^{-1}} (N_{\text{PMT}}/500)$, 
where $N_{PMT}$ is the number of PMTs. 

\section{Neutron background}

\subsection{Intrinsic background sources}

Neutrons inside the target may be produced by spontaneous fission of heavy elements and 
by ($\alpha$,n)-reactions. For the rate of both contributions the concentrations of U and Th in 
the liquid are the relevant parameters. 
The neutron rate in the target region can be written as  
$n_{int}\simeq 0.4~{\rm s^{-1}} (c_\text{U,Th}/10^{-6})$. 
Hence, for the aimed concentration values as described above the intrinsic contribution to 
the neutron background is negligible.

\subsection{External background sources}

Several sources contribute to the external neutron background. 
We first discuss external cosmic muons which produce neutrons in the target region via spallation and muon capture. 
Those muons intersect the detector and should be identified by the veto. 
However, some neutrons may be captured after the veto time window. 
Therefore we estimate the  rate of neutrons, which are 
generated by spallation processes of through going muons and by stopped negative muons which 
are captured by nuclei.\\

The first contribution is estimated by calculating the muon flux for different shielding values 
and taking into account a $E^{0.75}$ dependence for the cross section of neutron production, 
where $E$ is the depth dependent mean energy of the total muon flux. 
The absolute neutron flux is finally obtained by considering measured values in several experiments 
(LVD \cite{lvd}, MACRO \cite{Ambrosio:1998wu}, CTF \cite{CTF}) in the Gran Sasso underground laboratory
and extrapolating these results by comparing muon fluxes and mean energies for the different shielding factors. 
Table~\ref{t2} gives the expected neutron rate depending on the shielding.
\begin{table}[h]
\begin{center}

\begin{tabular}{lrrr}
\hline
\multicolumn{1}{c}{Overburden}  & \multicolumn{1}{c}{Muon rate}  & \multicolumn{1}{c}{Mean muon energy}  & \multicolumn{1}{c}{Neutrons}  \\
(m.w.e.) & (s$^{-1}$) & (GeV) & (s$^{-1}$) \\
\hline
40      &       $1.1 \cdot 10^{3}$    &  14  & 2        \\
60    & $5.7 \cdot 10^{2}$    &  19  & 1.4       \\
80    & $3.5 \cdot 10^{2}$    &  23  & 1        \\
100   & $2.4 \cdot 10^{2}$    &  26  & 0.7  \\
300   & $2.4 \cdot 10^{1}$    &  63  & 0.15     \\
\hline
\end{tabular}
\caption[Estimated neutron rate in the active detector region due to
    through going cosmic muons.]
{\label{t2} Estimated neutron rate in the active detector
  region due to through going cosmic muons.}
\end{center}
\end{table}

Negative muons which are stopped in the target region can be captured by nuclei where a  
neutron is released afterwards. 
The rate can be estimated quite accurately by 
calculating the rate of stopped muons as a function of the depth of shielding and 
taking into account the ratio between the $\mu$-life time and $\mu$-capture times.
As the capture time in Carbon is known to be around 25 $\mu$s
($\approx$1~ms in H) 
only about 10~\% of captured muons may create a neutron.
Since the concentration in Gd is so low, its effect can be neglected here.
The estimated results are shown in Table~\ref{t3}.
The neutron generation due to through going muons dominates.

\begin{table}[h]
\begin{center}

\begin{tabular}{lrr}
\hline
\multicolumn{1}{c}{Overburden}  & \multicolumn{1}{c}{Muon stopping rate}   &  \multicolumn{1}{c}{Neutrons}   \\
(m.w.e.) & (t$^{-1}$~s$^{-1}$) & (s$^{-1}$) \\
\hline
40      &       $5 \cdot 10^{-1}$     &  0.7      \\
60    & $3 \cdot 10^{-1}$     &  0.4       \\
80    & $1.2 \cdot 10^{-1}$   &  0.2    \\
100   & $6 \cdot 10^{-2}$     &  0.08  \\
300   & $2.5 \cdot 10^{-3}$   &  0.003    \\
\hline
\end{tabular}
\caption[Estimated neutron rate in the target region due to stopped negative muons]{\label{t3} Estimated neutron rate in the target region due to stopped negative muons.}
\end{center}
\end{table}

\subsection{Beta-neutron cascades}

Muon spallation on $^{12}$C nuclei in the organic liquid scintillator 
may generate
$^8$He, $^9$Li, and $^{11}$Li which may undergo beta decay with a 
neutron emission.
In that case those background events show the same signature as a 
neutrino event.
For shallow shielding depths the muon flux is too high to allow tagging 
by the muon veto,
as the lifetimes of these isotopes are between 0.1~s and 1~s.
The cross sections for the production of $^8$He, $^9$Li have been 
measured by a group of TUM at the SPS at CERN
with muon energies of 190~GeV (NA54 experiment \cite{NA54}). In this 
experiment only the combined production  $^8$He + $^9$Li where obtained 
without ability to separate each isotope.
An estimate for the background rates for shallow depth experiments like CHOOZ can be 
obtained from results of the
KamLAND experiment by calculating the muon flux for energies above
$\sim$500~GeV \cite{horthonsmith}.
With this assumption an event rate of about 0.4 per day in the 
target region can be estimated for a 300~m.w.e. shielding.
A more conservative estimate is obtained assuming a E$^{0.75}$ scaling as we did
in calculating the neutron flux.
Then the rate should be around 2 events per day.
In Table~\ref{tab:na54choozfar} all radioactive $^{12}$C-spallation products including the 
beta-neutron cascades are
shown with the estimated event rates in both detectors. \\
 
The Q-values of the beta-neutron cascade decays
is 8.6~MeV, 11.9~MeV, 20.1~MeV for $^8$He, 
$^9$Li, and $^{11}$Li, respectively.
In the experiment the $^8$He production rate might be measured if we 
set a dedicated trigger after a muon event in the target region looking 
for the double cascade of energetic betas 
($^8$He $\rightarrow$ $^8$Li $\rightarrow$ $^8$Be) occuring 
in 50~\% of all $^8$He decays (see Figures \ref{fig:8He} and \ref{fig:9Li}).
Nothing similar exist in the case of the $^9$Li, but the beta endpoint 
is here above the endpoint of positron induced by reactor
antineutrinos.
Nevertheless, from the NA54 experiment \cite{NA54} results the total 
cross section of  $^8$He + $^9$Li is known, and if the $^8$He is
evaluated separately,  some redundancy on the total $\beta$-neutron cascade will be available. 
Figure \ref{fig:8He} shows the relevant branching ratios of the $^8$He isotope,
normalized to 100~\%. The neutrons emitted in these decays are typically around 
1~MeV. 
Figure \ref{fig:9Li} shows the relevant branching ratios of the $^9$Li isotope,
 normalized to 100~\%. 
%
%
%
\begin{figure}
\begin{center}
\includegraphics[width=0.8\textwidth]{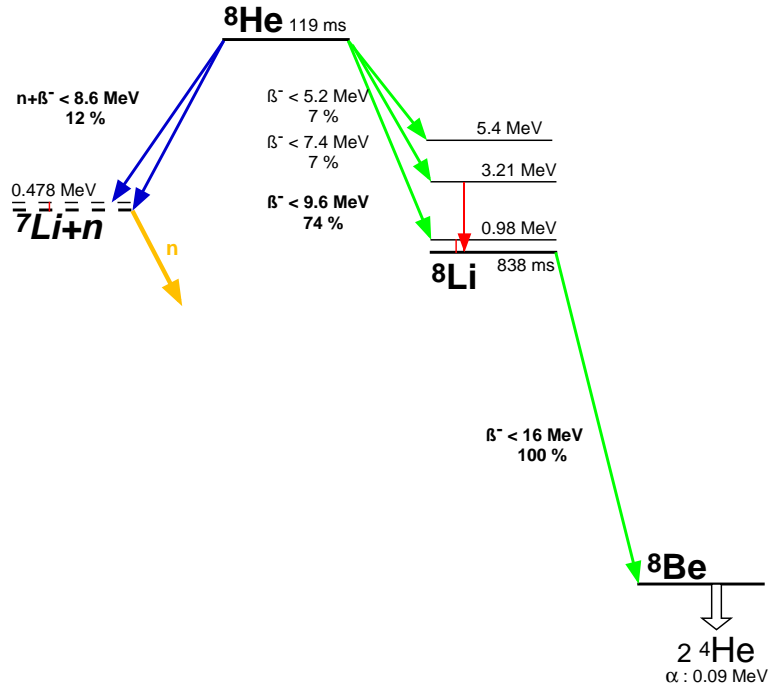}
\caption[Relevant branching ratios for the decay of the $^8$He isotope.]
{Relevant branching ratios for the decay of the $^8$He isotope,
 normalized to 100~\%. Half-lives are quoted, as well as the end-point of the $\beta$ decays. 
Neutrons emitted in these decays are typically around 1~MeV.
The double cascade decay to the $^8$Be offer a possibility to measure 
, {\it in situ}, the production rate.}
\label{fig:8He}
\end{center}
\end{figure}
\begin{figure}
\begin{center}
\includegraphics[width=0.7\textwidth]{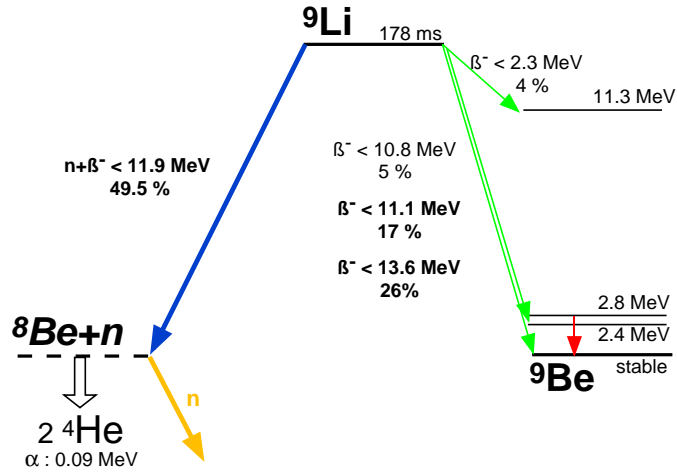}
\caption[Relevant branching ratios for the decay of the $^9$Li isotope.]
{Relevant branching ratios for the decay of the $^9$Li isotope,
 normalized to 100~\%. Half-lives are quoted, as well as the end-point of the $\beta$ decays. 
Neutrons emitted in these decays are typically around 
1~MeV. In case of $\beta$ decay to $^8$Be, the latter transform immediately to two low energy $\alpha$ particles.}
\label{fig:9Li}
\end{center}
\end{figure}
\begin{table}[h]
\begin{center}
\begin{tabular}{c|rr|rr}
\noalign{\bigskip}
\hline
           & \multicolumn{2}{c}{Near detector} & \multicolumn{2}{c}{Far detector} \\
\hline
  Isotopes & $R_{\mu}$ &  $R_{\mu}$  & $R_{\mu}$  &  $R_{\mu}$   \\
           & ($E^{0.75}$ scaling) & ($E>500$~GeV) & ($E^{0.75}$
  scaling) &  ($E>500$~GeV)\\             & \multicolumn{4}{c}{per day} \\ 
\hline 
$^{12}$B & \multicolumn{4}{c}{not measured} \\
$^{11}$Be & $<18$  &  $<3.8$ &  $<2.0$  &  $<0.45$ \\
$^{11}$Li & \multicolumn{4}{c}{not measured} \\
$^{9}$Li  & $17 \pm 3$ & $3.6$ &  $1.7 \pm 0.3$ & $0.36$ \\
$^{8}$Li  & $31 \pm 12$ & $6.6$ &  $3.3 \pm 1.2$ & $0.7$\\ 
$^{8}$He  & \multicolumn{4}{c}{$^{8}$He \& $^{9}$Li measured together}\\
$^{6}$He  & $126 \pm 12$ & $26.8$ &  $13.2 \pm 1.3$ & $2.8$\\
$^{11}$C & $7100\pm455$ & $1510$  & $749\pm48$ & $159.3$ \\
$^{10}$C  & $904\pm114$ & $192$ & $95\pm12$ & $20.2$ \\
$^{9}$C   & $38\pm12$ & $8.1$   & $4.0\pm1.2$ & $0.85$\\
$^{8}$B   & $60\pm11$ & $12.7$  & $5.9\pm1.2$ & $1.25$\\
$^{7}$Be  & $1800\pm180$ & $382.9$ & $190\pm19$ & $40.4$\\
\noalign{\smallskip} 
\hline
\end{tabular}
\caption[Radioactive isotopes induced by muons in liquid scintillator
  targets at the CHOOZ near and far detectors.]{Radioactive isotopes produced
  by muons and their secondary shower particles in liquid scintillator
  targets at the CHOOZ near and far detectors. 
The rates $R_{\mu}$ (events/d) are given for a target of 
$4.4 \times 10^{29}$ $^{12}$C (For a mixture of 80~\% Dodecane and 20~\%
PXE, 12.7~m$^3$)  at a depth of 60~m.w.e. for the near detector 
and 300~m.w.e. for the far detector. Because of the positron annihilation 
the visible energy in  $\beta^+$ decays is shifted by 1.022~MeV.
$^9$Li and $^8$He could not be evaluated separately. 
Columns 3 and 5 correspond to an estimate of the
number of events assuming that the isotopes are produced only by high
energy muon showers $E>500$~GeV \cite{horthonsmith}. A neutrino signal
rate of 85 events per day is expected at CHOOZ-far, without
oscillation effect (for a power plant running at nominal power, both
  dead  time and detector efficiency are not taken into account here).}
\label{tab:na54choozfar}
\end{center}
\end{table}
}
\subsection{External neutrons and correlated events}
Very fast neutrons, generated by cosmic muons outside the detector, may penetrate into the target 
region.
As the neutrons are slowed down through scattering, recoil protons may give rise to a visible signal 
in the detector. This is  followed by a delayed neutron capture event.
Therefore, this type of background signal gives the right time correlation and can mimic a 
neutrino event.
Pulse shape discrimination in order to distinguish between $\beta$ events and recoil protons is 
in principle possible, but should not be applied in the analysis as additional statistical and 
systematic uncertainties should be avoided in the experiment.
As the muon is not seen by the veto, those correlated events may be dangerous for 
the experiment. \\

Therefore a Monte-Carlo program has been written to estimate the correlated background rate for 
a shielding depth of 100~m.w.e. and flat topology.
In order to test the code the correlated background for the old Chooz experiment 
(different detector dimensions, 300~m.w.e. shielding)
has been calculated with the same program.
The most probable background rate was determined to be 0.8 counts per day.
A background rate higher than 1.6 events per day is excluded by 90~\%~C.L.
This has to be compared with the measured rate of 1.1 events per day.
We conclude that the Monte-Carlo program reproduces the real correlated background value within 
roughly a factor~2. \\

For Double-Chooz we calculated the correlated background rate for 100~m.w.e. shielding and estimated
the rates for other shielding values by taking into account the different muon fluxes and assuming
a $E^{0.75}$ scaling law for the probability to produce neutrons.  
The neutron capture rate in the Gd~loaded scintillator for an overburden of 100~m.w.e. is about 300/h.
However, only 0.5~\% from those neutrons create a signal in the scintillator within the neutrino window
(i.e. between 1~MeV and 8~MeV), because most deposit in total much more energy  
during the multiple scattering processes until they are slowed down to thermal energies.
The quenching factors for recoil protons and carbon nuclei has been taken into account.
In addition around 75~\% from those events generate a signal in the muon veto above 4~MeV (visible 
$\beta$ equivalent energy).
In total the correlated background rate is estimated to be about 3.0~counts per day for
100~m.w.e. shielding.
In Table~\ref{t4} the estimated correlated background rates are shown for different shielding 
depths.\\
\begin{table}[h]
\begin{center}

\begin{tabular}{lrr}
\hline
\multicolumn{1}{c}{Overburden}  & \multicolumn{1}{c}{Total neutron rate}  &  \multicolumn{1}{c}{Correlated background rate} \\

    (m.w.e.)     &      in $\nu$-target (h$^{-1}$)        &   (d$^{-1}$) \\ 
\hline
40      & 829       &      8.4        \\
60      & 543       &      5.4    \\
80      & 400       &      4.2    \\
100     & 286       &      3.0    \\
300     &  57       &      0.5    \\
\hline
\end{tabular}
\caption[Limits on the estimated neutron rate and the correlated
  background rate due to fast neutrons]
{\label{t4} Estimated neutron rate in the target region and the correlated background rate due to fast neutrons generated outside the detector by cosmic muons.}
\end{center}
\end{table}

The correlated background rates can be compared with accidental rates,
where by chance a neutron signal falls into the time window opened by a 
$\beta^+$-like~event.
The background contribution due to accidental delayed coincidences can
be determined {\it in situ} by measuring the single counting rates of neutron-like and $\beta^+$-like events.
Therefore the accidentals are not so dangerous as correlated background events.
Taking for granted we reach reasonably low concentrations of radioactive elements in the detector materials, 
especially in the scintillator itself (see discussion above), the beta-gamma rate above 1~MeV can 
be expected to be about a few counts per second.
If the time window for the delayed coincidence is $\sim$200~$\mu$s 
(this should allow a highly efficient neutron detection in Gd~loaded scintillators),
and the veto efficiency is at 98~\%
the accidental background rates can be estimated as depicted in Table~\ref{t5}.
The rate of neutrons which cannot be correlated to muons (``effective neutron rate'') 
is calculated by $ n_{\text{eff}} = n_{\text{tot}} \cdot (1 - \epsilon ) $, 
where $n_\text{tot}$ is the total neutron rate (sum of the numbers given in Table~\ref{t2} and 
\ref{t3}) 
and $\epsilon$ is the veto efficiency.
If the veto efficiency is 98~\% or better, the accidental background for the far
detector is far below one event per day (see following Table~\ref{t5}).\\

\begin{table}[h]
\begin{center}

\begin{tabular}{lrr}
\hline
\multicolumn{1}{c}{Overburden}  & \multicolumn{1}{c}{Effective neutron rate}  & \multicolumn{1}{c}{Accidental background rate} \\
            (m.w.e.)     &      ($\rm h^{-1}$)           &  ($\rm d^{-1}$)        \\
\hline
40    &  97   & 2.4            \\
60    &  65   & 1.6       \\
80    &  43   & 1.0           \\
100   &  28   & 0.7       \\
300   &   6   & 0.15           \\
\hline
\end{tabular}
\caption[Example of estimated accidental event rates for different
  shielding depths.] {\label{t5} Example of estimated accidental event
  rates for different shielding depths. The rates scale with the total
  beta-gamma rate above 1~MeV 
(here $b_{tot} = b_{ext} + b \approx 2.5~\rm s^{-1}$), the time window 
(here $\tau = 200 \, \mu $s) and the effective neutron background rate 
(here a muon veto efficiency of 98~\% was assumed).}
\end{center}
\end{table}
\subsection{Conclusion}
We conclude that correlated events are the most severe background source for the experiment.
Two processes mainly contribute: $\beta$-neutron cascades and very fast external neutrons.
Both types of events are coming from spallation processes of high energy muons.
In total the background rates for the near detector will be between 9/d and 23/d if a shielding
of 60~m.w.e. is choosen.
For the far detector a total background rate between 1/d and 2/d can be estimated.

\cleardoublepage
\cleardoublepage
\chapter{Experimental Errors}
\label{sec:systematics}
\section{From CHOOZ to Double-CHOOZ}
In the first CHOOZ experiment, the total systematic error amounted to
2.7~\%. The goal of Double-CHOOZ is to reduce the overall systematic 
uncertainty to 0.6~\%. 
A summary of the CHOOZ systematic errors is given in Table \ref{choozto2chooz}
\cite{choozlast}. The right column presents the new experiment goals. Lines 1,4, 
and 5 correspond to systematic uncertainties related  to the reactor 
flux and the cross section of neutrinos on the target protons. 
These errors become negligible if one uses two antineutrino
detectors located at different baselines. 
In order to improve the systematic uncertainties related to the
detector and to the $\nuebar$ selection cuts, the Double-CHOOZ
experiment will take advantage of the latest technical developments
achieved by the recent scintillator detector CHOOZ
\cite{choozlast}, CTF \cite{CTF}, KamLAND \cite{Eguchi:2002dm}, 
Borexino \cite{BorexReactor}, and the LENS R\&D phase \cite{lens}.
\begin{table}[h]
\begin{center}
\begin{tabular}{lrr}
\hline
 & \multicolumn{1}{c}{CHOOZ} & \multicolumn{1}{c}{Double-CHOOZ}  \\
\hline
 Reactor cross section & 1.9~\% & ---    \\
 Number of protons     & 0.8~\% & 0.2~\%  \\
 Detector efficiency   & 1.5~\% & 0.5~\%  \\
 Reactor power         & 0.7~\% & ---    \\
 Energy per fission    & 0.6~\% & ---    \\
\hline
\end{tabular}
\caption{\label{choozto2chooz} Overview of the systematic errors of
  the CHOOZ and Double-CHOOZ experiment.}
\end{center}
\end{table}
\section{Relative normalization of the two detectors}
The goal of Double-CHOOZ is to use two $\nuebar$ detectors in
order to cancel or decrease significantly the systematic
uncertainties that limit the  $\t13$ neutrino mixing angle measurement. 
However, beside those uncertainties, the
relative normalization between the two detectors is the most
important source of error and must be carefully controlled.
This section covers the uncertainties related to the $\nuebar$
interaction and selection in the analysis, as well as the
electronics and data acquisition dead times.
\section{Detector systematic uncertainties}
\subsection{Solid angle}
The distance from the CHOOZ detector to the cores of the nuclear plant
have been measured to within $\pm 10$~cm by the CHOOZ experiment. 
This translates into a systematic error of 0.15~\% in Double-CHOOZ, because the
effect becomes relatively more important for the near detector located 
 100-200~meters away from the reactor.
Specific studies are currently ongoing to guarantee this 10~cm error. 
 Furthermore, the ``barycenter'' of the neutrino emission in the
 reactor core must be monitored with the same precision. 
In a previous experiment at Bugey~\cite{Bugey}, a 5~cm change of
this barycenter was measured and monitored, using the instrumentation
of the nuclear power plant~\cite{garciaz1992}. Our goal is to
confirm that this error could be kept below 0.2~\%.
\subsection{Number of free protons in the target}
\subsubsection{Volume measurement}
In the first CHOOZ experiment, the volume measurement was done with an 
absolute precision of  0.3~\%~\cite{choozlast}.
The goal is to reduce this uncertainty by a factor of two, but only on the 
relative volume measurement between the two inner acrylic vessels (the
other volumes do not constitute the $\nuebar$ target).
An R\&D program has already started in order to find the optimal solution
for the relative volume determination (See
Section~\ref{subsec:volumemes}). 
Among some ideas under study,
we plan to use the same mobile tank to fill both targets; a pH-based
measurement is being studied as well.  
A more accurate measurement could be performed by combining a
traditional flux measurement with a weight measurement of the quantity
of liquid entering the acrylic vessel.
Furthermore we plan to build both inner acrylic targets at
the manufacturer and to move each of them as a single unit into the detector
site. A very precise calibration of both inner vessels is thus foreseen
at the manufacturer (filling tests).
\subsubsection{Density}
The uncertainty of the density of the scintillator is $\sim$0.1~\%.
The target liquid will be prepared in a large single batch,
so that they can be used for the two detector fillings. 
The same systematic effect will
then occur in both detectors and will not contribute to the
overall systematic error (this effect will be included automatically in the
absolute normalization error, see Chapter~\ref{sec:sensitivity}).  
However, the measurement and control of the temperature will be
mandatory to guarantee the stability of the density in both targets
(otherwise it would contribute to the relative uncertainty, 
see Chapter~\ref{sec:sensitivity}). 
To thermalize both $\nuebar$ targets, the temperature control and
circulation of the liquid in the external veto is foreseen. 
\subsubsection{Number of hydrogen atoms per gramme}
This quantity is very difficult to measure, and the error is of the
order of 1~\%; however, the target liquid will be prepared in a large 
single batch (see above). 
This will guarantee that, even if the absolute value is not known to
a high precision, both detectors will have the same number of hydrogen atoms per
gramme. This uncertainty, which originates in the presence of unknown 
chemical compounds in the liquid, does not change with time.
\subsection{Neutron efficiency}
The thermal neutron is captured either on hydrogen or on Gadolinum
(other reactions such as Carbon captures can be neglected). 
We outline here the systematical errors related to the neutron signal.
\subsubsection{Gadolinium concentration}
Gd concentration can be extracted from a time capture measurement
done with a neutron source calibration (see Chapter~\ref{sec:calibration}). 
A very high precision can be reached on the neutron efficiency (0.3~\%)
by measuring the detected neutron multiplicity from a
Californium source (Cf).
This number is based on the precision quoted in~\cite{choozlast}, but
taking away the Monte-Carlo uncertainty, since we work with
two-identical detectors.
This precision is expected to be better by a factor of two in the
Double-CHOOZ experiment because it is easier to compare two experimental
measurements in identical detectors than to compare a
 theoretical spectrum with a measurement. 
We can increase our sensitivity to very small differences in the
response from both detectors by using the same calibration source for
the measurements.
The Californium source calibration can be made all along the z-axis
of the detector, and is thus snsitive to spatial effects due to the variation of 
Gd concentration (staying far enough from the boundary of the target,
and searching for a top/down assymetry). A difference between the time capture of both
detectors could also be detected with a sensitivity slightly less 
than 0.3~\%. 
\subsubsection{Spatial effects}
We consider here the spill in/out effect, i.e the
edge effect associated with neutron capture close to the acrylic
vessel surrounding the inner target~\cite{choozlast}, and the
angle between the neutron direction and the edge of the acrylic target that
is slightly different between the two detectors. 
The $\sim$1~\% spill in/out effect oberved in the first CHOOZ
experiment~\cite{choozlast} cancels by using a set of two identical
detectors (same effect).  
Nevertheless the second effect (angle) persists, but is considered to be
negligible.
%
\subsection{Positron efficiency}
The simulation of the Double-CHOOZ detectors confirms that 
a 500~keV energy cut induces a positron inefficiency smaller than 0.1~\% 
(see Chapter~\ref{sec:fulldetsimul}). 
The relative uncertainties between both detectors lead thus  to an
even smaller systematic error and is therefore negligible.
\section{Selection cuts uncertainties}
The analysis cuts are potentially important sources of systematic errors. In
the first CHOOZ experiment, this amounted in total to 1.5~\%~\cite{choozlast}. 
The goal of the new experiment is to reduce this error by a factor of three.
The CHOOZ experiment used 7 analysis cuts to select the $\nuebar$ 
(one of them had 3 cases, see Section~8.7 of~\cite{choozlast}). 
In Double-CHOOZ we plan to reduce the number of selection cuts
 to 3 (one of them will be very loose, and may not even be used). 
This can be achieved because of reduction of the number of 
accidentals background events, only possible with the new detector
design (see Chapter~\ref{sec:overview}).
To select $\nuebar$ events we have to identify the prompt positron 
followed by the delayed neutron (delayed in time and separated in space).
The trigger will require two local energy depositions of more than
500~keV in less than 200~$\mu$s. 
\subsection{Identifying the prompt positron signal}
Since any $\nuebar$ interaction deposits at least 1~MeV (slightly less
due to the energy resolution effect) the energy cut at 500~keV does not 
reject any  $\nuebar$ events. 
As a consequence, there will not be any systematic error associated with
 this cut (see Figure \ref{fig:positronspectrum}). 
The only requirement is the  stability of the energy selection cut, 
which is related to the energy calibration (see Chapter~\ref{sec:calibration}). 
\begin{figure}[h]
\begin{center}
\includegraphics[width=0.6\textwidth]{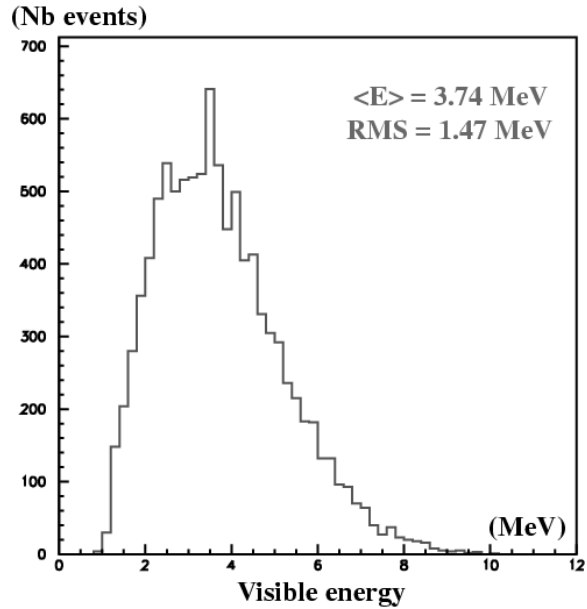}
\caption[Simulation of the positron energy spectrum  measured with the 
Double-CHOOZ  detector]{Simulation of the positron energy spectrum (in~MeV) measured with the 
Double-CHOOZ  detector (10,000 events, without backgrounds). Positron energy is fully contained 
  with a probability of 99.9~\%, as a consequence  of the 60~cm scintillating buffer.}
\label{fig:positronspectrum}
\end{center}
\end{figure}
\begin{figure}[h]
\begin{center}
\includegraphics[width=0.6\textwidth]{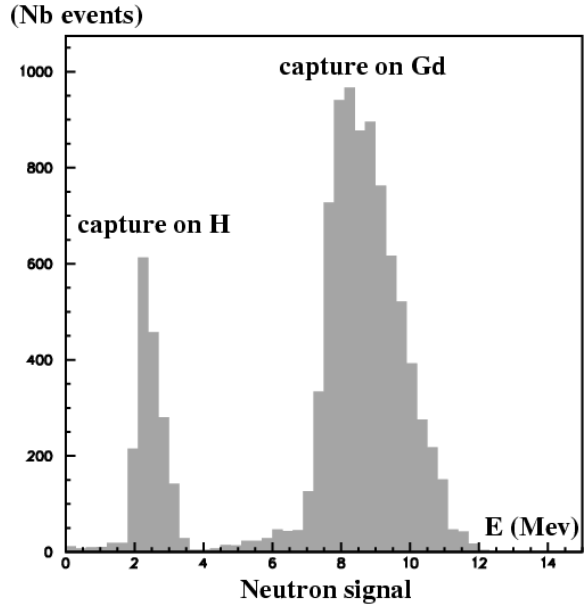}
\caption[Simulation of the neutron energy spectrum  measured with the 
Double-CHOOZ detector]{Simulation of the neutron energy spectrum (in~MeV) measured with the 
Double-CHOOZ detector (10,000 events, without backgrounds). There are two energy
peaks for the neutron capture on hydrogen (releasing 2.2~MeV) and on gadolinium 
(releasing about 8~MeV). The Double-CHOOZ experiment will
select all neutron events with an energy greater than 6~MeV.
The resulting systematic uncertainty thus depends on the relative 
calibration between the near and far detectors.}
\label{fig:neutronspectrum}
\end{center}
\end{figure}
\subsection{Identifying the neutron delayed signal}
The energy spectrum of a neutron capture has two peaks, the first peak 
at 2.2~MeV tagging the neutron capture on hydrogen, 
and the second peak at around 8~MeV tagging the neutron capture on Gd
(see Figure~\ref{fig:neutronspectrum}). 
The selection cut that identifies the neutron will be set at about 6~MeV,
which is above the energy of neutron capture on hydrogen and
all radioactive contamination.
At this energy of 6~MeV, an error of $\sim$100~keV on the selection
cut changes the number of neutrons by $\sim$0.2~\%. 
This error on the relative calibration is achievable by using
 identical Cf calibration source for both detectors (see Chapter~\ref{sec:calibration}).
\subsection{Time correlation}
The neutron time capture on Gd in the CHOOZ detector is displayed in 
Figure~\ref{fig:amdel}. But since the exact analytical behaviour
 describing the neutron capture time on Gd is not known, 
the absolute systematic error for a single detector
cannot be significantly improved with respect to
 CHOOZ~\cite{choozlast}. However, the uncertainty originating from
the liquid properties disappears by comparing the near and far
 detector neutron time capture. The remaining effect deals with the
 control of the electronic time cuts. 
For completeness, a redundant system will be designed  in order to control
perfectly these selection cuts (for example
time tagging in a specialized unit and using Flash-ADC's).
\subsection{Space correlation}
The distance cut systematic error (distance between prompt and
delayed events) was published as 0.3~\% in the CHOOZ
experiment~\cite{choozlast}. This cut is very difficult to calibrate, 
 since the rejected events are typically $\nuebar$ 
candidates badly reconstructed. In Double-CHOOZ, this cut
 will be either largely relaxed (two meters instead of one meter for instance) 
or totally suppressed, if the accidentals event rate is low enough, as
expected from current simulations (see Chapter~\ref{sec:backgrounds}).
\begin{figure}[h]
\begin{center}
\includegraphics[width=0.65\textwidth]{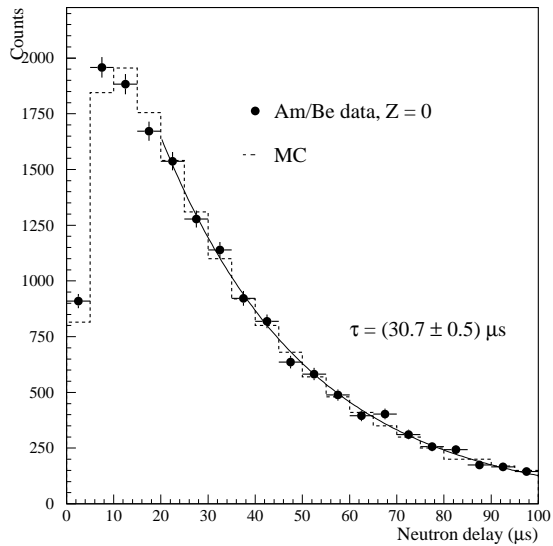}
\caption[Neutron delay distribution measured with the Am/Be source at
  the detector centre in the CHOOZ detector]
{Neutron delay distribution measured with the Am/Be source at
  the detector centre in the CHOOZ detector~\cite{choozlast}. 
The time origin is defined by the $4.4$~MeV $\gamma$-ray.}
\label{fig:amdel}
\end{center}
\end{figure}
\subsection{Veto and dead time}
The Double-CHOOZ veto will consist of a liquid scintillator and 
have a thickness of 60~cm liquid scintillator at the far site, and
even larger at the near detector site. 
The veto inefficiency comes from the 
through going cables and the supporting structure material. This inefficiency was 
low enough in the first experiment, and should be acceptable for the 
CHOOZ-far detector. 
However, it must be lowered for the near detector because the muon
flux is a factor 30 higher for a shallower overburden of 60~m.w.e..
A constant dead time will be applied in coincidence with each through
going muon. This has to be measured very carefully since the resulting dead time
will be very different for the two detectors: a few percent at the far
detector, and at moreless~30~\% at the near detector. 
A  1~\% precision on the knowledge of this dead time is mandatory. 
This will require the use of several independent methods:
\begin{itemize}
\item{the use of a synchronous clock, to which the veto will be applied,}
\item{a measurement of the veto gate with a dedicated flash ADC,}
\item{the use of an asynchronous clock that randomly generates two
  particles mimicking the antineutrino tag 
  (with the time between them characteristic of the neutron capture on
  Gd). With this method, all dead times (originating from the veto as
  well as from the data acquisition system) will be measured simultaneously.
  The acquisition of a few thousands such events per day would achieve
  the required precision,}
\item{the generation of sequences of veto-like test pulses 
(to compare the one predicted dead time to the actually measured).}
\end{itemize}
\subsection{Electronics and acquisition}
The trigger will be rather simple. It will use only the total analog sum
of  energy deposit in the detector. 
Two signals of more than 500~keV in 200~$\mu$s will be required.
\subsection{Summary of the systematic uncertainty cancellations}
A summary of the systematic errors associated with $\nuebar$ event
selection cuts is given in Table \ref{tab:cuterrors}.
\begin{table}[h]
\begin{center}
\begin{tabular}{lrrc}
  \hline
  & \multicolumn{1}{c}{CHOOZ}          & \multicolumn{2}{c}{Double-CHOOZ} \\
  \hline
  selection cut &   rel. error $(\%)$ & rel. error $(\%)$ & Comment\\
  \hline
  positron energy$^\ast$ &  $0.8$ & $0$ & not used \\
  positron-geode distance &  $0.1$ & $0$ & not used \\
  neutron capture &  $1.0$ &  $0.2$ &  Cf calibration\\
  capture energy containment & $0.4$ & 0.2 & Energy calibration \\
  neutron-geode distance &  $0.1$ & 0 & not used  \\
  neutron delay &  $0.4$ & 0.1 &  --- \\
  positron-neutron distance &  $0.3$ & $0-0.2$ & 0 if not used\\
  neutron multiplicity$^\ast$ &  $0.5$ & $0$ & not used\\
  combined$^\ast$ & $1.5$ & $0.2$-$0.3$ & ---\\
    \hline
    \multicolumn{3}{l}{$^\ast${\small average values}} 
\end{tabular}
\caption[Summary of the neutrino selection cut uncertainties]
{Summary of the neutrino selection cut uncertainties. CHOOZ
values have been taken from~\cite{choozlast}.}
\label{tab:cuterrors}
\end{center}
\end{table}
}
We summarize in Table \ref{tab:syscancels} the systematic
uncertainties that totally cancel, or to a large extent, in the
Double-CHOOZ experiment.
\begin{table}[h]
\begin{center}
\begin{tabular}{lrr}
\hline
 & \multicolumn{1}{c}{CHOOZ} & \multicolumn{1}{c}{Double-CHOOZ} \\
\hline
Reactor power             &         0.7~\%   &   negligible \\
Energy per fission        &         0.6~\%   &   negligible \\
$\nuebar$/fission         &         0.2~\%   &   negligible \\
Neutrino cross section    &         0.1~\%   &   negligible \\
Number of protons/$\mathrm{cm^3}$  &         0.8~\%   &   0.2~\% \\
Neutron time capture      &         0.4~\%   &   negligible \\       
Neutron efficiency        &         0.85~\%  &   0.2~\% \\       
Neutron energy cut \footnotemark  & 0.4~\%   &   0.2~\%\\
\hline
\end{tabular}
\caption[ Summary of systematic errors
  that cancel or are significantly decreased in Double-CHOOZ]
{\label{tab:syscancels} Summary of systematic errors
  that cancel or are significantly decreased in Double-CHOOZ.} 
\end{center}
\end{table}
\footnotetext{Energy cut on gamma spectrum from a Gd neutron capture.}
The error on the absolute knowledge of the chemical composition of the
Gd scintillator disappears. There  remains only the measurement error on the 
volume of target (relative between two detectors).
The error on the absolute knowledge of the gamma spectrum from a Gd
neutron capture disappears. However, there will be a calibration error 
on the difference between the 6~MeV energy cut in both detectors.   
\subsection{Systematic uncertainties outlook}
Table \ref{systematics2} summarizes the identified systematic errors
that are currently being considered for the Double-CHOOZ experiment.
\begin{table}[h]
\begin{center}
\begin{tabular}{lrr}
\hline
 & \multicolumn{1}{c}{After CHOOZ} & \multicolumn{1}{c}{Double-CHOOZ Goal} \\
\hline
 Solid angle & 0.2~\%         & to confirm \\
 Volume      & 0.2~\%             & to confirm \\
 Density     & 0.1~\%             & 0.1~\% \\
 Ratio H/C   & 0.1~\%             & 0.1~\%  \\
 Neutron efficiency & 0.2~\%      & 0.1~\%   \\
 Neutron energy & 0.2~\%          & 0.2~\% \\ 
 Spatial effects    & neglect ?  & to confirm \\
 Time cut    & 0.1~\%               & 0.1~\% \\
 Dead time(veto) & 0.25~\%        & to improve  \\
 Acquisition & 0.1~\%              &  0 .1~\% \\
 Distance cut  & 0.3~\%      & 0-0.2~\%  \\
\hline 
 Grand total & 0.6~\%&  $<$ 0.6~\% (to confirm) \\ 
\hline
\end{tabular}
\caption[Systematic errors that can
be achieved without improvement of the CHOOZ published  
systematic uncertainties]{\label{systematics2}
The column ``After CHOOZ'' lists the systematic errors that can
be achieved without improvement of the CHOOZ published systematic
uncertainties \cite{choozlast}.
In Double-CHOOZ, we estimate the total systematic error on the 
normalization between the detectors to be less than 0.5~\%. 
The aim of the work prior the final proposal is to confirm this
number, and thus increase the safety margin of the experiment.}
\end{center}
\end{table}
\section{Background subtraction error}
The design of the detector will allow a Signal/Background (S/B)
ratio of about 100 to be achieved 
(compared to 25 at full reactor power in the first experiment \cite{choozlast}). 
The knowledge of the background at a level around 30-50~\% will reduce
the background systematic uncertainties to an acceptable level.
In the Double-CHOOZ experiment, two background components have been
identified, uncorrelated and correlated (see
Chapter~\ref{sec:backgrounds}).  Among those backgrounds, one has:
\begin{itemize}
\item{The accidental rate, that can be computed from the single event
  measurements, for each energy bin.}
\item{The fast neutrons creating recoil protons, and then a neutron
  capture. This background was dominant in the first experiment
  \cite{choozlast}.  The associated energy spectrum is relatively flat
  up to a few  tens of~MeV.}
\item{The cosmogenic muon induced events, such as $^9$Li and $^8$He, 
 that  have been studied and measured at the NA54 CERN experiment 
 \cite{NA54} in a muon beam as well as in the KamLAND experiment \cite{Eguchi:2002dm}. 
 Their energy spectrum goes well above 8~MeV, and follows a well defined shape.}
\end{itemize}
The backgrounds that will be measured are:
\begin{itemize}
\item{Below 1~MeV (this was not possible in the first experiment, due
  to the different detector design and the higher energy threshold)}
\item{Above 8~MeV (where there remains only 0.1~\% of the neutrino
  signal).}
\item{By extrapolating from the various thermal power of the plant
     (refueling will result in two months per year at half power).}
\end{itemize}
From the measurement of the accidental events energy shape, and from
the extraction of the cosmogenic events shape, the shape of the
fast neutron events can be obtained with a precision greater than what
is required.
\section{Liquid scintillator stability and calibration}
The experiment has some sensitivity to a slight distortion induced by 
neutrino oscillations.  A rate only analysis  would 
only provide a sensitivity that is twice the quoted value of 0.03 on
$\mathrm{sin^2(2\theta _{13})}$. 
From the simulation,  identical energy scales at the 1~\% level is
necessary. 
The specification of no more than 100~keV scale difference at 6~MeV is 
achieved if this 1~\% level is obtained.
This relative calibration is easier than an absolute linearity, but
still very important to consider in the detector design. 
We can, for example, move the same calibration radioactive sources from
one detector to the other, and directly
compare the position of the well defined calibration peaks.  
\cleardoublepage
\cleardoublepage
\chapter{Sensitivity and discovery potential}
\label{sec:sensitivity}
We describe here the details of the simulation of the 
Double-CHOOZ experiment. The sensitivity to \ssqtt is presented in
Section~\ref{sec:sensitivitylimit}, and we present the discovery 
potential of the experiment in Section~\ref{sec:discovery}. 
The statistical analysis (systematic error 
handling) introduced here is based on the work of~\cite{huberreactor2003}.
\section{The neutrino signal}
\label{sec:parameters}
In this section we describe the set of parameters used in the simulation.
\subsection{Reactor \nuebar spectrum}
The \nuebar spectrum above detection threshold is the result of
$\beta^-$ decays of \atom{U}{235}{}, \atom{U}{238}{},
\atom{Pu}{239}{} and \atom{Pu}{241}{} fission
products. Measurements for \atom{U}{235}{}, \atom{Pu}{239}{} and
\atom{Pu}{241}{} and theoretical
calculations for \atom{U}{238}{} are used to evaluate the \nuebar spectrum
\cite{Schreckenbach:1985ep,Hahn:1989zr}. 
While a nuclear reactor operates, the fission products
proportions evolve in time; as an approximation in this evaluation, we use a typical averaged
fuel composition during a reactor cycle corresponding to  
 55.6~\% of \atom{U}{235}{}, 32.6~\% of \atom{Pu}{239}{}, 7.1~\% of \atom{U}{238}{} and 
4.7~\% of \atom{Pu}{241}{}.
The mean energy release per fission is then 
203.87~MeV and the energy weighted cross section for $\nuebar \, p
\rightarrow n \, e^+$ amounts to $\left<\sigma\right>_{\rm fission} = 5.825 \cdot
10^{-43}$~cm$^2$ per fission.
\subsection{Detector and power station features}
Table~\ref{tab:choozb} contains the principal features of the CHOOZ
power station nuclear cores, as well as their distances from the near and
far detectors.
\begin{table}[h!]
  \begin{center}
    \begin{tabular}{lrr}
      \hline
      & \multicolumn{1}{c}{CHOOZ-B-1}  & \multicolumn{1}{c}{CHOOZ-B-2}\\
      \hline
      Electrical Power
      (raw/net $\text{GW}_{\text{e}}$) & 1.516/1.455 & 1.516/1.455 \\
      Thermal power
      ($\text{GW}_{\text{th}}$)          & 4.2   & 4.2 \\
      Global load factor                 & 80~\%      & 80~\%  \\
      Near detector distance             & 100-200~m      & 100-200~m \\
      Far detector  distance             & 1,000~m     & 1,100~m \\
      \hline
    \end{tabular}
  \end{center}
  \caption[Chooz power station main features]{Chooz power station main
  features~\cite{CEAelecnuc}.
  \label{tab:choozb}}
\end{table}
Table~\ref{tab:detectorsim} presents the characteristics of the
detectors used in the simulation. We considered a target scintillator 
composition  of 20~\% of PXE and 80~\% of dodecane
(see Chapter~\ref{sec:scintillator}). 
This translates into  $8.33 \cdot 10^{29}$~free
protons in the $12.7~\mathrm{m}^3$ inner acrylic vessel. For simplicity
we assume that the two cores are equivalent to a single core of
8.4~$\text{GW}_{\text{th}}$ 
located 150~m away from the near detector and 1,050~m from the
far detector. We checked that a full simulation with two separated
cores at CHOOZ does not change the results presented here.
\begin{table}[h!]
  \begin{center}
    \begin{tabular}{lrr}
      \hline
                    & \multicolumn{1}{c}{Near Detector}                &  \multicolumn{1}{c}{Far Detector}  \\      
      \hline
      Distance      &  100 m                       &  1,050 m \\
      Target volume &  $12.7~\text{m}^3$           &  $12.7~\text{m}^3$\\
      Target mass   &  10.16 tons                  &  10.16 tons  \\
      Free H        &  $8.33\,10^{29}$             & $8.33\,10^{29}$  \\
      \hline
      Detection efficiency   &  80~\%                & 80~\%\\      
      Reactor efficiency    & 80~\%                 & 80~\%\\
      Dead time     & 50~\%                         & a few~\% \\ 
      Overall efficiency & 32~\%                     & 64~\%\\
      \hline
      \nuebar events after 3~years & 3,213,000       & 58,000 \\
      \hline
    \end{tabular}
  \end{center}
  \caption[Detector parameters used in the simulation]
{Detector parameters used in the simulation. As an example we take
  here the near detector distance at 100~m. Results presented
  in this chapter don't change if this distance is increased to 200~m.
  \label{tab:detectorsim}}
\end{table}
The global load factor of the CHOOZ nuclear reactor is taken to be 80~\%.
We assume that the detection efficiency for both detectors is 80~\%
(69.8~\% in CHOOZ~\cite{choozlast}). We neglect the dead time for the
far detector (300~m.w.e. overburden). Since the CHOOZ near site will
be shallower, between 60 to 80 m.w.e, we apply a dead time
of 50~\% to be conservative (a 500~$\mu\text{sec}$ cut to each muon
crossing the detector leads to a dead time around 30~\% at
60~m.w.e). 
The overall efficiencies used in the simulation for the near and 
far detectors are thus respectively 32~\% and 64~\%. 
\subsection{Expected number of events}
Neglecting the correction terms of order
$\alpha=\left(\frac{\sdm2}{\adm2}\right)^2\approx (2\cdot 10^{-2})^2$, we used
the following
\nuebar survival probability:
\begin{equation}
  \label{eq:survivalprobability}
  P_{\nuebar \rightarrow \nuebar} = 1-\ssqtt\sin^2\left(1.27\frac{\dmgui_{23}
  L[\text{m}]}{\Enu[\text{MeV}]}\right).
\end{equation}
The expected number of antineutrino events in the near ($N_i^N$) and far
detector ($N_i^F$), in the energy bin $[E_i,E_{i+1}]$, is
\begin{equation}
  \label{eq:eventrates}
  N_i^A  = 
  {\cal F}^A\int_{E_i}^{E_{i+1}}\int_0^{+\infty} S(\Enu,\Enu')\sigma(\Enu)\phi_i(\Enu,L^A)
  P_{\nuebar\rightarrow\nuebar}\left(\Enu,L^A\right)\dd\Enu\dd
  \Enu', 
\end{equation}
where $A=N,F$. The cross section~$\sigma$ is given in 
equation~\ref{eq:ibdecaycrosssection}, and the $\nuebar$ flux
is computed according to Figures~\ref{fig:nuspe} and \ref{fig:burnup} . 
The normalization factor~${\cal F}$ includes the global load factor~$G$
(fraction of running time of the reactors over a year), the reactor thermal
power~$P$, the detector efficiency~$\varepsilon^A$, the dead time
fraction ~$D^A$, the target volume~$V$ and the exposure time~$T$:
\begin{equation}
  \label{eq:norm}
  {\cal F}^A = G \times P \times V \times T \times (1-D^A) \times \varepsilon^A
\end{equation}
The energy resolution effect is taken into account as follows:
\begin{equation}
  \label{eq:energyresolution}
  S(E,E') = {\cal N}\left(E-E',\frac{8~\%}{\sqrt{E}}\right),
\end{equation}
where ${\cal N}$ is a Gaussian distribution.
In practice, we have used an energy bin size at least four times larger 
than the energy resolution effect and thus we neglected it in first 
approximation for this analysis. We checked this approximation by
comparing our results with the work of~\cite{huberreactor2003,theta13globalana}.
\section{Systematic errors handling}
\label{sec:chi2analysis}
\subsection{$\chi^2$ analysis}
In this section we describe the $\chi^2$-analysis of the near-far
detector set and how we implemented the systematic errors previously
discussed. We write $O_i^A$ the computed number of events observed in
$i^{\text{th}}$ energy bin in near (\mbox{$A=N$}) and far
(\mbox{$A=F$}) detectors. The theoretical predictions for the detector $A$
in the $i^{\text{th}}$ bin is
\begin{equation}
  \label{eq:tia}
  T_i^A = \left(1+a+b^A+c_i\right)
  \sum_{j=1}^{N_{\text{cores}}}(1+f_j)N_{i,j}^A + g^A M_i^A
\end{equation}
where $a$, $b^A$, $c_i$, $f_j$, $g^A$ will be the fitted
parameters. $M_i^A$ is the first order correction term to take into
account the energy scale uncertainty, obtained by replacing
$E_{\text{vis}}$ by $(1+g^A)E_{\text{vis}}$:

\begin{equation}
  \label{eq:mia}
  M_i^A = \sum_{j=1}^{N_{\text{cores}}}\left.\frac{\dd
  N_{i,j}^A(g^A)}{\dd g^A}\right|_{g^A=0}
\end{equation}
where $N_{i,j}^A$ is the computed number of events in
$i^{\text{th}}$ bin in detector $A$ coming from the $j^{\text{th}}$
reactor core:
\begin{eqnarray}
  \label{eq:nija}
  N_{i,j}^A & = &
  {\cal F}^A\int_{E_i}^{E_{i+1}}\int_0^{+\infty} 
S(\Enu,\Enu')\sigma(\Enu)\phi_i(\Enu,L_j^A)
  P_{\nuebar\rightarrow\nuebar}\left(\Enu,L_j^A\right)\dd\Enu\dd 
\Enu'\nonumber\\
\end{eqnarray}
which depends on the oscillation parameters through the survival
probability. The observed number of
events $O_i^A$ has been chosen to be the computed event number for
given ``true~values'' of the oscillation parameters:
$O_i^A=\sum_{j=1}^{N_{\text{cores}}}N_{i,j}^A(\ssqtt,\dmgui)$.
We used a $\chi^2$ function including the full spectral information from
both detectors:
\begin{eqnarray}
  \label{eq:chi2}
  \chisq & = &
\sum_{i=1}^{N_{\text{bins}}}\sum_{A=N,F}\frac{\left(T_i^A-O_i^A-e^AB_i^A\right)^2}%
  {O_i^A + (\sbtb O_i^A)^2 + B_i^A + (\sbkg B_i^A)^2}\nonumber\\
  & + & \left(\frac{a}{\sabs}\right)^2
  + \sum_{i=1}^{N_{\text{bins}}} \left(\frac{c_i}{\sshp}\right)^2
  + \left(\frac{d-\dmgui_{\text{Best}}}{\sdmt}\right)^2
  + \sum_{j=1}^{N_{\text{cores}}} 
\left(\frac{f_j}{\scfl}\right)^2\nonumber\\
   & + &
  \sum_{A=N,F} \left[
    \left(\frac{b^A}{\srel}\right)^2 +
    \left(\frac{g^A}{\sscl}\right)^2 +
    \left(\frac{e^A}{\sbkg}\right)^2\right]
\end{eqnarray} \\

For each point in the oscillation parameters space, the $\chi^2$
function has to be minimized with respect to the parameters $a$,
$b^N$, $b^F$, $c_i$, $g^N$, $g^F$, $d$, $e^N$, $e^F$, $f_j$ modeling
the systematic errors. The parameter $a$ refers to the error on the
overall normalization of the number of events common to both
detectors. Parameters $b^N$ and $b^F$ relate to the uncorrelated
normalization uncertainties of the two detectors. The energy scale
uncertainty is taken into account through parameters $g^N$ and $g^F$
in the expression of $T_i^A$ in equation~\ref{eq:tia}.
We assumed here a flat background distribution:
\begin{equation}
  \label{eq:background}
  B_i^A = \alpha \sum_{j=1}^{N_{\text{bins}}} 
\frac{O_j^A}{N_{\text{bins}}}
\end{equation}
The numerical minimization has been performed with the  MINUIT
package~\cite{Minuit}. We now discuss all the relevant terms of
Equation~\ref{eq:background} in turn.
\subsection{Absolute normalization error: \sabs}
We include a common overall normalization error for the event rate of the
near and far detectors. This error accounts for the uncertainty on the
\nuebar flux of the reactor, the detection cross section, or any
bias that could affect both detectors in the same way\footnote{For
  instance,  a bias in the volume measurement affecting
the two detectors is equivalent to an uncertainty in the reactor
\nuebar flux.}. This error is of the order of a few percent; one
has for instance 1.4~\% in~\cite{Declais}, 2~\% in~\cite{Eguchi:2002dm}.
The overall normalization error has almost no impact on the
sensitivity to an oscillation effect in the
Double-CHOOZ experiment since two detectors will be used (see Figures~
\ref{fig:doublechoozdiscovery1} and~\ref{fig:doublechoozdiscovery2}).
Nevertheless, we included an absolute normalization error 
$\sabs=2~\%$ in the simulation. 
\subsection{Relative normalization error: \srel}
We take into account an uncorrelated normalization uncertainty between 
the near and far detectors. This is the dominant experimental error for
the Double-CHOOZ experiment. There are contributions from
uncertainties on the detector design (fiducial volume, stability of
the liquid scintillator, relative dead time measurement) and the 
uncertainties related to the \nuebar event selection cuts (relative 
detector efficiency). 
According to the results presented in Chapter~\ref{sec:systematics},
we take the relative normalization error $\srel=0.6~\%$ as our
default value. 
\subsection{Spectral shape error: \sshp}
To take into account the \nuebar spectrum shape uncertainty, we 
introduce an error \sshp on the theoretical prediction 
for each energy bin which we take to be fully uncorrelated 
between different energy bins.
Since this error is induced by the physical uncertainty on the fission
 product beta decay spectra, it is fully correlated between the 
corresponding bins in the near and far detector.
In the simulation we use the shape error value
$\sshp=2~\%$, as measured in~\cite{Bugey}.
\subsection{Energy scale error: \sscl}
We take into account the energy scale calibration uncertainty by
introducing a parameter $g^A$ for each detector ($A=N,F$), and
replacing the observed energy $E_{\text{obs}}$ by
$(1+g^A)E_\text{obs}$. We assume that the energy calibration is
known with an error of $\sscl\sim0.5~\%$.
We found that, as long as no detailed background simulation is performed
on the data, this error can be neglected in first approximation for
the sensitivity computations. 
This is understandable since the Double-CHOOZ experiment is mostly 
sensitive to the number of events integrated over the whole positron
spectrum. 
Nevertheless, a careful study of this error is going on to better
understand its influence on the discovery potential of Double-CHOOZ. 
\subsection{Individual core power fluctuation error: \scfl}
Since the Double-CHOOZ power station has two nuclear cores, we
introduced an independent error of $\scfl=0.5~\%$ mimicking a 
thermal power fluctuation of each nuclear core. Indeed, depending on 
the exact location of the near detector site,
the near and far detectors will not receive the same \nuebar
contribution from both cores. In that case, an independent fluctuation
in the two cores could lead to a relative systematic error between the
detectors. However, we found this error to be negligible and we do not 
consider it further.
\subsection{Background subtraction error}
We considered two different ways to introduce an error on the
background subtraction procedure. 
\subsubsection{Reactor \nuebar shape background: \sbtb}
 This is modeled as an uncorrelated error \sbtb in 
 the background subtraction step.  
 This error is bin-to-bin uncorrelated,
 uncorrelated between the near and far detectors, and proportional to
 the bin content (i.e. the background has the same shape as the
 positron spectrum).  
 Typically we used values ranging from $\sbtb=0.5~\%$ to 
 $\sbtb=1.5~\%$.
\subsubsection{Flat background: \sbkg}
 This background is closer in shape to the background of fast
 neutrons  created in the rocks close to the detector. It 
  was dominating in CHOOZ~\cite{choozlast}, and is expected to play 
 an important role in Double-CHOOZ as well (see Chapter~\ref{sec:backgrounds}). 
 We assume that it amounts typically for $R_N=1~\%$ \& $R_F=1~\%$
 of the total \nuebar signal. To be conservative we  consider an error on
 those rates of  $\sbkg^N=100~\%$ $\sbkg^F=100~\%$, in the near
 and far detectors.
A careful study of the impact of the background on the sensitivity and
on the discovery potential as well is going on.
\section{Sensitivity in the case of no oscillations}
\label{sec:sensitivitylimit}
\begin{figure}[h!]
\begin{center}
\includegraphics[angle=-90,width=0.75\textwidth]{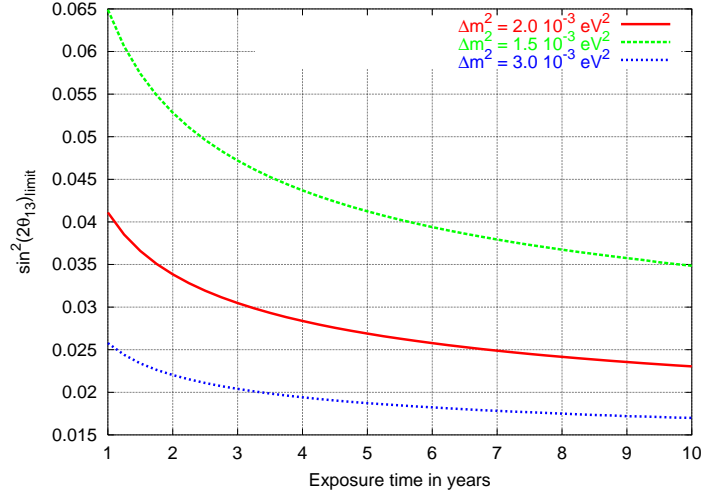}
\caption[Evolution of \ssqtt sensitivity with the exposure time
  (with  \dmgui interval taken from the oscillation analysis of
  Super-Kamiokande data in July 2003)]%
{Evolution of \ssqtt sensitivity with the exposure time.
 The three curves shown here are for different values of \dmgui as shown in
  the legend.} 
\label{fig:senstimesk-1}
\end{center}
\end{figure}

We present our results for the current best fit
value of the atmospheric mass splitting
\mbox{$\dmgui_{23}=2.0^{+1.0}_{-0.7}\cdot 10^{-3}~\text{eV}^2$}
\cite{SK_atm_nu2002} as our default value. 
Nevertheless, we also used the recent analysis of the Super-Kamiokande data leading to 
\mbox{$\dmgui_{23}= 2.4^{+0.6}_{-0.5}\cdot
  10^{-3} \text{eV}^2$} \cite{SkAtmLoverE2004}, for completeness.
We also assume that a forthcoming accelerator experiment will provide a precise
measurement  of $\dmgui_{23}$, with an error better than 20~\%: 
\mbox{$\sdmt=0.2\cdot\dmgui$} prior to the Double-CHOOZ
result~\cite{Minos1, Minos2}.
Figure~\ref{fig:senstimesk-1} displays the expected sensitivity of
Double-CHOOZ in the case of no-oscillations, 
as a function of time. In this case we  have a sensitivity of
$\ssqtt<0.045$ (90~\% C.L.) after one year of data taking, and
$\ssqtt<0.03$ after three years.
\begin{figure}[h!]
\begin{center}
\includegraphics[angle=-90 , width=0.75\textwidth]{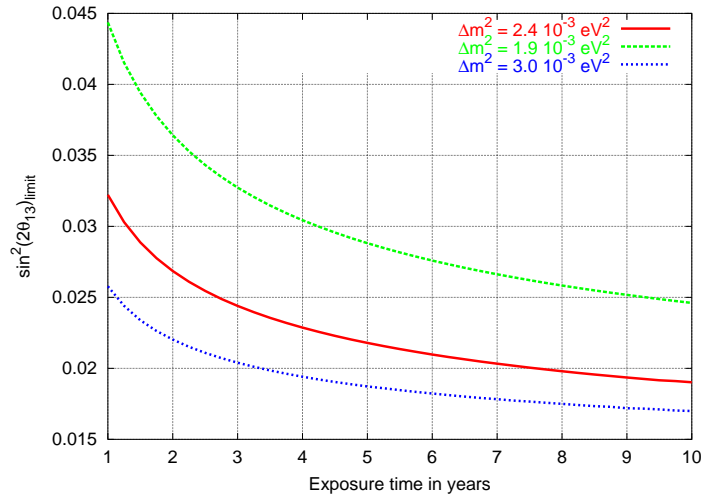}
\caption[Evolution of \ssqtt sensitivity with the exposure time
  (with  \dmgui 90~\% C.L. interval taken from the second analysis
  (L/E) of the same Super-Kamiokande data]
{Evolution of \ssqtt sensitivity with the exposure time. The
  three curves shown here are for different values of \dmgui as shown in
  the legend. These values have been chosen from the second analysis
  (L/E) of the same Super-Kamiokande data ~\cite{SkAtmLoverE2004}.}
\label{fig:senstimesk-2}
\end{center}
\end{figure}
%
%
%
%
 The sensitivity dependence with respect to the atmospheric mass
 splitting $adm2$ value is shown in Figure~\ref{fig:sensdm2}. 
\begin{figure}[h!]
\begin{center}
\includegraphics[angle=-90 , width=0.75\textwidth]{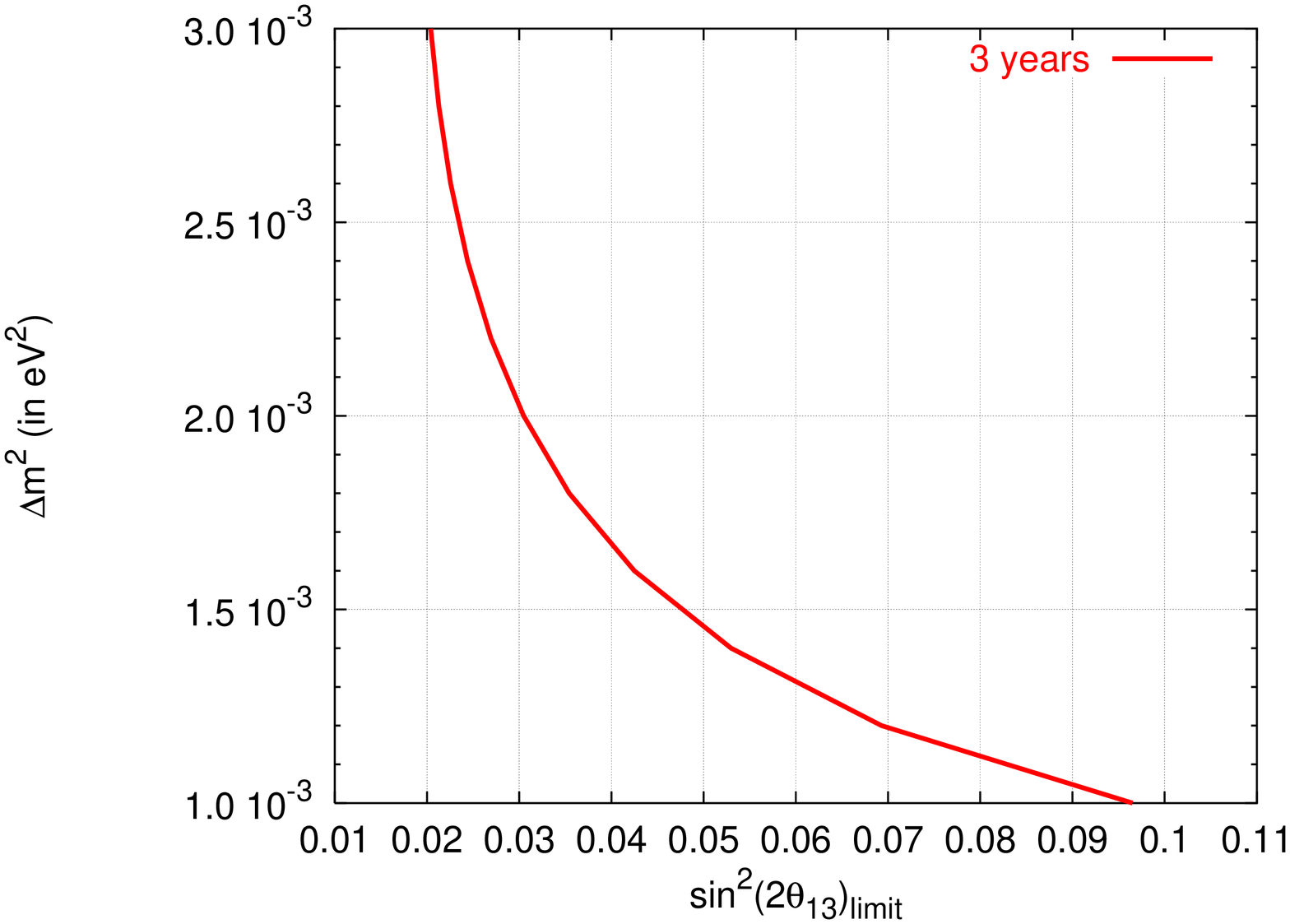}
  \caption[Double-CHOOZ sensitivity limit at 90~\% C.L. (for 1
  d.o.f).]{Double-CHOOZ sensitivity limit at 90~\% C.L. (for 1 d.o.f).}
  \label{fig:sensdm2}
\end{center}
\end{figure}
Figure~\ref{fig:relativenorm1} displays the effect of $\sigma_{rel}$
on the sensitivity of Double-CHOOZ in the  $(\ssqtt,\adm2)$ plane.
The relative normalization influence on the \ssqtt limit as a
function of the exposure time is shown in Figure~\ref{fig:relativenorm2}. 
\begin{figure}[h!]
\begin{center}
\includegraphics[angle = -90 , width=0.75\textwidth]{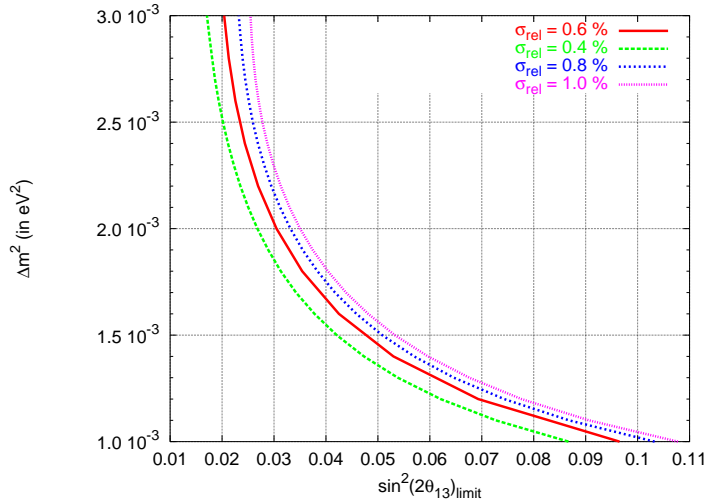}
  \caption[Influence of the relative normalization uncertainty on the
  \ssqtt limit in the $(\ssqtt,\dmgui)$ plane in the case of no oscillations]%
{Influence of the relative normalization uncertainty on the \ssqtt limit in the
    $(\ssqtt,\dmgui)$ plane in the case of no oscillations 
    (for three years of operation).}
  \label{fig:relativenorm1}
\end{center}
\end{figure}
\begin{figure}[h!]
\begin{center}
\includegraphics[angle = -90 , width=0.75\textwidth]{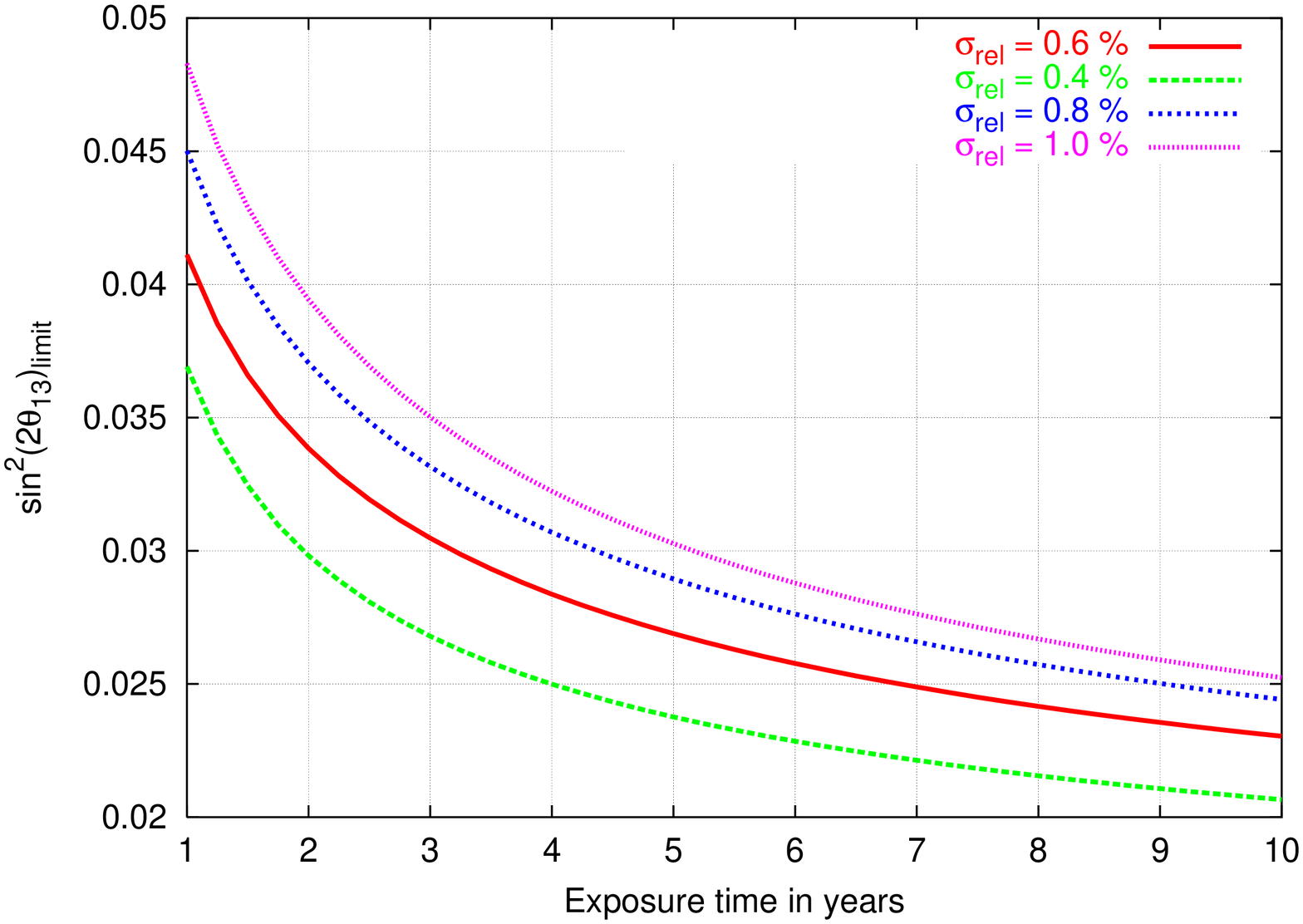}
  \caption[Influence of the relative normalization uncertainty on the \ssqtt limit as a
  function of the exposure time in the case of no oscillations.]
{Influence of the relative normalization uncertainty on the \ssqtt limit as a
  function of the exposure time (in years) in the case of no oscillations.}
  \label{fig:relativenorm2}
\end{center}
\end{figure}
\subsection{Comparison of Double-CHOOZ and the T2K
  sensitivities}
\label{sec:2choozjparklim}
We compute both the Double-CHOOZ and the T2K
sensitivities,  in the $\s2t13$-$\delta$ plane, for 
three dates: January~2009, January~2011, and January~2015.
We assume that the Double-CHOOZ experiment will start to take data
with two detectors on January~2008, while the T2K
experiment will start exactly two years later, on January~2010, 
with the nominal beam intensity (since the T2K neutrino line is 
expected to be completed within the year 2009, we assume that the 
accelerator commissioning will be finished by the end of~2009 
\cite{noon04jparcbeam, noon04jparcphysic}). 
For the computation of the Double-CHOOZ sensitivity we assume here a 
relative normalization error of 0.6~\% for both detectors. 
The correlated backgrounds considered here amount to 1.5~\% of the signal for
both the near and far detectors. Several background components of
known shape have been included (proton recoil, accidental,
spallation, see Chapter~\ref{sec:backgrounds}). An additional
uncorrelated background component of 0.5~\% is also considered
here. All backgrounds are supposed to be known with a 50~\% error.
Details of the analysis procedure are given in~\cite{huberreactor2003,
  theta13globalana}. 
For the simulation of the T2K experiment, the experimental 
parameters are taken from~\cite{jparc1, jparc2}.
We used nominal  1~year and  5~year running times for T2K, and 1, 
3, and 7~years for the reactor setup (with 20,000~events/year). 
We compute the two-dimensional allowed fit regions (i.e.,
the parameters on the axes are the fitted parameters, in the
$\s2t13$-$\delta$ plane) 
for three dates: January 2009, January 2011,
and January~2015. The curves for T2K include all correlations and 
degeneracies and are obtained as projections of the fit manifolds onto 
the $\s2t13$-$\delta$ plane~\cite{huberreactor2003, theta13globalana}.
\begin{figure}[h!]
\begin{center}
\includegraphics[width=\textwidth]{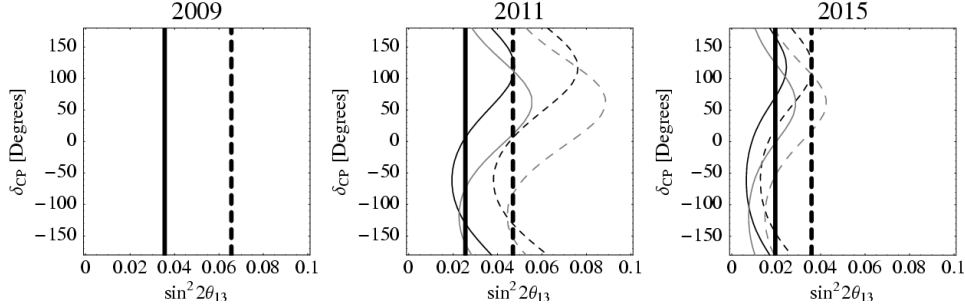}
\caption[Limit at 90~\% C.L. in the  $\ssqtt$-$\delta$ plane
for Double-CHOOZ and T2K]{Limit at 90~\% C.L. in the  $\s2t13$-$\delta$
plane
for Double-CHOOZ and T2K~\cite{huberreactor2003, theta13globalana}.
The following oscillation parameters have been used: 
$\Delta{m}_{31}^2=2 \cdot 10^{-3}~\text{eV}^2$, 
$\Delta{m}_{21}^2=7 \cdot 10^{-5}~\text{eV}^2$,
$\sin^2(2\theta_{23})=1.0$, $\sin^2(2\theta_{12})=0.8$, 
and $\ssqtt=0$.
We have considered 1 d.o.f for the analysis of the Double-CHOOZ experiment,
but 2 d.o.f. for the analysis of T2K that is sensitive to both 
\ssqtt and $\delta$ simultaneously.
90~\% C.L. intervals are shown with solid lines, and 3$\sigma$
  intervals are displayed with dashed lines. The  thick curves 
describe the Double-CHOOZ setup, and the thin curves the T2K
  experiment, with black curves for best-fit solution, and gray curves
  for the $\text{sgn}(\dmgui_{31})$-degeneracy.}
\label{fig:doublechoozjparclim}
\end{center}
\end{figure}
\section{Discovery potential}
\label{sec:discovery}
\subsection{Impact of the errors on the discovery potential}
\begin{figure}[hb]
\begin{center}
\includegraphics[angle = -90 ,width=0.9\textwidth]{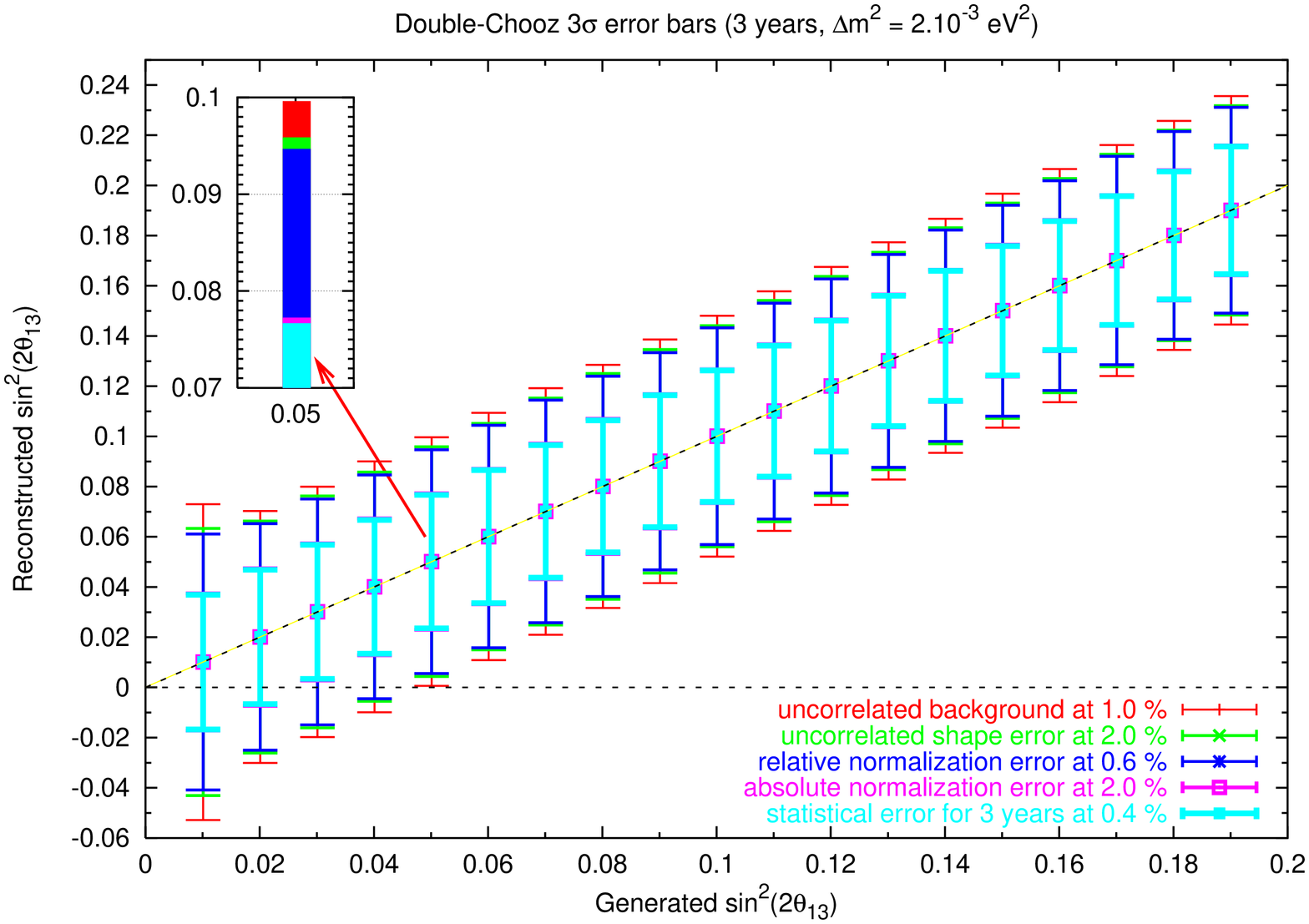}
\caption[Statistical and systematic errors contributions to \ssqtt
  measurement ($\dmgui_{31}=2.0~10^{-3}~\text{eV}^2$)]%
{Statistical and systematic errors contributions to \ssqtt measurement.  
We assumed here SK-I analysis best fit value
$\dmgui_{31}=2.0~10^{-3}~\text{eV}^2$, 3 years of data taking for Double-CHOOZ
with 64~\% (expecting around 58,000~events in the case of no oscillations)
of efficiency in the far detector and 32~\% in the near one. We also set the
systematic errors to the standard ones: the absolute normalization to
2~\%, the relative to 0.6~\%, the shape uncertainty to 2~\% and the
background to 1~\%. The different error intervals are plotted at
with a 3~$\sigma$ confidence level. We see here that the discovery potential limit of
Double-CHOOZ to detect a non-vanishing value of \ssqtt is around
0.05. We also see here that struggling harder than the level of 0.6~\% on 
the relative normalization could lower this discovery potential
  limit. \label{fig:doublechoozdiscovery1}}
\includegraphics[angle = -90 ,
width=0.9\textwidth]{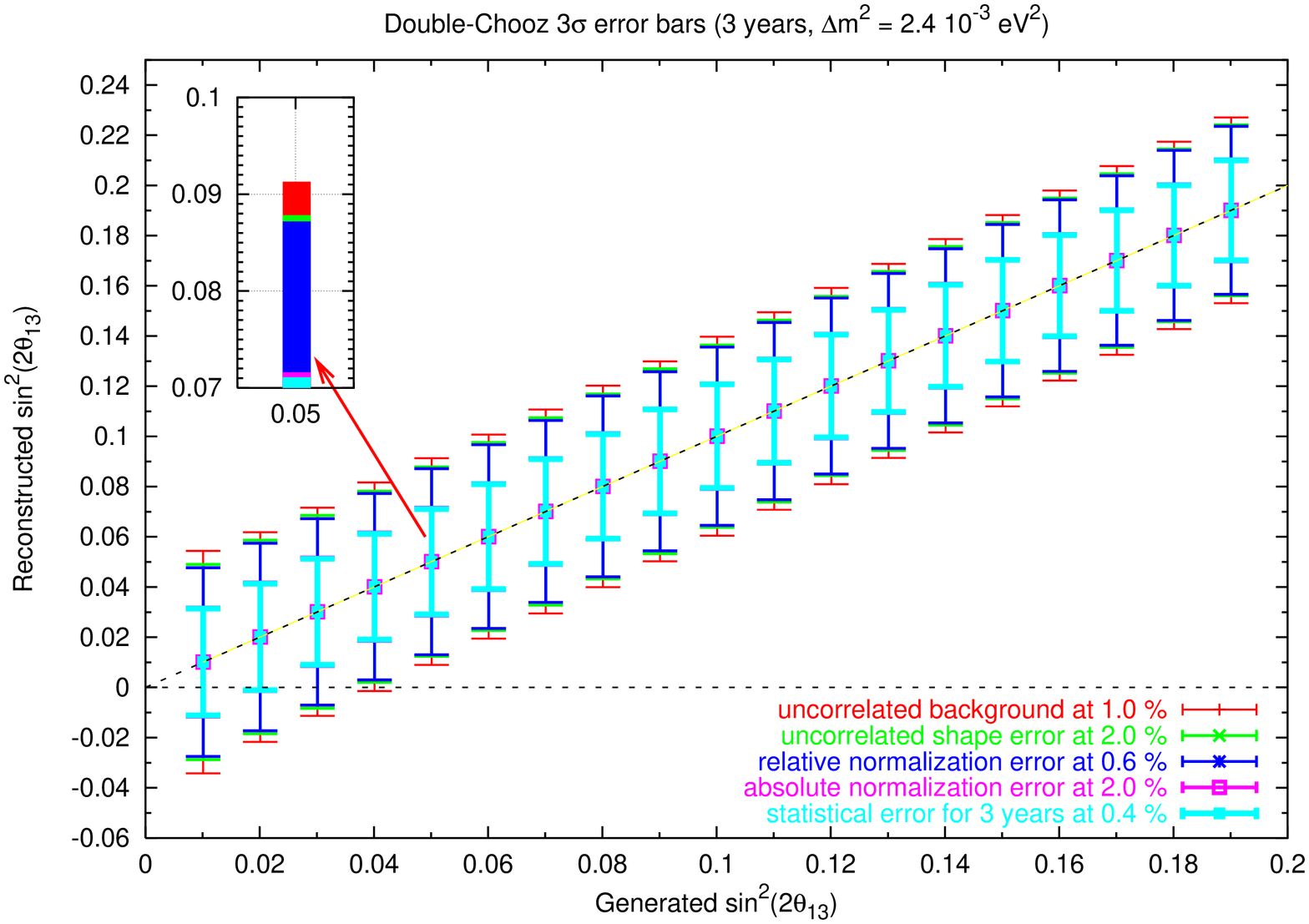}
\caption[Statistical and systematic errors contributions to \ssqtt
  measurement ($\dmgui_{31}=2.4~10^{-3}~\text{eV}^2$)]{Same as
Figure~\ref{fig:doublechoozdiscovery1} but for
  $\dmgui_{31}=2.4~10^{-3}~\text{eV}^2$.
\label{fig:doublechoozdiscovery2}}
\end{center}
\end{figure}

The 3$\sigma$ discovery potential of Double-CHOOZ is displayed on
Figures~\ref{fig:doublechoozdiscovery1} and
\ref{fig:doublechoozdiscovery2}, for respectively 
$\Delta{m}_{31}^2 = 2.0~\text{and} 2.4\cdot10^{-3}~\text{eV}^2$. 
In the first case, a non-vanishing value of $\s2t13=0.05$    
could be detected at 3~$\sigma$ after three years of
data taking. For the second case, this value becomes $\s2t13=0.04$.
\subsection{Comparison of Double-CHOOZ and the T2K discovery potential}
\begin{figure}[hb]
  \begin{center}
\includegraphics[width=\textwidth]{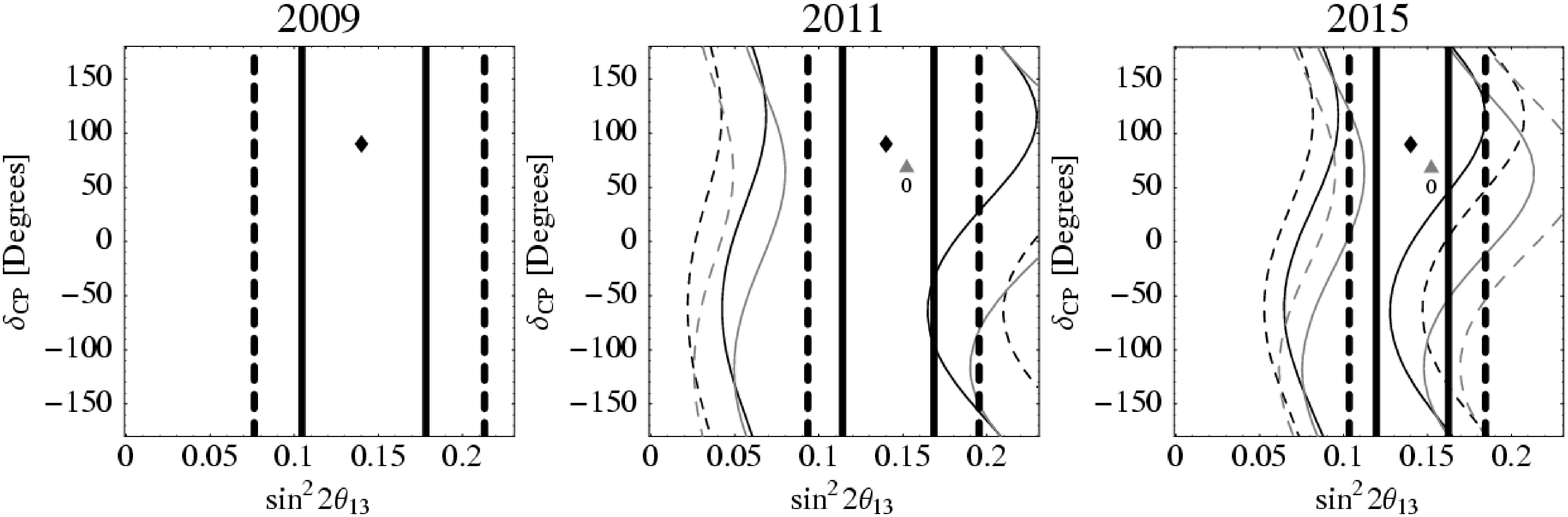}
    \caption[Measurement of $\ssqtt$ and $\delta$ with
    Double-CHOOZ and T2K experiments ($\ssqtt = 0.14$)]%
{Measurement of \ssqtt and $\delta$ with
  Double-CHOOZ and T2K~\cite{huberreactor2003,theta13globalana}.
  The following oscillation parameters have been used: 
  $\Delta m_{31}^2=2 \cdot 10^{-3} \, \mathrm{eV}^2$, 
  $\Delta m_{21}^2=7 \cdot 10^{-5} \, \mathrm{eV}^2$,
  $\sin^2(2\theta_{23})=1.0$, $\sin^2(2\theta_{12}) = 0.8$. 
  The $\theta_{13}$ mixing angle was generated as $\ssqtt=0.14$ 
  and the CP-$\delta$ phase has been fixed at  $\delta=\pi/2$.
  We considered 1 d.o.f. for the analysis of the Double-CHOOZ experiment,
  but 2 d.o.f. for the analysis of T2K that is sensitive to both 
  \ssqtt \& $\delta$ simultaneously.
  90~\% C.L. interval are shown with solid lines, and 3$\sigma$
  intervals are displayed with dashed lines. The thick curves 
  describe the Double-CHOOZ setup, and the thin curves the T2K
  experiment, with black curves for best-fit solution, and gray curves
  for the $\text{sgn}(\Delta m_{31}^2)$-degeneracy. The minimum
  $\chi^2$ is drawn at marked points.\label{fig:doublechoozjparcprec1}}
\includegraphics[width=\textwidth]{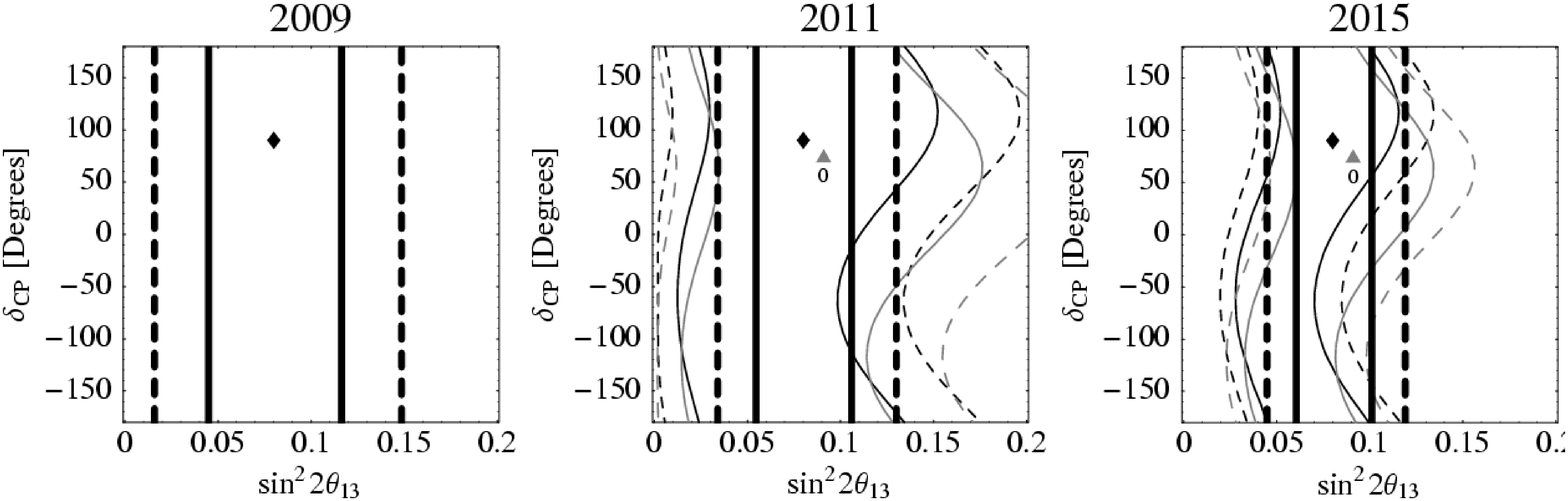}
\caption[Measurement of \ssqtt and $\delta$ with
Double-CHOOZ and T2K experiments ($\ssqtt = 0.08$)]%
{Same as Figure~\ref{fig:doublechoozjparcprec1}, but for the
$\theta_{13}$ mixing angle was generated at $\ssqtt = 0.08$. 
\label{fig:doublechoozjparcprec2}}
\includegraphics[width=\textwidth]{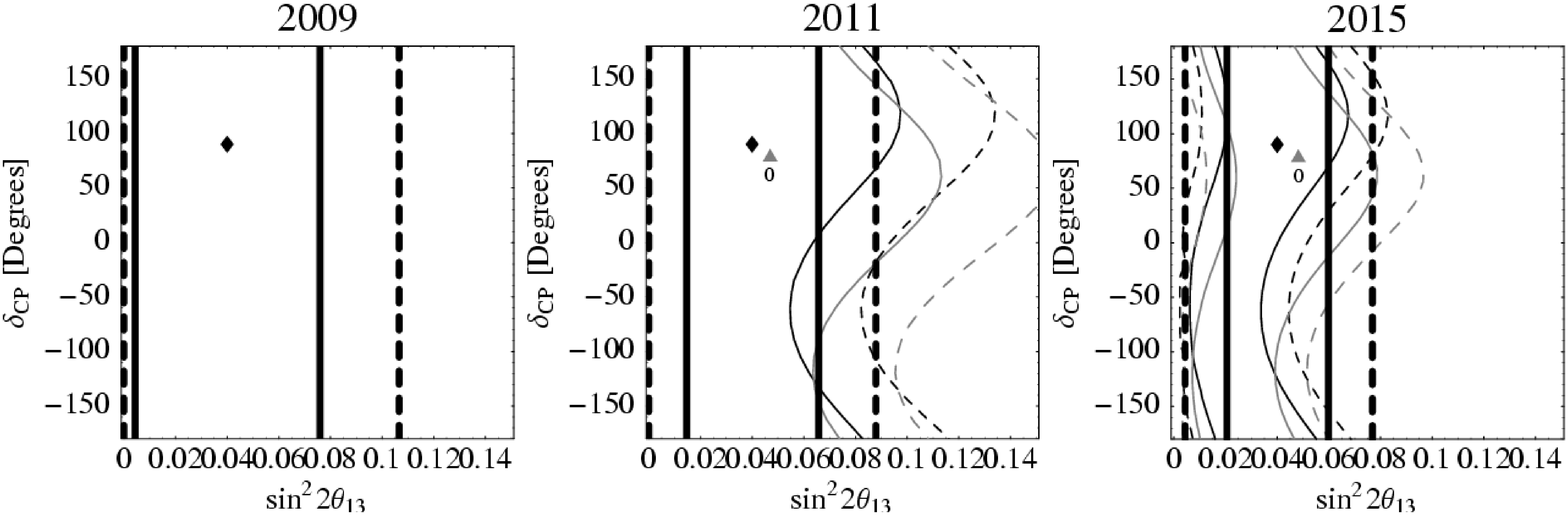}
\caption[Measurement of \ssqtt and $\delta$ with
  Double-CHOOZ and T2K experiments ($\ssqtt = 0.04$)]%
{Same as Figure~\ref{fig:doublechoozjparcprec1}, but for the
$\theta_{13}$ mixing angle was generated at $\ssqtt = 0.04$.
\label{fig:doublechoozjparcprec3}}
\end{center}
\end{figure}

The computation is done as presented in Section~\ref{sec:2choozjparklim}, for both
the Double-CHOOZ and the T2K experiments taken at three dates: 
January~2009, January~2011, and January~2015. To investigate the discovery
potential of both experiments, we used three benchmark values 
$\ssqtt=0.14$,~0.08,~0.04. Results are presented respectively  
in the $\ssqtt$-$\delta$ plan in
Figures~\ref{fig:doublechoozjparcprec1},~\ref{fig:doublechoozjparcprec2}
and~\ref{fig:doublechoozjparcprec3}.
%
%

%
\cleardoublepage
\appendix
\renewcommand{\thepage}{\Alph{chapter}-\arabic{page}}
\cleardoublepage
\chapter{$\nuebar$ and safeguards applications}
\label{sec:IAEA}
The International Atomic Energy Agency (IAEA) is the United Nations
agency in charge of the development of peaceful use of atomic energy~\cite{IAEA}.
 In particular IAEA is the verification authority of the Treaty on the 
Non-Proliferation of Nuclear Weapons (NPT). To do that job
inspections of civil nuclear installations and related facilities 
under safeguards agreements are made in more than 140~states.
IAEA use many different tools for these verifications, like neutron
monitors, gamma spectroscopy, but also bookkeeping of the isotopic 
composition at the fuel element level before and after their use in 
the nuclear power station. In particular it verifies that weapon-origin 
and other fissile materials that Russia and USA have released from
their defense programs are used for civil applications. \\

The existence of a $\nuebar$ signal sensitive to the power and
isotopic composition of a reactor core could provide a mean to
address certain safeguards applications. Thus the IAEA very recently
asked member states to make a feasibility study to determine
whether antineutrino detection methods might provide practical
safeguards tools for selected applications.  If this method proves to
be useful, IAEA has the power to decide that any new nuclear power
plant to be built has to include an $\nuebar$ monitor. \\

The high penetration power of antineutrinos and the detection
capability might provide a mean to make ``remote''and non-intrusive
measurements of plutonium content in reactors and in large inventories
of spent fuel. The antineutrino flux and energy spectrum depend upon
the thermal power and the fissile isotopic composition of the reactor
fuel.  Because the antineutrino signal from the reactor decreases as
the  square of the distance from the reactor to the detector the
"remote"  measurement is really only practical at distances of a few
tens of meters if one is constrained to ``small'' detectors of the order
of few cubic meters in size. 
Based on predicted and observed $\beta$~spectra, the number of
$\nuebar$  per fission from $^{239}$Pu is known to be less than the
number from $^{235}$U. This variation has been directly measured in
 reactor antineutrino experiments. This may offer a mean to monitor
 changes in the relative amounts of $^{235}$U and $^{239}$Pu in the core and in
 freshly discharged spent fuel. If made in conjunction with accurate
 independent measurements of the thermal power (including the ambient
 reactor temperature and the flow rate of cooling water), antineutrino
 measurements might provide an estimate of the isotopic composition of
 the core, in particular plutonium inventories. The shape of the
 antineutrino  spectrum can provide additional information about core
 fissile  isotopic composition. \\
\begin{table}[h]
\begin{center}
\begin{tabular}{lrrr}
\hline
                    & $^{235}$U & $^{239}$Pu & $^{241}$Pu \\
\hline
$\nuebar$/fission   & 6.2     & 5.6      & 6.4\\
End point (MeV)     & 9.0     & 7.4      & 9.3\\
\hline
\end{tabular}
\caption{\label{tab:aieaupu} 
Number of $\nuebar$ emitted per fission and end points of U and Pu 
fissile isotopes.}
\end{center}
\end{table}
In order to determine the feasibility of antineutrino detection for
safeguards applications, a series of scenarios involving antineutrino
 detectors should be defined, both for reactors and for spent fuel
 inventories. The effectiveness, sensitivity, and possible
 vulnerabilities  of antineutrino detection should be examined for
 these  scenarios. 
For the IAEA, the proposed feasibility study should seek to establish or
refute the utility of antineutrino detection methods as a new
safeguards   tool, and serve as a guide for future efforts. Additional
lab tests and theoretical calculations should also be performed to
more  precisely estimate the underlying $\beta$~spectra of plutonium and
uranium fission products, especially at low energies, corresponding
to the most energetic antineutrinos.  \\

The appropriate starting point for this scenario is a representative
PWR. For this reactor type, simulations of the evolution of the
antineutrino  flux and spectrum over time should be provided, and the
required  precision of the antineutrino detector and independent power
measurements should be estimated.
In that respect the measurement performed by the Double-CHOOZ 
experiment with its near detector, as it is explained in the proposal, 
will constitute the most precise determination of the antineutrinos 
emitted by a PWR. In particular, the follow-up of the spectrum and rate 
after refueling with fresh $^{235}$U, would allow a precision study of the 
correlation between plutonium content and the measured spectrum. If it 
is possible in addition to have a detailed follow-up of the evolution 
of the fuel burn-up, by the use of fission chambers, the data gathered 
by these experiments will constitute an excellent experimental basis for 
the above feasibility studies of potential monitoring and for 
bench-marking fuel management codes.
This measurement will help to meet another important point of the IAEA concern,
 linked to the verification of provisions of the US-Russian Plutonium
Management and Disposition Agreement (PMDA). 
This  agreement concerns MOX fuel made using weapon origin plutonium. \\

Verifying core burn up while the reactors
are operating would provide a mean to determine whether or not
the  disposition criteria have been met. 
From the present knowledge of the antineutrino spectrum emitted by the
fission products, we see that the most energetic part offers the best
possibility to disentangle fission from $^{235}$U and $^{239}$Pu. Unfortunately
the   present uncertainty in that region of energy is rather large,
due to  the difficulties of measuring the corresponding low
energy~$\beta^{-}$.

Thus, in relation to this feasibility studies, new measurements of
the $\beta$~spectrum for the various fissile elements are mandatory. A
group of  nuclear physicists has developed tools, in the frame of
MiniINCA collaboration~\cite{miniinca}, which can be modified to perform these
measurements at ILL.  
Needless to say that a more precise knowledge of
the antineutrinos  emitted in the reactor core would also benefit
the  physics measurements of $\t13$.
The overall IAEA feasibility studies are larger than the topics briefly
described  above. It is also of interest to study other present
reactor types,  like BWRs, FBRs, and possibly CANDU reactors. Future
reactors  (e.g., PBMRs, Gen IV reactors, accelerator-driven
sub-critical assemblies  for transmutation), especially reactors using
carbide, nitride, metal or molten salt fuels must also be considered.
IAEA seeks also to the possibility of monitoring large spent-fuel
elements.  For this application, the likelihood is that antineutrino
detectors  could only make measurements on large quantities of
$\beta$~emitters,  e.g., several cores worth of spent fuel. In the time
of the  experiment the discharge of parts of the core will happen and
the  Double-CHOOZ experiment will quantify the sensitivity of such monitoring.
More generally the techniques developed for the detection of
antineutrinos  could be applied for the monitoring of nuclear
activities at  the level of a country. For example a KamLAND type
detector~\cite{Eguchi:2002dm} deeply submerged off the coast of the
country, would offer the  sensitivity to detect a new underground
reactor located at several hundreds of kilometers. In that respect, 
the progress in term of  detecting medias (Gd doped liquid
scintillators) would be greatly helpful.
\cleardoublepage
\chapter[Nuclear reactor $\beta$ spectra]{Nuclear reactor $\mathbf{\beta}$ spectra}
\label{sec:sphn}
New measurements of the $\beta$ spectrum for various fissile elements
present in a nuclear reactor will be very important for the Double-CHOOZ
experiment to understand the physics at the near detector. Of course,
it is less important for the oscillation analysis, since the absolute
normalization error on the $\nuebar$ flux is absorbed if two detectors
are used simultaneously at different baselines. These new integral
measurements deal with a complete characterization of the $\beta$
spectrum produce in the fuel element by taking into  account the
evolution  of the fuel. This information is
important to characterize the antineutrino spectra at the Double-CHOOZ
experiment but is also unavoidable for the feasibility studies of using
antineutrino detection methods as a new safeguards tool. 

In the frame of the Mini-INCA project~\cite{miniinca}, the group has developed a set
of experimental  tools to perform quasi online $\alpha$- and
$\gamma$-spectroscopy  analyzes on irradiated isotopes and to monitor online
the neutron flux in the high flux reactor of the ILL reactor. It has also
developed  competences on the Monte-Carlo simulations of complex
systems and in particular nuclear reactors. These competences will be
used to  provide to the community a set of integral $\beta$  energy
spectra relevant  for the Double-CHOOZ experiment and for safeguards
studies  and to understand and monitor all the fluctuations in the
antineutrino spectra originated from the reactor source.
\section[New $\beta$ energy spectra measurements at ILL]
{New $\mathbf{\beta}$ energy spectra measurements at ILL}
The $\alpha$ and $\gamma$ spectroscopy station, connected to an
irradiation  channel of the ILL reactor, offer the possibility to
perform irradiations in a quasi thermal neutron flux up to 20 times
the  nominal value in a PWR. This irradiation can be followed by
measurements and repeated as many time as needed. It offers then the
unique possibility to characterize the evolution of the beta spectrum
as a function of the irradiation time and the irradiation cooling.
The expected modification of the $\beta$ spectrum as a function of the
irradiation time is connected to the transmutation induced by neutron
capture  of the fissile and fission fragment elements. It is thus
related to the natural evolution of the spent-fuel in the reactor. 
The modification of the $\beta$ spectrum as a function of the cooling
time is  connected to the decay chain of the fission products and
is then  a mean to select the emitted fragments by their time of live.
This information is important because long-lived fission fragments
accumulate  in the core and after few days mainly contribute to the
low  energy part of the antineutrino-spectra. We propose to modify the
spectroscopy station by adding a large dynamic
$\beta^{-}$~spectrometer  and to measure  the $\beta$~spectra for $^{235}$U,
$^{239}$Pu, $^{241}$Pu  and $^{243}$Cm for different irradiation and
cooling times. Due to the mechanical transfer of the sample from the
irradiation  spot to the measurement station an irreducible delay time
of 30 mn  is imposed leading to the loss of short-live fragments.
To characterize the $\beta$~prompt emissions online measurements will
be  done on a neutron guide where cold neutrons are available.
\section{Reactivity monitoring}
Micro-fission chambers developed for high neutron fluxes are used in
core in the ILL reactor.  They provide very precise neutron flux
measurements  and allow to monitor in line the reactivity fluctuations
of the core.  Due to their small dimensions (4~mm in diameter and 4~cm
in length)  and the low fissile deposit, they should allow to measure
very  precisely the gravity center of the core, with a negligible flux
perturbation, if placed out core of the Chooz reactor.
\section{Double-CHOOZ reactor core simulation and follow-up}
By the mean of Monte-Carlo and deterministic codes developed for neutron
flux calculation and evolution at ILL and for various type of
transmutation scenario,  we propose to model the complete history
of Chooz reactor core to study the sensitivity of the neutrino
spectrum to the isotopic composition and fuel burn up. 
\cleardoublepage
\chapter{Some numbers from the CHOOZ experiment}
\label{sec:choozfirst}
The CHOOZ experiment \cite{chooz1, chooz2, chooz3, choozlast} was located close to the
CHOOZ nuclear power plant, in  the North of France, 10~km from the
Belgian border. The power plant consists of
 two twin pressurized water reactors (PWR), the first of a series of the newly developed
 N4 PWR generation in France~\cite{CEAelecnuc}. The thermal power of each reactor is
 4.25~GW$_\text{e}$ (1.3~GW$_\text{e}$). 
These reactors started respectively in May and August 1997, just after
 the start of the data taking of the CHOOZ detector (April 1997). This opportunity
 allowed a measurement of the reactor-off background, and a separation
 of individual reactors contributions.

 The detector was located in an underground laboratory about 1~km from the
 neutrino source. The 300~m.w.e. rock overburden reduced the external cosmic ray
 muon flux, by a factor of about 300, to a value of 0.4~m$^{-2} \, \text{s}^{-1}$. 
This was  the main criterion to select this site. Indeed, the
 previous experiment at the Bugey reactor power plant \cite{Bugey} showed the
 requirement of reducing by two orders of magnitude the flux of fast neutrons
 produced by muon-induced nuclear spallations in the material surrounding the
 detector. 
 The neutron flux was measured at energies greater than 8~MeV 
 and found to be about 1/day, in good agreement with the prediction.

 The detector envelope consisted of a cylindrical steel vessel, 5.5~m diameter and
 5.5~m height.  The vessel was placed in a pit (7~m diameter and
7~m deep), and was surrounded  by 75~cm of low activity sand. It was
composed  of three concentric regions, from inside to outside:
\begin{itemize}
\item{a central 5 tons target in a transparent Plexiglas container filled with a
 0.09~\% Gd-loaded scintillator}
\item{an intermediate 70~cm thick region, filled with non-loaded scintillator and
 used to protect the target from PMT radioactivity and to contain the
gammas from  neutron capture on Gd. These 2 regions were viewed by 192~PMTs}
\item{an outer veto, filled with the same scintillator.}
\end{itemize}
 
 The scintillator showed a degradation of the transparency over time, which
 resulted in a decrease of the light yield (live time around 250
 days). The event position was reconstructed by fitting the charge
 balance, with a  typical precision of 10~cm for the positron and
 20~cm for the neutron.  
Source and laser calibrations found that due to the small size of the
 detector the time reconstruction was less precise than expected. 
 The reconstruction became more difficult
 when the event was located near the PMTs, due to the $1/r^2$ divergence
 of the  light collected (see Figure~31 of \cite{choozlast}).
   
 The final event selection used the following cuts:
 \begin{itemize}
\item{positron energy smaller than 8~MeV (only 0.05~\% of the positrons have a
 higher energy)}
\item{neutron energy between 6 and 12~MeV}
\item{distance from the PMT support structure larger than 30~cm for both positron and
 neutron}
\item{distance between positron and neutron smaller than 100~cm}
\item{low particles multiplicity: when a third particle is detected in the
  time window between the positron and neutron candidates, a complicated
  cut must be applied (see 8.7 of \cite{choozlast})}.
\end{itemize}
%
%

The neutron capture on Gd is identified by a 6~MeV  cut on the
 total energy emitted. This cut induce a systematic error of 0.4~\%, 
due to the poor knowledge of the emission spectrum of the gammas
 released after the neutron capture. 

The scintillating buffer around the target was important
 enough to reduce  the gammas escape. This cut was calibrated with a
 neutron source.
 The 3 cuts on the distances were rather difficult to calibrate, due to the
 the reconstruction problems described above. This created a tail
 of badly reconstructed  events,  which was very difficult to simulate 
(0.4~\% systematic  error on the positron-neutron distance cut). 
 The positron threshold was carefully calibrated, as shown in Figure~39 of \cite{choozlast}.
 The value of the threshold depends upon the position of the event, due to the 
 variation of solid angle and to the shadow of some mechanical pieces such as
 the neck of the detector (0.8~\% systematic error).
 The time cut relied on Monte-Carlo simulation. 
 The corresponding systematic error was estimated  to be 0.4~\%.
 The final result was given as the ratio of the number of measured events versus the
 number of expected events, averaged on the energy spectrum. It was
 found to be:
\begin{center}
           R = 1.0 $\pm$ 2.8~\% (stat) $\pm$ 2.7~\% (sys).
\end{center}
Two components were identified in the background:
  \begin{itemize}
 \item{Correlated events: which had a flat distribution for energies greater than
 8~MeV, and were due to the recoil protons from fast spallation neutrons. It
 was extrapolated to 1~event/day.}
\item{Accidental events: which were obtained from the measure of the singles rates.}
\end{itemize}
The total noise was measured during the reactor-off, and by extrapolating the
 signal versus power  straight line (see Figure~49 of \cite{choozlast}). It is in good
 agreement  with the sum of the correlated and accidental components. 
These numbers have to be compared to a signal of 26~events/day at full
 reactor power. 
The systematic error was due mainly to the reactor uncertainties~(2~\%), 
 the detector efficiency~(1.5~\%), and to the normalisation of the
 detector dominated by the error on the proton number from the H/C ratio in the liquid~(0.8~\%).
The resulting  exclusion plot is shown in Figure~58 of \cite{choozlast}. The corresponding
 limit on $\s2t13$ is 0.14 for $\Dm2 = 2.6 \, 10^{-3}~\text{eV}^2$, and 0.2 for
 $\Dm2= 2.0 \, 10^{-3}~\text{eV}^2$. This limit disappears for $\Dm2 < 0.8
 \, 10^{-3}~\text{eV}^2$, due to the $\sim$1~km distance between the cores and the CHOOZ detector.   
%
%

%
\renewcommand{\thepage}{\roman{page}}
\pagestyle{empty}
\newpage
\vspace*{\stretch{1}}
\begin{center}
{\Large \bf Acknowledgments}
\end{center}
\vspace{5mm}
\'Electricit\'e de France (E.D.F.) is contributing to this project and
studying the possibility of the near detector laboratory construction.
The local authorities (Mairie de Chooz and Conseil g\'en\'eral des
Ardennes) have been supporting this project.
Special thanks are due to F.~Bobisut and  B.~Vallage for their very careful
reviews of the experiment. 
Warm thanks are due to B.~Svodoba and our american colleagues for fruitflul
discussions on the experimental issues. 
We thank C.~Bemporad, J.~Bouchez, C.~Cavata, Y.~D\'eclais,
M.~Froissart, J.~Mallet, and S.~Petcov for very useful discussions
on the Double-CHOOZ experiment and neutrino physics.
Finally, we would like to ackowledge the reactor working group members for the high quality
LENE workshops, and M.~Goodman for editing the useful Reactor Neutrino
White Paper. 
\vspace*{\stretch{1}}
\newpage
\listoftables
\newpage
\listoffigures
%
%
\newpage

%
\end{document}